\documentclass[aps,pra,twocolumn,nobibnotes]{revtex4}
\usepackage[utf8]{inputenc}
\usepackage[T1]{fontenc}
\usepackage{bm}
\usepackage{ulem}
\usepackage{epsfig}
\usepackage{graphicx}
\usepackage{amssymb,amsmath,amsbsy,amsgen,amsfonts}
\usepackage{dcolumn}
\usepackage{amsthm}
\usepackage{mathtools}
\usepackage{mathrsfs}
\usepackage{latexsym}
\usepackage{array}
\usepackage{color}
\usepackage{amstext}
\usepackage{multirow}
\usepackage{bbold}
\usepackage{color}
\usepackage{xcolor}
\usepackage{listings}
\usepackage{tikz}
\usepackage{lipsum}
\usepackage{textcomp}
\usepackage{qcircuit}

\makeatletter
\allowdisplaybreaks[1]

\makeatletter
\let\old@makecaption=\@makecaption
\usepackage{caption}
\usepackage{subcaption}
\let\@makecaption=\old@makecaption
\makeatother

\usepackage{epstopdf} %for texshop on mac

\DeclareSymbolFont{symbols}{OMS}{cmsy}{m}{n}

\newcommand{\ket}[1]{\ensuremath{\left|#1\right>}}
\newcommand{\bra}[1]{\ensuremath{\left<#1\right|}}

\newcommand{\secref}[1]{Sec.~\ref{#1}}
\newcommand{\appendref}[1]{Appendix~\ref{#1}}
\newcommand{\equatref}[1]{Eq.~(\ref{#1})}
\newcommand{\figref}[1]{Fig.~\ref{#1}}
\newcommand{\tabref}[1]{Table~\ref{#1}}

\begin{document}

\title{Implementation of single-qubit measurement-based t-designs using IBM processors}

\author{Conrad Strydom}
\email{conradstryd@gmail.com}
\author{Mark Tame}
\affiliation{Department of Physics, Stellenbosch University, Matieland 7602, South Africa}

\begin{abstract}
Random unitary matrices sampled from the uniform Haar ensemble have a number of important applications both in cryptography and in the simulation of a variety of fundamental physical systems.  Since the Haar ensemble is very expensive to sample, pseudorandom ensembles in the form of $t$-designs are frequently used as an efficient substitute, and are sufficient for most applications.  We investigate $t$-designs generated using a measurement-based approach on superconducting quantum computers.  In particular, we implemented an exact single-qubit 3-design on IBM quantum processors by performing measurements on a 6-qubit graph state.  By analysing channel tomography results, we were able to show that the ensemble of unitaries realised was a 1-design, but not a 2-design or a 3-design under the test conditions set, which we show to be a result of depolarising noise during the measurement-based process.  We obtained improved results for the 2-design test by implementing an approximate 2-design, in which measurements were performed on a smaller 5-qubit graph state, but the test still did not pass for all states.  This suggests that the practical realisation of measurement-based $t$-designs on superconducting quantum computers will require further work on the reduction of depolarising noise in these devices.
\end{abstract}

%\pacs{}

\maketitle

%%%%%%%%%%%%%%%%%%%%%%%%%%%%
%%%%%%%%%%%%%%%%%%%%%%%%%%%%
%%%%%%%%%%%%%%%%%%%%%%%%%%%%
%%%%%%%%%%%%%%%%%%%%%%%%%%%%
\section{Introduction}\label{sec:introduction} 

Random unitary matrices have a number of important applications, which include estimating noise~\cite{application1}, realising private channels~\cite{application2}, modeling thermalisation~\cite{application3} and formulating quantum mechanical models of black holes~\cite{application4}.  However, the generation of uniformly distributed random unitaries is very resource intensive, since the resources required to sample randomly with respect to the Haar measure on $U(2^n)$, the group of unitary transformations on a $n$-qubit system, scale exponentially with $n$~\cite{inefficient}.  A $t$-design is a pseudorandom ensemble of which the statistical moments match those of the uniform Haar ensemble up to some finite order $t$. Hence, a $t$-design is by definition a $(t-1)$-design.  These $t$-designs can rarely be distinguished from the true random ensemble, and so they are often used as a substitute.  Even approximate $t$-designs are sufficient for many applications, for example approximate 1-designs can be used for encrypting quantum data~\cite{approx1}, approximate 2-designs can be used for estimating channel fidelities~\cite{approx2} and approximate 3-designs can be used for solving black-box problems~\cite{approx3}.

In random circuit constructions, $t$-designs on $n$ qubits are realised by applying gates selected randomly from a universal set to qubits from a $n$-qubit system.  Approximate $n$-qubit $t$-designs can be realised efficiently, since the resources required (the number of gates and random bits) scale polynomially with $n$ and $t$~\cite{RCC1, RCC2, RCC3, RCC4}.  Very efficient random circuit constructions for exact $n$-qubit 2-designs, where the resources required scale almost linearly with $n$, have also been devised~\cite{RCC5}.  More recently, random circuit constructions for general exact $n$-qubit $t$-designs were proposed~\cite{RCC6}.  However, they are only feasible for small systems, since the number of gates required scales exponentially with $n$ and $t$ for large $n$.  Random circuit constructions have two major disadvantages, namely that they require a source of classical randomness, which can be expensive if it needs to be reliable, and that they require the reconfiguring of physical quantum gates, which is bound to introduce noise.

A measurement-based approach~\cite{MBT1, MBT2}, inspired by measurement-based quantum computing~\cite{Raussendorf, Briegel}, avoids both these problems at the cost of additional qubits.  Measurement-based quantum computing is an alternative method to perform quantum computing, where the computation is carried out by performing single-qubit measurements on an entangled resource state instead of by applying unitary operations (or gates) as with the circuit model.  Reducing the entire computation to single-qubit measurements has the benefit of avoiding the swapping of qubits around a large circuit, which would introduce additional noise.  Measurement-based quantum computing is advantageous in physical systems, such as photonic or superconducting systems, or cold atoms, where the qubits to be entangled are spatially close to each other so that the entangled resource state can be generated efficiently.

Measurement-based $t$-designs are realised by performing a deterministic sequence of single-qubit measurements on a highly entangled graph state.  Turner and Markham present a measurement-based protocol for realising an exact single-qubit 3-design, which requires a 6-qubit graph state, and discuss measurement-based protocols for realising higher order approximate single-qubit $t$-designs, which require larger graph states~\cite{MBT1}.  Efficient approximate $n$-qubit measurement-based $t$-designs, where the number of qubits in the entangled resource state scale polynomially with $n$ and $t$, have also been found~\cite{MBT2}.  It is still unknown whether exact single-qubit measurement-based $t$-designs exist for $t > 3$ or whether exact multi-qubit measurement-based $t$-designs exist at all.

In previous experiments, multi-qubit pseudorandom ensembles, in which the expected distribution of matrix elements of unitary operators sampled from the uniform Haar ensemble is reproduced, have been realised using a nuclear magnetic resonance quantum processor~\cite{experiment1} and single-qubit 1-designs and 2-designs have been realised using photons~\cite{experiment2}.  In this paper, we implement the exact single-qubit measurement-based 3-design of Ref.~\cite{MBT1} and our own approximate single-qubit measurement-based 2-design on IBM superconducting quantum computers, accessible through their website~\cite{IBM}.  These were implemented by performing single-qubit measurements on 6-qubit and 5-qubit graph states respectively.  Since measurement errors are responsible for a significant amount of noise on IBM quantum processors, and since this noise is predominantly classical, we performed quantum readout error mitigation to improve results~\cite{QREM1}.  Both the exact 3-design implementation and the approximate 2-design implementation passed our test for a 1-design, but not for a 2-design or a 3-design.  Further investigations, presented in appendices, suggest that depolarising noise is likely what prevented these implementations from passing the test for a 2-design and a 3-design.

This paper is structured as follows.  In \secref{sec:background}, we discuss measurement-based processing using graph states and how it can be used to generate $t$-designs.  We also give an overview of the channel tomography technique used to analyse results and the quantum readout error mitigation technique used to improve results.  In \secref{sec:experiments}, we describe the implementations of the exact 3-design and approximate 2-design and present the results obtained.  Some concluding comments are given in \secref{sec:conclusion}.  Supplementary appendices follow, in which further discussion of the implementations and analysis of the results is presented.

%%%%%%%%%%%%%%%%%%%%%%%%%%%%
%%%%%%%%%%%%%%%%%%%%%%%%%%%%
%%%%%%%%%%%%%%%%%%%%%%%%%%%%
%%%%%%%%%%%%%%%%%%%%%%%%%%%%
\section{Background}\label{sec:background}

%%%%%%%%%%%%%%%%%%%%%%%%%%%%
%%%%%%%%%%%%%%%%%%%%%%%%%%%%
\subsection{Measurement-based t-designs}\label{sec:mbbackground}

Quantum graph states are a fundamental resource for measurement-based quantum computing~\cite{Raussendorf, Briegel}, and a wide range of other protocols, including quantum secret sharing~\cite{Markham, Bell}, quantum sensing~\cite{Friis, Shettell} and quantum games~\cite{Paternostro, Prevedel}.  A $n$-qubit graph state is defined in relation to a connected graph with $n$ vertices.  Such a graph state is made by preparing each qubit in the state $\ket{+}=\left(\ket{0}+\ket{1}\right)/\sqrt{2}$ and then applying controlled phase gates $CZ=\text{diag}(1,1,1,-1)$, between a pair of qubits whenever their corresponding vertices are connected by an edge in the corresponding graph~\cite{Nielsen}.  Linear cluster states are graph states corresponding to a graph in which the degree of each vertex is less than or equal to 2 (excluding rings).

\begin{figure}
    \centering
    \begin{tikzpicture}
    \draw (0, 0) circle[radius=0.25] node {1};
    \draw (1, 0) circle[radius=0.25] node {2};
    \node[] at (2, 0) {$\cdots$};
    \draw (3, 0) circle[radius=0.25] node {$n$};
    \node[] at (4.5, 0) {Step 1};
    \node[] at (0, -0.6) {$\rho_{\text{in}}$};
    \node[] at (1, -0.6) {$\ket{+}$};
    \node[] at (3, -0.6) {$\ket{+}$};
    \draw (0, -1.5) circle[radius=0.25] node {1};
    \draw (1, -1.5) circle[radius=0.25] node {2};
    \node[] at (2, -1.5) {$\cdots$};
    \draw (3, -1.5) circle[radius=0.25] node {$n$};
    \node[] at (4.5, -1.5) {Step 2};
    \draw (0.25, -1.5) -- (0.75, -1.5);
    \draw (1.25, -1.5) -- (1.5, -1.5);
    \draw (2.5, -1.5) -- (2.75, -1.5);
    \node[] at (0, -2.1) {$\phi_{1}$};
    \node[] at (1, -2.1) {$\phi_{2}$};
    \draw (0, -3) circle[radius=0.25] node {1};
    \draw (1, -3) circle[radius=0.25] node {2};
    \node[] at (2, -3) {$\cdots$};
    \draw (3, -3) circle[radius=0.25] node {$n$};
    \node[] at (4.5, -3) {Step 3};
    \node[] at (3, -3.6) {$\rho_{\text{out}}$};
    \end{tikzpicture}
    \caption{Summary of measurement-based processing with a $n$-qubit linear cluster state.  Step 1 shows the initialisation of the qubits.  Step 2 shows the entangled cluster state (after application of controlled phase gates between adjacent qubits) as well as the measurements performed on each qubit.  Step 3 shows the state resulting from these measurements.}
    \label{fig:mbp}
\end{figure}
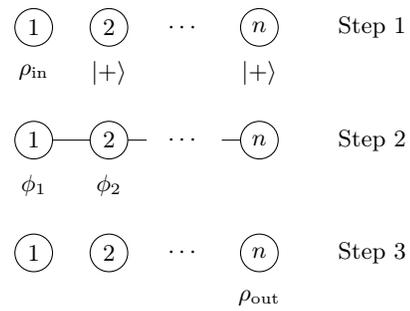

Unitary operations can be implemented by performing single-qubit measurements on linear cluster states~\cite{Nielsen}, as illustrated in \figref{fig:mbp}.  The first qubit is prepared in the input state, $\rho_{\text{in}}=\ket{\psi_{\text{in}}}\bra{\psi_{\text{in}}}$, to which the implemented unitary operation is to be applied.  The remaining qubits are prepared in the state $\ket{+}$ (Step~1), and qubits are then entangled via controlled phase gates applied between adjacent qubits (Step~2).  Each of the qubits 1 to $n-1$ are then measured in the basis $\left\{\ket{\prescript{+}{\ }{\phi}},\ket{\prescript{-}{\ }{\phi}}\right\}$ with $\ket{\prescript{\pm}{\ }{\phi}}=(\ket{0}\pm e^{-\text{i}\phi}\ket{1})/\sqrt{2}$, which we will refer to as a measurement in the $\phi$-direction, and which reduces to a measurement in the Pauli $X$-basis, $\{\ket{+},\ket{-}\}$, when $\phi=0$.  This results in the $n^{\text{th}}$ qubit being prepared in the output state, $\rho_{\text{out}}$ (Step~3).

Each measurement in the $\phi$-direction is logically equivalent to applying the random unitary,
\begin{equation}
U_m(\phi)=HZ^mR_{z}(\phi),
\label{eq:unitary}
\end{equation}
to $\ket{\psi_{\text{in}}}$ where $m\in\{0,1\}$ is the random measurement outcome, $H$ is a Hadamard, $Z$ is the Pauli $Z$ operation and $R_{z}(\phi)=e^{-\text{i}Z\phi/2}$ is a $z$-rotation by the angle $\phi$.  Hence, a $n$-qubit linear cluster implements the random unitary
\begin{equation}
U_{\boldsymbol{m}}(\boldsymbol{\phi})=U_{m_{n-1}}(\phi_{n-1})\cdots U_{m_{1}}(\phi_{1}),
\label{eq:unitarycluster}
\end{equation}
where qubit $i$ is measured in the $\phi_i$-direction with $\phi_i\in [0,\pi]$ and $m_i\in\{0,1\}$ is the result of this measurement.  Here $\boldsymbol{\phi}$ and $\boldsymbol{m}$ denote ordered lists of angles and measurement outcomes respectively.  To present lists of measurement outcomes, we use little endian encoding, that is, they are presented as bit strings in which the leftmost bit is the outcome of the measurement on qubit $n-1$ and the rightmost bit is the outcome of the measurement on qubit 1.  This is in correspondence with how measurement outcomes are presented on IBM processors.

Now, given $\boldsymbol{\phi}$, consider the ensemble of unitaries $\{p_{\boldsymbol{m}},U_{\boldsymbol{m}}(\boldsymbol{\phi})\}$ for all $\boldsymbol{m}$.  Note that by the linearity of the cluster, $p_{\boldsymbol{m}}=\frac{1}{2^{n-1}}$ for all $\boldsymbol{m}$, so that the distribution is uniform.  An ensemble of unitaries $\{p_i,U_i\}$ is an $\epsilon$-approximate $t$-design if there exists an $\epsilon$ such that for all $\rho\in\mathcal{B}(H^{\otimes t})$, with $H=\mathbb{C}^2$, we have
\begin{equation}
(1-\epsilon)\mathbb{E}^t_H(\rho)\leq\sum_{i}p_iU_i^{\otimes t}\rho\left(U_i^{\otimes t}\right)^{\dagger}\leq(1+\epsilon)\mathbb{E}^t_H(\rho),
\label{eq:def}
\end{equation}
where the matrix inequality $A\leq B$ holds if $B-A$ is positive semidefinite and
\begin{equation}
\mathbb{E}^t_H(\rho)=\int U^{\otimes t}\rho\left(U^{\otimes t}\right)^{\dagger}\,dU
\end{equation}
is the expectation of the uniform Haar ensemble.  For exact $t$-designs $\epsilon=0$.   Turner and Markham~\cite{MBT1} show that for the 6-qubit cluster state and measurement angles $\phi_1=0$, $\phi_2=\frac{\pi}{4}$, $\phi_3=\arccos{\sqrt{1/3}}$, $\phi_4=\frac{\pi}{4}$ and $\phi_5=0$, the ensemble $\{p_{\boldsymbol{m}},U_{\boldsymbol{m}}(\boldsymbol{\phi})\}$, which consists of the 32 unitaries corresponding to the 32 possible measurement outcomes $\boldsymbol{m}$, is an exact 3-design.

%%%%%%%%%%%%%%%%%%%%%%%%%%%%
%%%%%%%%%%%%%%%%%%%%%%%%%%%%
\subsection{Channel tomography}\label{sec:ctbackground}

Channel tomography can be used to determine the extent to which the predicted unitary operations are realised by cluster state implementations in an experiment.  We consider a method proposed for single-qubit channels by Nielsen and Chuang~\cite{ChannelTomo1, ChannelTomo2}.  Given any input state $\rho_{\text{in}}$, we write the output state as
\begin{equation}
\varepsilon(\rho_{\text{in}})=\sum_{mn}E_m\rho_{\text{in}} E_n^{\dagger}\chi_{mn},
\end{equation}
where $E_0=I$, $E_1=X$, $E_2=-\text{i}Y$, $E_3=Z$ and $\chi$ is a 4 by 4 matrix.  Since the operators $E_i$ are fixed, the channel is fully characterised by $\chi$, and so channel tomography amounts to determining $\chi$.  The entries of $\chi$ depend on the action of the channel on the probe input states, $\ket{0}$, $\ket{1}$, $\ket{+}=\left(\ket{0}+\ket{1}\right)/\sqrt{2}$ and $\ket{+_{y}}=\left(\ket{0}+\text{i}\ket{1}\right)/\sqrt{2}$, which is determined by performing state tomography on the output state for these input states.  In particular,
\begin{equation}
\chi=\frac{1}{4}\begin{pmatrix}I&X\\X&-I\\\end{pmatrix}\begin{pmatrix}\rho_{1}'&\rho_{2}'\\\rho_{3}'&\rho_{4}'\\\end{pmatrix}\begin{pmatrix}I&X\\X&-I\\\end{pmatrix},
\end{equation}
where the submatrices of the middle matrix are defined by
\begin{align*}
\rho_{1}' &= \varepsilon\left(\ket{0}\bra{0}\right) \\
\rho_{4}' &= \varepsilon\left(\ket{1}\bra{1}\right) \\
\rho_{2}' &= \varepsilon\left(\ket{+}\bra{+}\right)+\text{i}\varepsilon\left(\ket{+_{y}}\bra{+_{y}}\right)-\frac{1+\text{i}}{2}\left(\rho_{1}'+\rho_{4}'\right) \\
\rho_{3}' &= \varepsilon\left(\ket{+}\bra{+}\right)-\text{i}\varepsilon\left(\ket{+_{y}}\bra{+_{y}}\right)-\frac{1-\text{i}}{2}\left(\rho_{1}'+\rho_{4}'\right), \\
\end{align*}
where $\varepsilon\left(\ket{0}\bra{0}\right)$, $\varepsilon\left(\ket{1}\bra{1}\right)$, $\varepsilon\left(\ket{+}\bra{+}\right)$ and $\varepsilon\left(\ket{+_{y}}\bra{+_{y}}\right)$ denote the output states determined for the respective probe input states.  Once constructed, we can use $\chi$ to quantify the reliability with which an expected channel is realised in an experiment, by calculating the channel fidelity,
\begin{equation}
F(\chi_e, \chi_c)= \text{Tr}\left(\sqrt{\sqrt{\chi_e}\chi_c\sqrt{\chi_e}}\right),
\end{equation}
where $\chi_e$ is the $\chi$ matrix which corresponds to the expected operation of the channel, and $\chi_c$ is the $\chi$ matrix of the actual channel obtained from channel tomography.  The channel fidelity ranges from 0 to 1, where 0 indicates that the channel deviates maximally from its expected operation and 1 indicates a perfect channel.

%%%%%%%%%%%%%%%%%%%%%%%%%%%%
%%%%%%%%%%%%%%%%%%%%%%%%%%%%
\subsection{Quantum readout error mitigation}\label{sec:qrembackground}

As a result of measurement errors, actual quantum states and channels are often more similar to expected states and channels than tomography results would suggest.  Since measurement errors on IBM quantum processors are mostly classical~\cite{QREM1}, quantum readout error mitigation can be used to obtain tomography results which more accurately reflect the prepared states and channels, as has been successfully done in a number of recent studies which also involved measurements on highly entangled states on IBM quantum processors~\cite{QREM2, QREM3, QREM4}.  To mitigate readout errors in a $n$-qubit experiment (the main experiment), we first use quantum detector tomography~\cite{QREM5} to construct a $2^n$ by $2^n$ calibration matrix $\Lambda$.  The entries of $\Lambda$ are the conditional probabilities of measuring each of the $2^n$ possible combinations of computational basis states, given that a specific combination of computational basis states was prepared, for all $2^n$ possible combinations of computational basis states.  In particular, each column of $\Lambda$ contains the $2^n$ conditional probabilities associated with one of the $2^n$ prepared combinations of computational basis states.  These conditional probabilities are determined in a series of separate experiments, in which each of the $2^n$ possible combinations of computational basis states is prepared on the $n$ qubits to be used in the main experiment and sufficient computational basis measurements are done to infer the associated conditional probabilities.  Once constructed, $\Lambda$ can be used to correct classical measurement errors in the main experiment by multiplying $\Lambda^{-1}$ by $\boldsymbol{p}_{\text{exp}}$, the column vector containing the relative frequencies obtained in the main experiment.  As a result of other noise, such as gate errors, the resulting vector, $\Lambda^{-1}\boldsymbol{p}_{\text{exp}}$, may be non-physical (relative frequencies may be negative or may not sum to one).  We therefore use qiskit's built-in method~\cite{IBMqrem}, which solves a constrained optimisation problem (least squares method), to find the closest physical relative frequency vector to $\Lambda^{-1}\boldsymbol{p}_{\text{exp}}$.

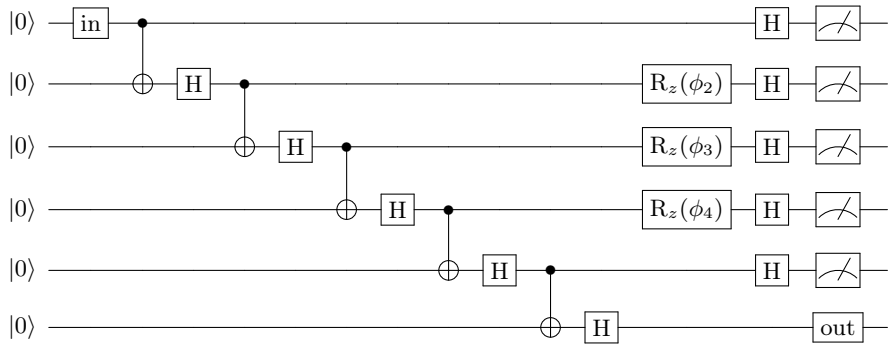
\begin{figure*}[t]
\centering
\hspace{1cm}
\Qcircuit @C=1em @R=1em {
\lstick{\ket{0}} & \gate{\text{in}} & \ctrl{1} & \qw & \qw & \qw & \qw & \qw & \qw & \qw & \qw & \qw & \qw & \gate{\text{H}} & \meter & \qw \\
\lstick{\ket{0}} & \qw & \targ & \gate{\text{H}} & \ctrl{1} & \qw & \qw & \qw & \qw & \qw & \qw & \qw & \gate{\text{R}_z(\phi_2)} & \gate{\text{H}} & \meter & \qw \\
\lstick{\ket{0}} & \qw & \qw & \qw & \targ & \gate{\text{H}} & \ctrl{1} & \qw & \qw & \qw & \qw & \qw & \gate{\text{R}_z(\phi_3)} & \gate{\text{H}} & \meter & \qw \\
\lstick{\ket{0}} & \qw & \qw & \qw & \qw & \qw & \targ & \gate{\text{H}} & \ctrl{1} & \qw & \qw & \qw & \gate{\text{R}_z(\phi_4)} & \gate{\text{H}} & \meter & \qw \\
\lstick{\ket{0}} & \qw & \qw & \qw & \qw & \qw & \qw & \qw & \targ & \gate{\text{H}} & \ctrl{1} & \qw & \qw & \gate{\text{H}} & \meter & \qw \\
\lstick{\ket{0}} & \qw & \qw & \qw & \qw & \qw & \qw & \qw & \qw & \qw & \targ & \gate{\text{H}} & \qw & \qw & \gate{\text{out}} & \qw
}
\caption{General quantum circuit for implementation of the exact measurement-based 3-design of Ref.~\cite{MBT1} on the \textit{ibmq\_toronto} quantum processor.  Here `in' represents the set of gates applied to construct the input state and `out' represents the set of gates applied and measurements done to perform tomography on the sixth qubit.  The angles for the $z$-rotation gates are $\phi_2=\phi_4=\frac{\pi}{4}$ and $\phi_3=\arccos{\sqrt{1/3}}$.}
\label{fig:qcircuit}
\end{figure*}

%%%%%%%%%%%%%%%%%%%%%%%%%%%%
%%%%%%%%%%%%%%%%%%%%%%%%%%%%
%%%%%%%%%%%%%%%%%%%%%%%%%%%%
%%%%%%%%%%%%%%%%%%%%%%%%%%%%
\section{Experiments}\label{sec:experiments}

%%%%%%%%%%%%%%%%%%%%%%%%%%%%
%%%%%%%%%%%%%%%%%%%%%%%%%%%%
\subsection{Implementation}\label{sec:implementation}

The exact measurement-based 3-design proposed by Turner and Markham~\cite{MBT1} and described in \secref{sec:mbbackground} was implemented on 6 physical qubits of the \textit{ibmq\_toronto} quantum processor.  \appendref{append:qubits3} provides more details on the \textit{ibmq\_toronto} quantum processor and how the logical qubits 1 to 6 were mapped onto the physical qubits of this processor.  The 4 channel tomography probe states were considered as input states.  For each input state, we prepared the 6-qubit linear cluster state and performed the appropriate single-qubit measurements.  Quantum state tomography was then done on the sixth qubit to construct the output state obtained for each of the 32 different measurement outcomes.  A general quantum circuit for the implementation is shown in \figref{fig:qcircuit}.

Qubits are initialised in the state $\ket{0}$ by default on IBM processors, and so the state $\ket{1}$ was prepared by applying the Pauli $X$ operation, the state $\ket{+}$ was prepared by applying a Hadamard and the state $\ket{+_y}$ was prepared by applying a Hadamard, followed by a $S$-gate.  Since IBM processors do not support controlled phase gates at the hardware level, we converted the Hadamards and controlled phase gates, needed to prepare the 6-qubit linear cluster state, into Hadamards ($H$) and controlled not ($CX$) gates using
\begin{equation}
CZ=(I\otimes H)CX(I\otimes H)
\end{equation}
and then eliminated redundant Hadamards using the fact that $H^2=I$.  Doing so ensured that redundant Hadamards, which would have increased noise in the results due to gate errors, were removed from the circuit.  Preparing the 6-qubit linear cluster state with controlled phase gates would have resulted in redundant Hadamards being introduced by the transpiler.  Since IBM processors can only perform measurements in the computational basis, $\{\ket{0}, \ket{1}\}$, measurements in the $\phi$-direction were realised by applying $R_{z}(\phi)$, followed by a Hadamard, and measuring in the computational basis.  Quantum state tomography was done using qiskit's built-in method~\cite{IBMtomo}, which uses maximum-likelihood estimation to ensure that the density matrices constructed from the data are physical (i.e. that they have a trace of 1 and are Hermitian).  The state tomography results were used to do channel tomography for each of the 32 different measurement outcomes to determine the extent to which the 32 corresponding unitary operations performed on the input states in the implementation matched the expected unitary operations.  The results are presented in \secref{sec:ctresults}.

Each of the 12 circuits needed for channel tomography (3 circuits for state tomography to construct the output state for each of the 4 input states) was run 5 times with 8000 shots on the \textit{ibmq\_toronto} quantum processor.  The counts obtained in the 5 different runs of the same circuit were then combined to obtain an effective run with 40000 shots for each of the 12 circuits.  This was done to decrease statistical noise in the tomography results.  The procedure was repeated 10 times to obtain 10 sets of channel tomography results ($\chi$ matrices) for each of the 32 different unitary operations.  Associated results, such as channel fidelities, quoted in the sections which follow, are an average of these 10 repetitions and the errors quoted are the standard deviations.

To obtain the conditional probabilities needed to construct the calibration matrix, required to mitigate readout errors in the tomography results, we prepared each of the 64 possible combinations of computational basis states on the same 6 qubits of the \textit{ibmq\_toronto} quantum processor as was used for the 3-design implementation and measured these qubits in the computational basis.  Each of the 64 circuits was run with 8000 shots and no combination of counts was done.  Readout errors in the raw tomography data (i.e. the counts obtained by running the various circuits) were then mitigated as described in \secref{sec:qrembackground}.  Results obtained using both the raw and the processed (error mitigated) data are presented in the sections which follow.

\begin{figure*}
    \centering
    \vspace{-0.5cm}
    \begin{minipage}{17.5cm}
    \begin{tikzpicture}
    \draw (0, 0) circle[radius=0.25] node {1};
    \draw (1, 0) circle[radius=0.25] node {2};
    \draw (2, 0) circle[radius=0.25] node {3};
    \draw (3, 0) circle[radius=0.25] node {4};
    \draw (4, 0) circle[radius=0.25] node {5};
    \draw (5, 0) circle[radius=0.25] node {6};
    \node[] at (0, -0.6) {$\phi_1$};
    \node[] at (1, -0.6) {$\phi_2$};
    \node[] at (2, -0.6) {$\phi_3$};
    \node[] at (3, -0.6) {$\phi_4$};
    \node[] at (4, -0.6) {$\phi_5$};
    \draw (0.25, 0) -- (0.75, 0);
    \draw (1.25, 0) -- (1.75, 0);
    \draw (2.25, 0) -- (2.75, 0);
    \draw (3.25, 0) -- (3.75, 0);
    \draw (4.25, 0) -- (4.75, 0);
    \end{tikzpicture}
    \end{minipage}
    
    \vspace{0.5cm}
    
    \begin{minipage}{1cm}
    Re($\chi$)
    
    \vspace{4.5cm}
    
    Im($\chi$)
    \end{minipage}
    \begin{minipage}{5.2cm}
    Raw
    \includegraphics[trim=3.5cm 1cm 2.5cm 1.4cm, clip, scale=0.5]{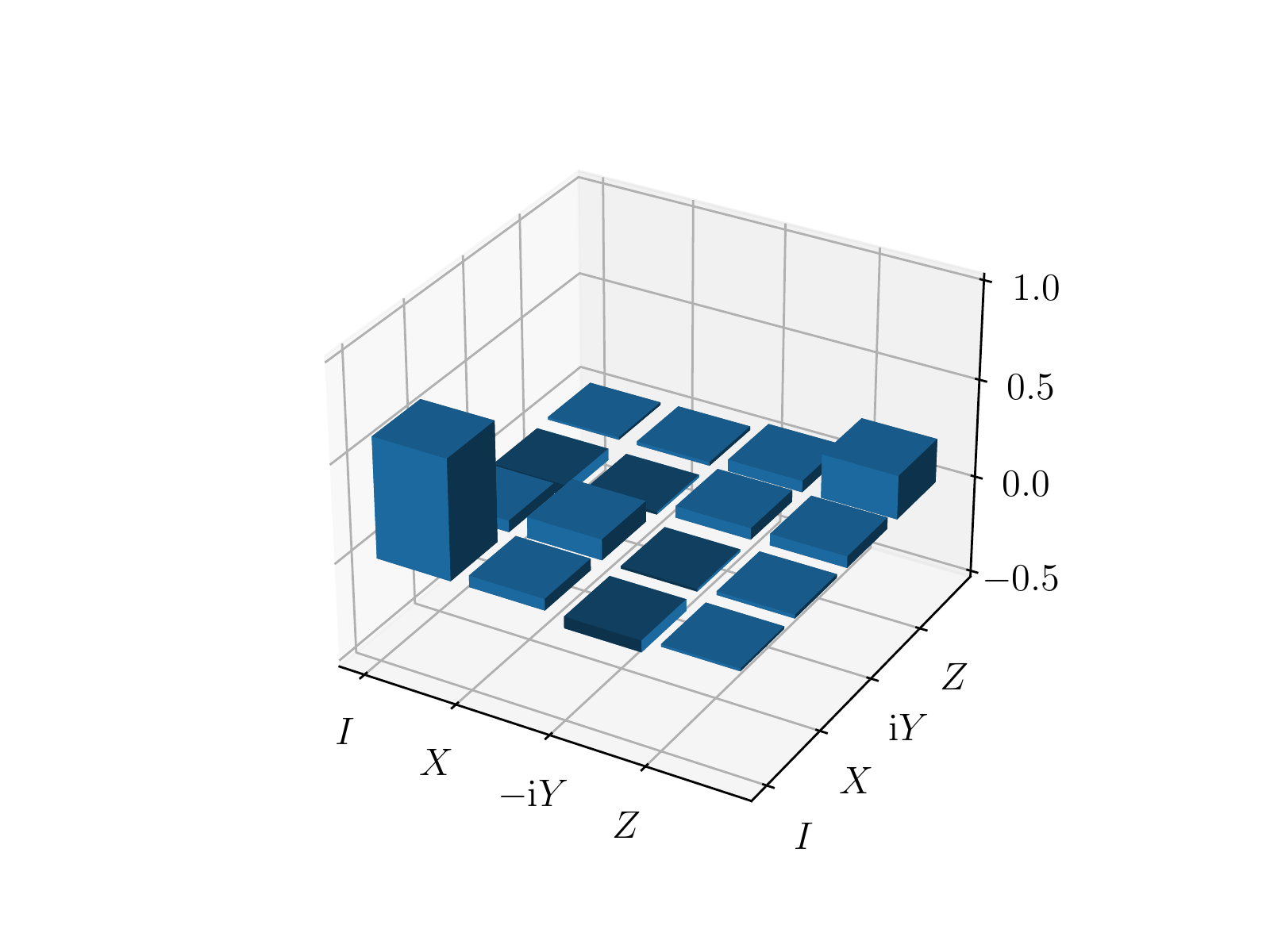}
    \includegraphics[trim=3.5cm 1cm 2.5cm 1.4cm, clip, scale=0.5]{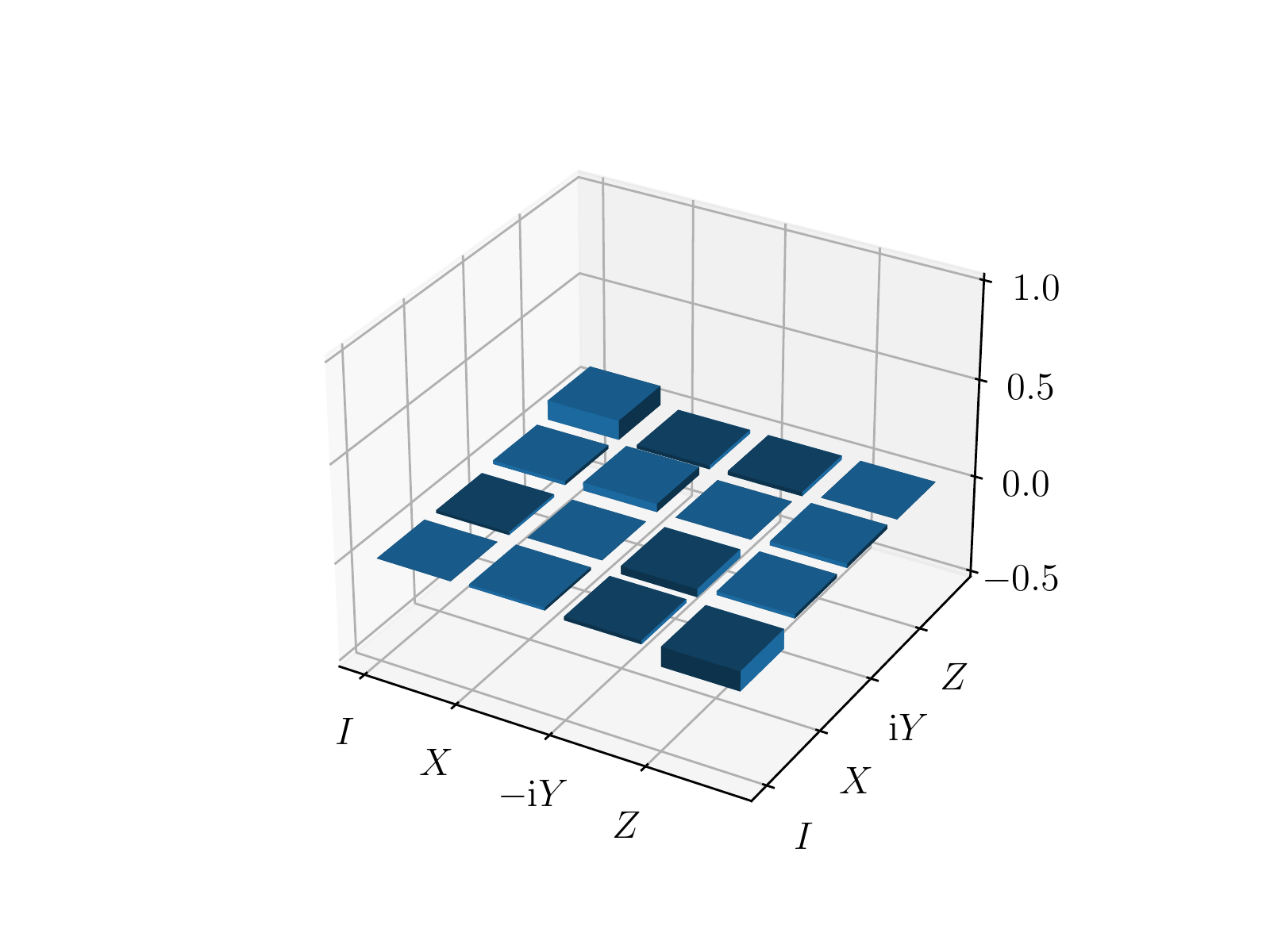}
    \end{minipage}
    \begin{minipage}{0.1cm}
    \end{minipage}
    \begin{minipage}{0.1cm}
    \begin{tikzpicture}
    \draw[gray, dashed] (0, 0) -- (0, -10);
    \end{tikzpicture}
    \end{minipage}
    \begin{minipage}{5.2cm}
    Processed
    \includegraphics[trim=3.5cm 1cm 2.5cm 1.4cm, clip, scale=0.5]{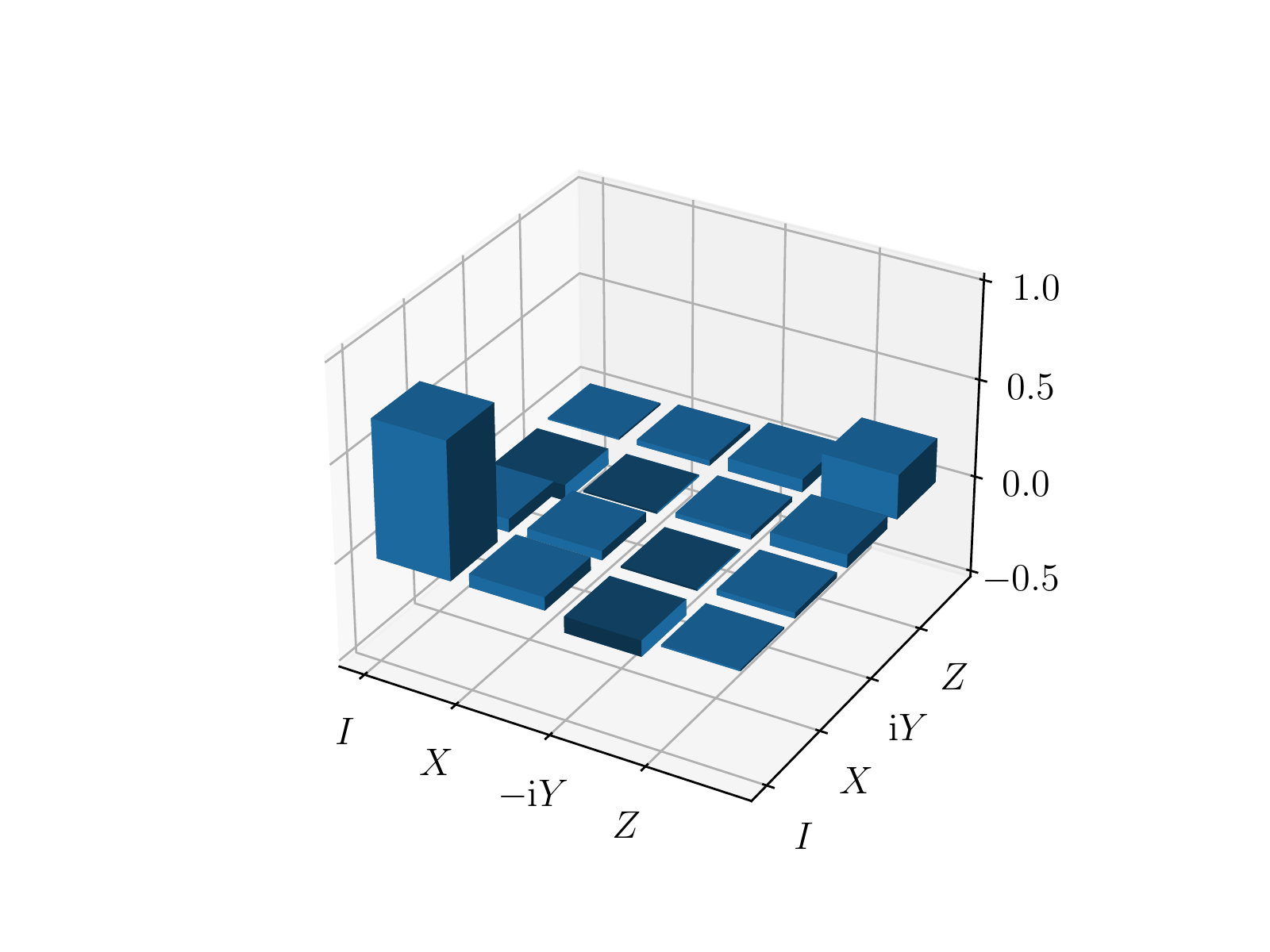}
    \includegraphics[trim=3.5cm 1cm 2.5cm 1.4cm, clip, scale=0.5]{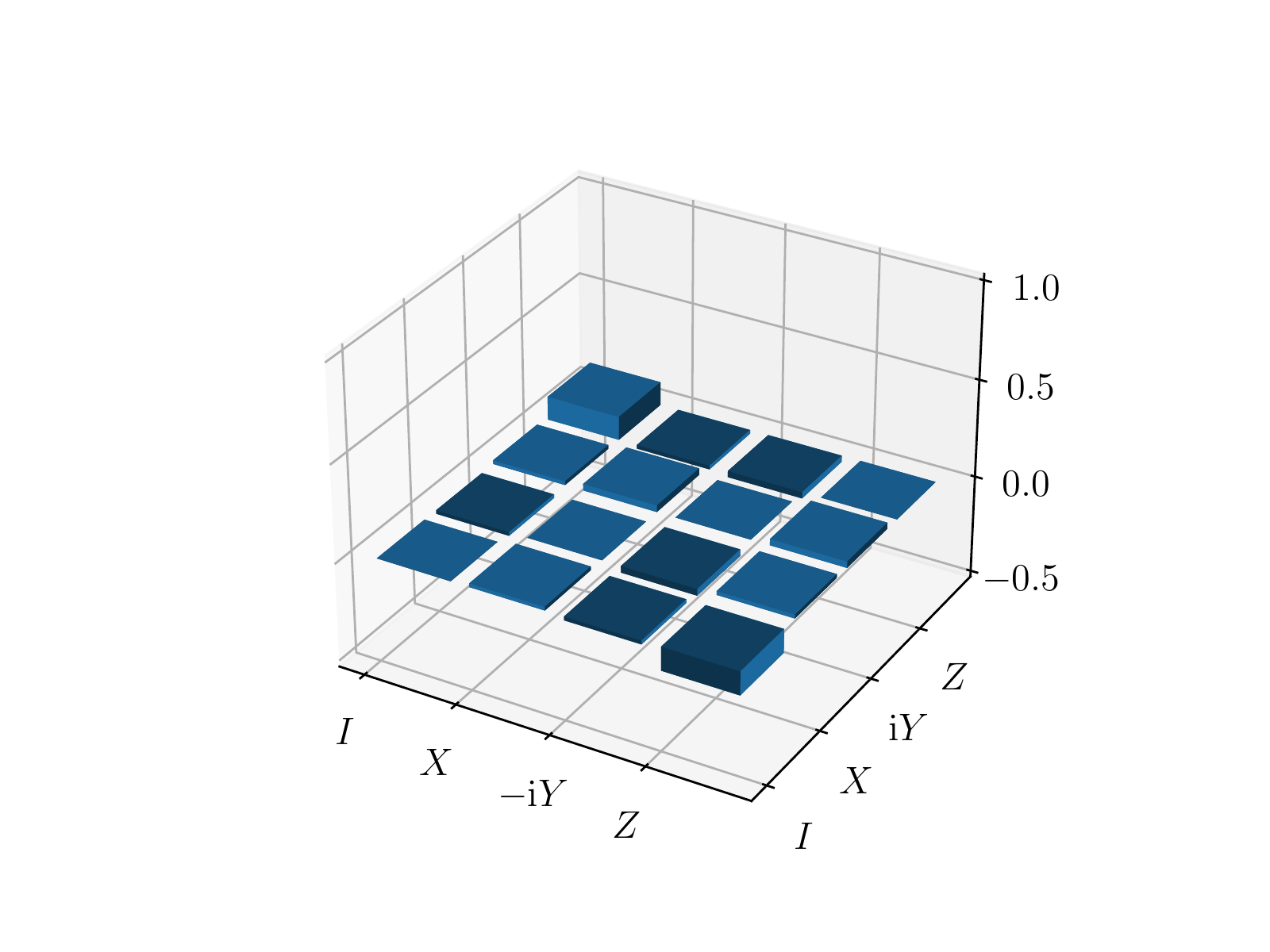}
    \end{minipage}
    \begin{minipage}{0.1cm}
    \end{minipage}
    \begin{minipage}{0.1cm}
    \begin{tikzpicture}
    \draw[gray, dashed] (0, 0) -- (0, -10);
    \end{tikzpicture}
    \end{minipage}
    \begin{minipage}{5.2cm}
    Ideal
    \includegraphics[trim=3.5cm 1cm 2.5cm 1.4cm, clip, scale=0.5]{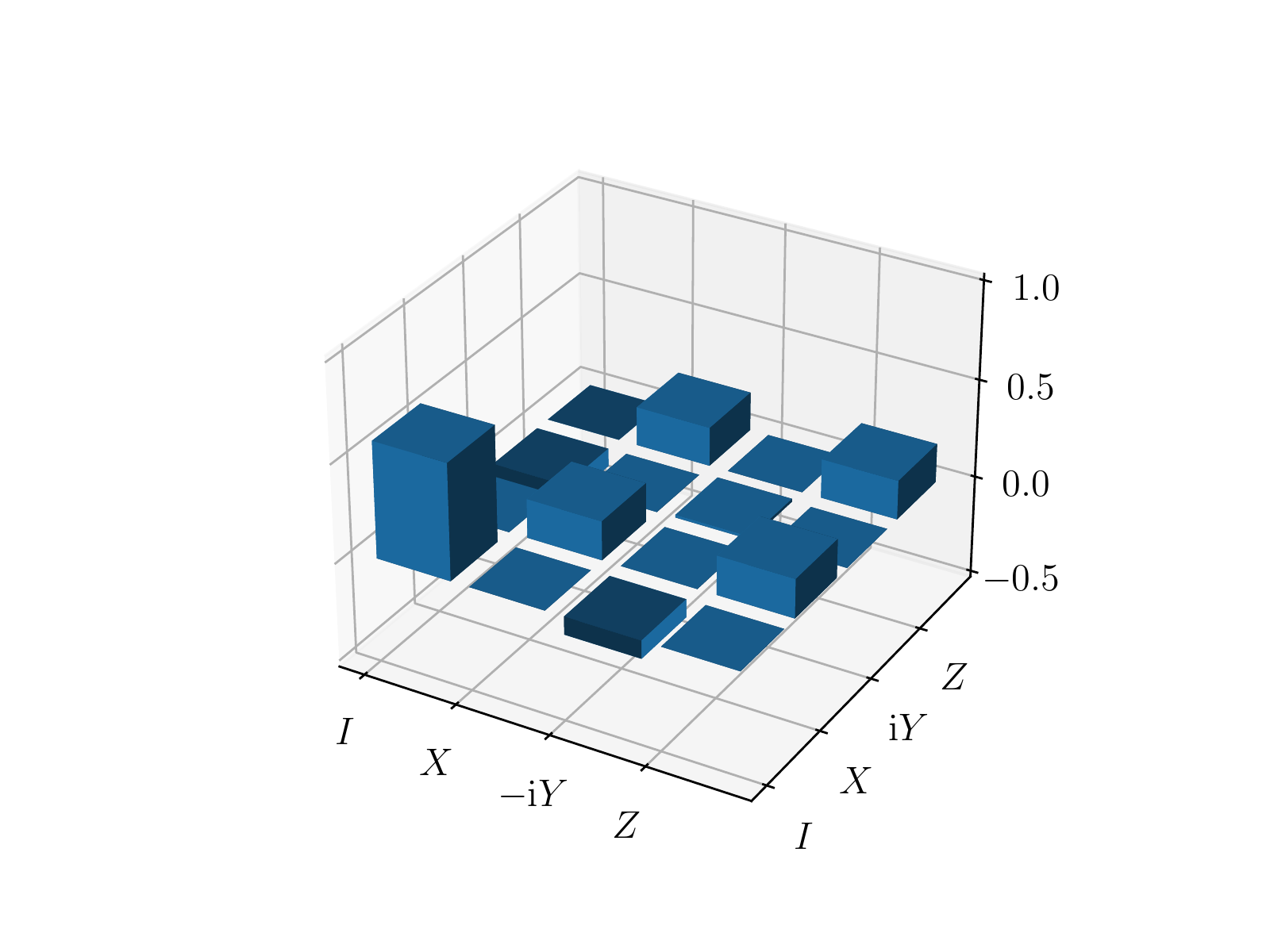}
    \includegraphics[trim=3.5cm 1cm 2.5cm 1.4cm, clip, scale=0.5]{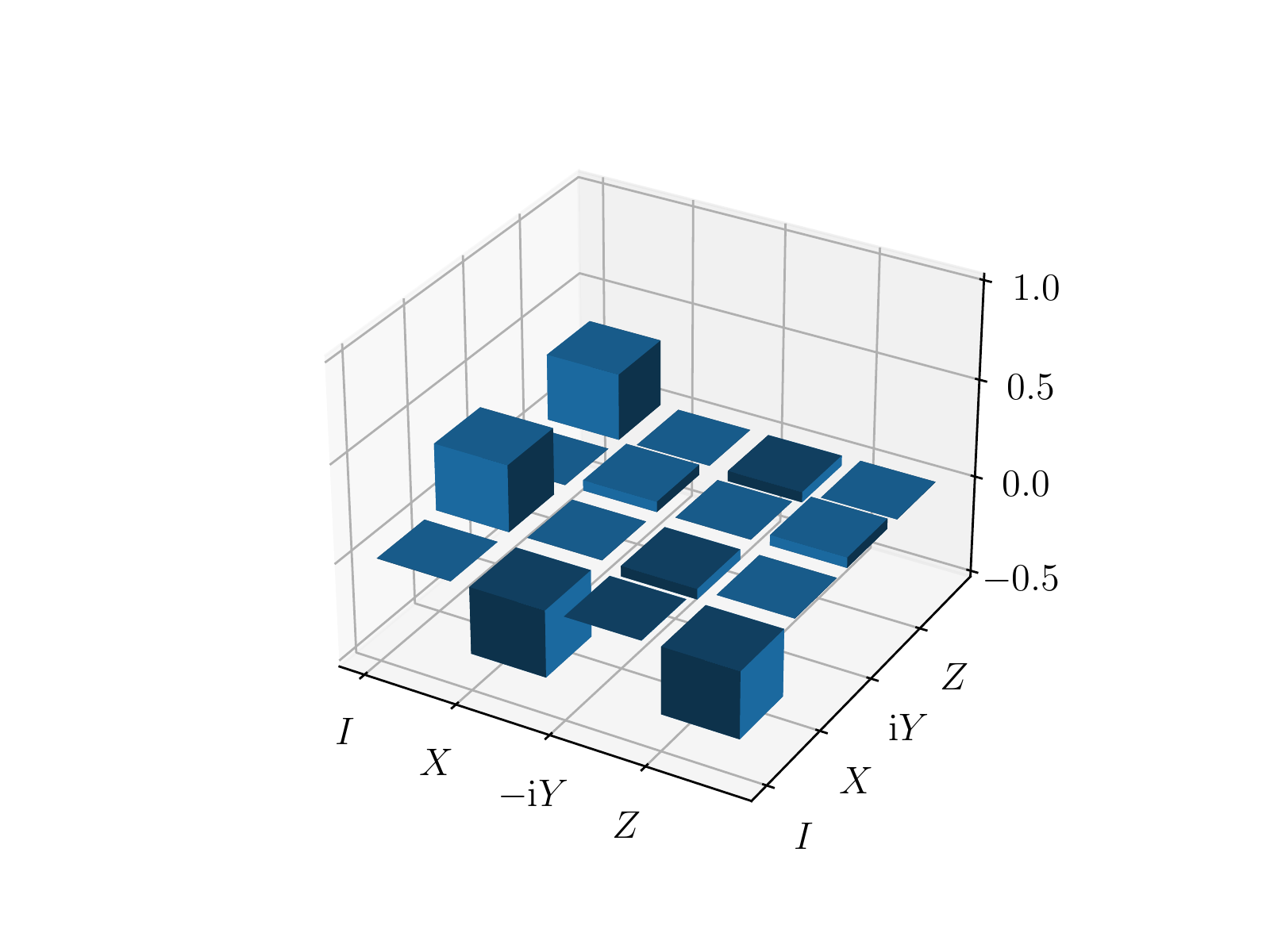}
    \end{minipage}
    \caption{Example of channel tomography results for the random unitary that agreed least with the ideal case generated for measurement outcome $\boldsymbol{m}=00000$ with the exact 3-design on the \textit{ibmq\_toronto} quantum processor.  The diagram at the top shows the entangled 6-qubit cluster state with the measurements performed on each qubit.  The $\chi$ matrix obtained without quantum readout error mitigation is shown on the left, the $\chi$ matrix obtained with quantum readout error mitigation is shown in the middle and the ideal $\chi$ matrix is shown on the right.  The real part of each matrix is shown above and the imaginary part of each matrix is shown below.}
    \label{fig:worst}
\end{figure*}
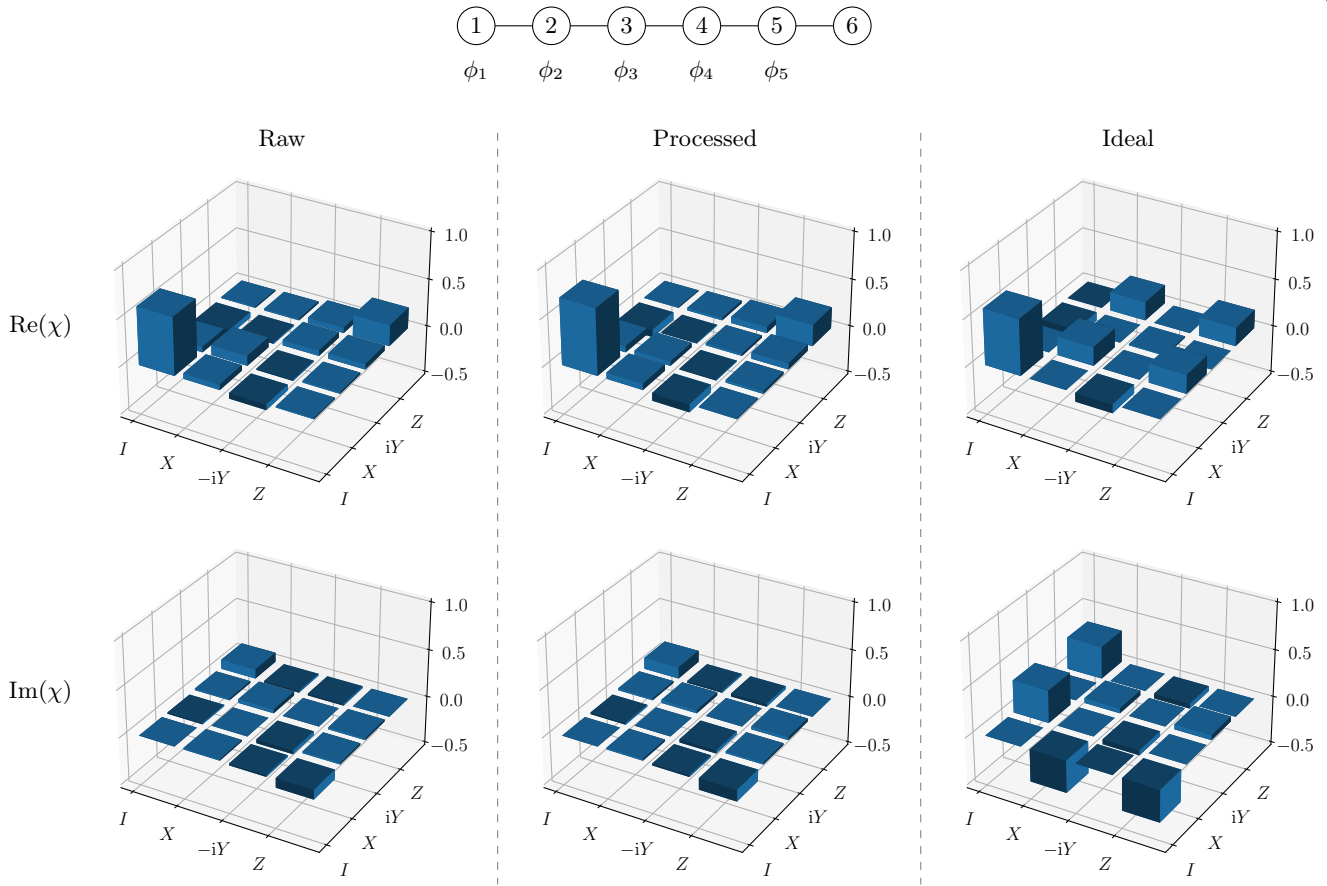

\begin{figure*}
    \centering
    \vspace{-0.5cm}
    \begin{minipage}{17cm}
    \begin{tikzpicture}
    \draw (0, 0) circle[radius=0.25] node {1};
    \draw (1, 0) circle[radius=0.25] node {2};
    \draw (2, 0) circle[radius=0.25] node {3};
    \draw (3, 0) circle[radius=0.25] node {4};
    \draw (4, 0) circle[radius=0.25] node {5};
    \draw (5, 0) circle[radius=0.25] node {6};
    \node[] at (0, -0.6) {$\phi_1$};
    \node[] at (1, -0.6) {$\phi_2$};
    \node[] at (2, -0.6) {$\phi_3$};
    \node[] at (3, -0.6) {$\phi_4$};
    \node[] at (4, -0.6) {$\phi_5$};
    \draw (0.25, 0) -- (0.75, 0);
    \draw (1.25, 0) -- (1.75, 0);
    \draw (2.25, 0) -- (2.75, 0);
    \draw (3.25, 0) -- (3.75, 0);
    \draw (4.25, 0) -- (4.75, 0);
    \end{tikzpicture}
    \end{minipage}
        
    \vspace{0.5cm}
    
    \begin{minipage}{1cm}
    Re($\chi$)
    
    \vspace{4.5cm}
    
    Im($\chi$)
    \end{minipage}
    \begin{minipage}{5.2cm}
    Raw
    \includegraphics[trim=3.5cm 1cm 2.5cm 1.4cm, clip, scale=0.5]{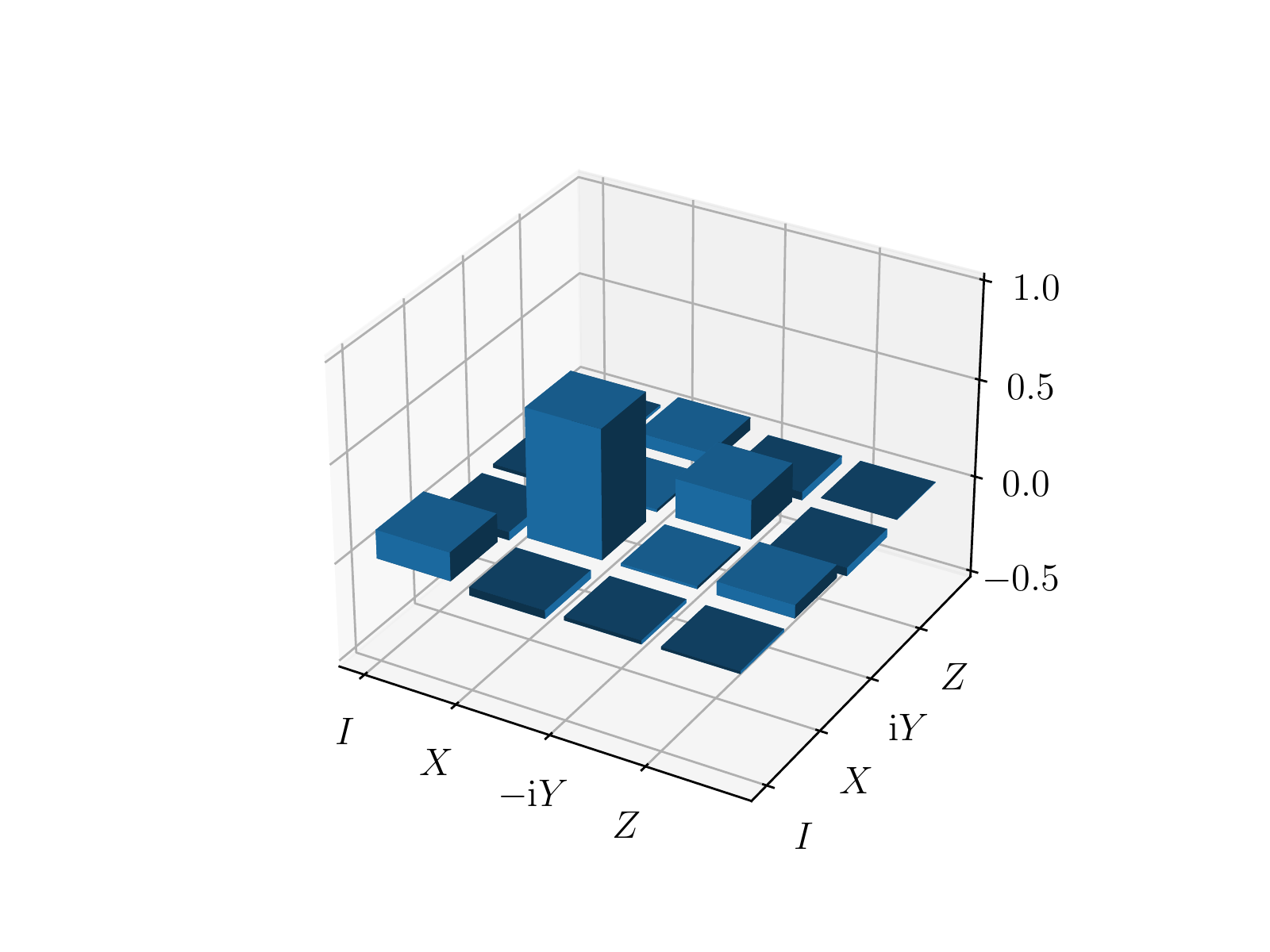}
    \includegraphics[trim=3.5cm 1cm 2.5cm 1.4cm, clip, scale=0.5]{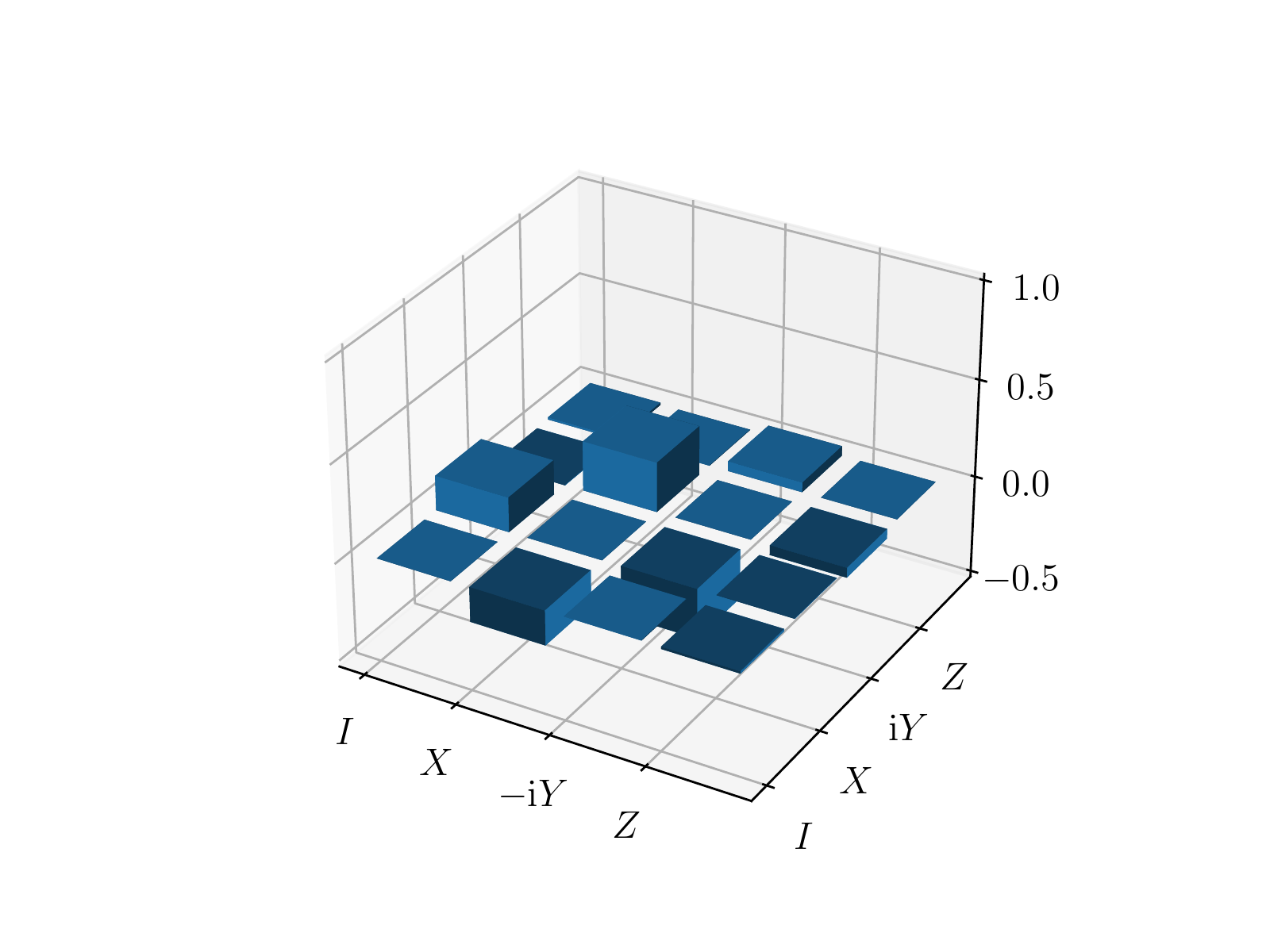}
    \end{minipage}
    \begin{minipage}{0.1cm}
    \end{minipage}
    \begin{minipage}{0.1cm}
    \begin{tikzpicture}
    \draw[gray, dashed] (0, 0) -- (0, -10);
    \end{tikzpicture}
    \end{minipage}
    \begin{minipage}{5.2cm}
    Processed
    \includegraphics[trim=3.5cm 1cm 2.5cm 1.4cm, clip, scale=0.5]{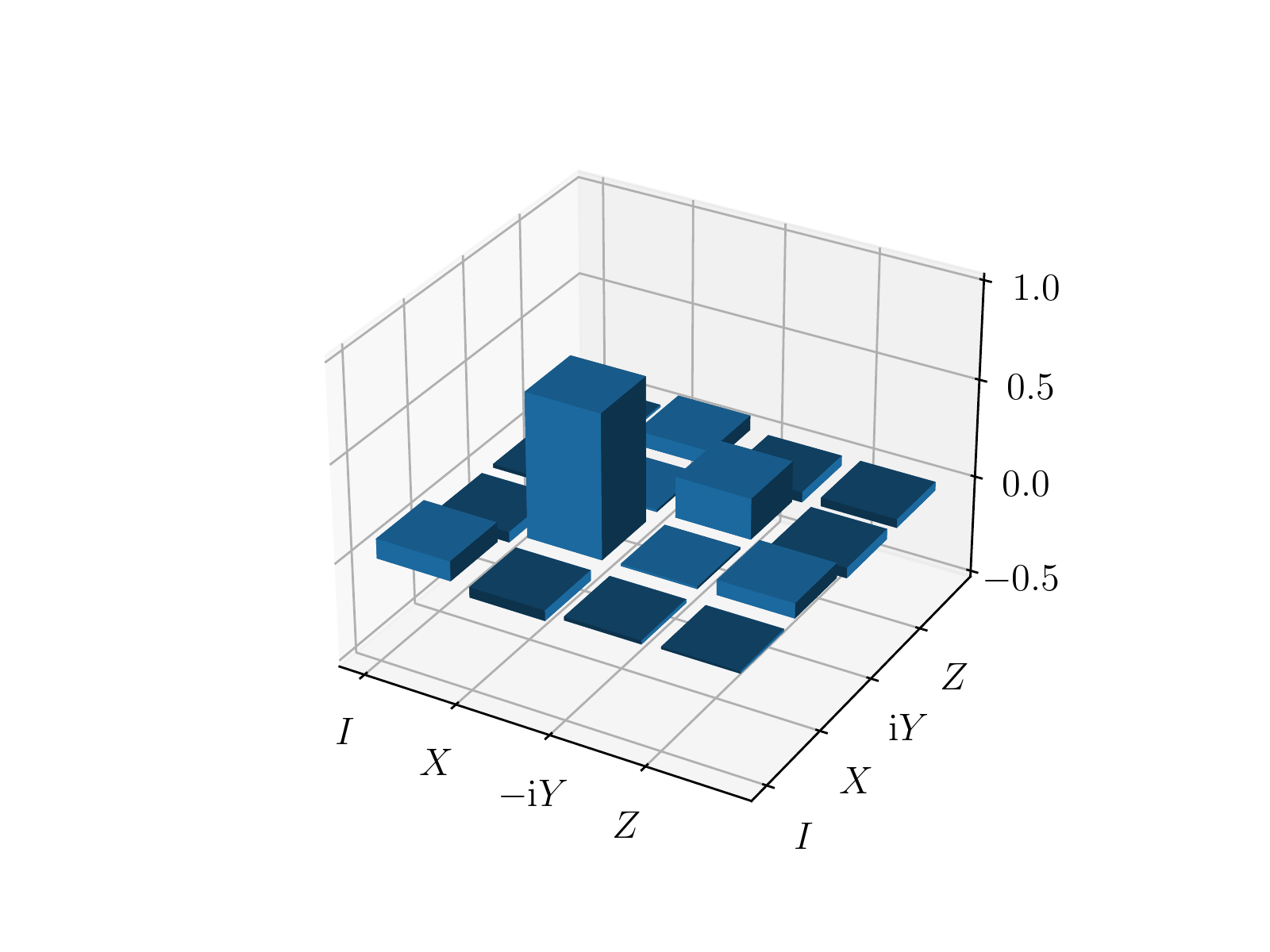}
    \includegraphics[trim=3.5cm 1cm 2.5cm 1.4cm, clip, scale=0.5]{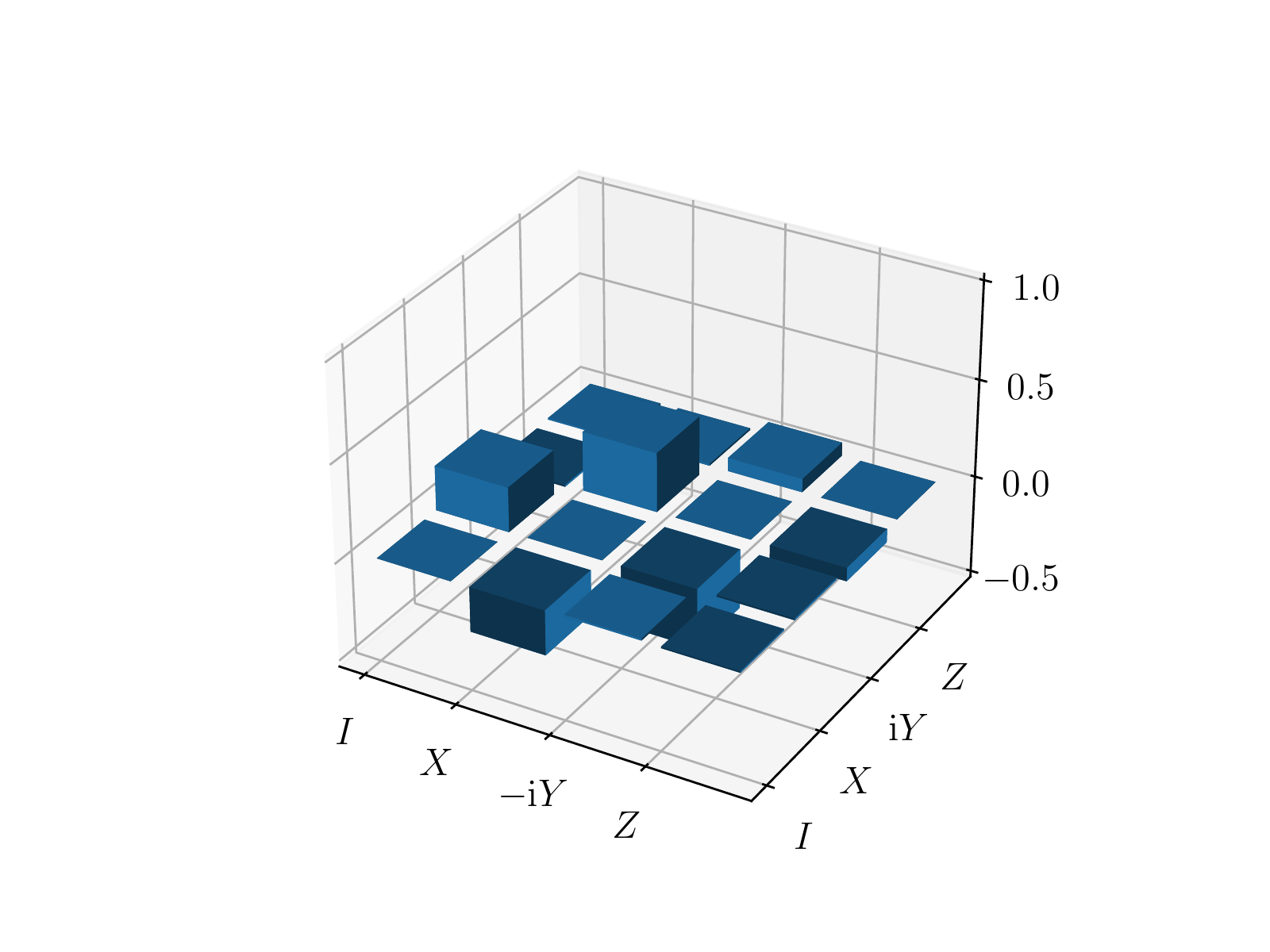}
    \end{minipage}
    \begin{minipage}{0.1cm}
    \end{minipage}
    \begin{minipage}{0.1cm}
    \begin{tikzpicture}
    \draw[gray, dashed] (0, 0) -- (0, -10);
    \end{tikzpicture}
    \end{minipage}
    \begin{minipage}{5.2cm}
    Ideal
    \includegraphics[trim=3.5cm 1cm 2.5cm 1.4cm, clip, scale=0.5]{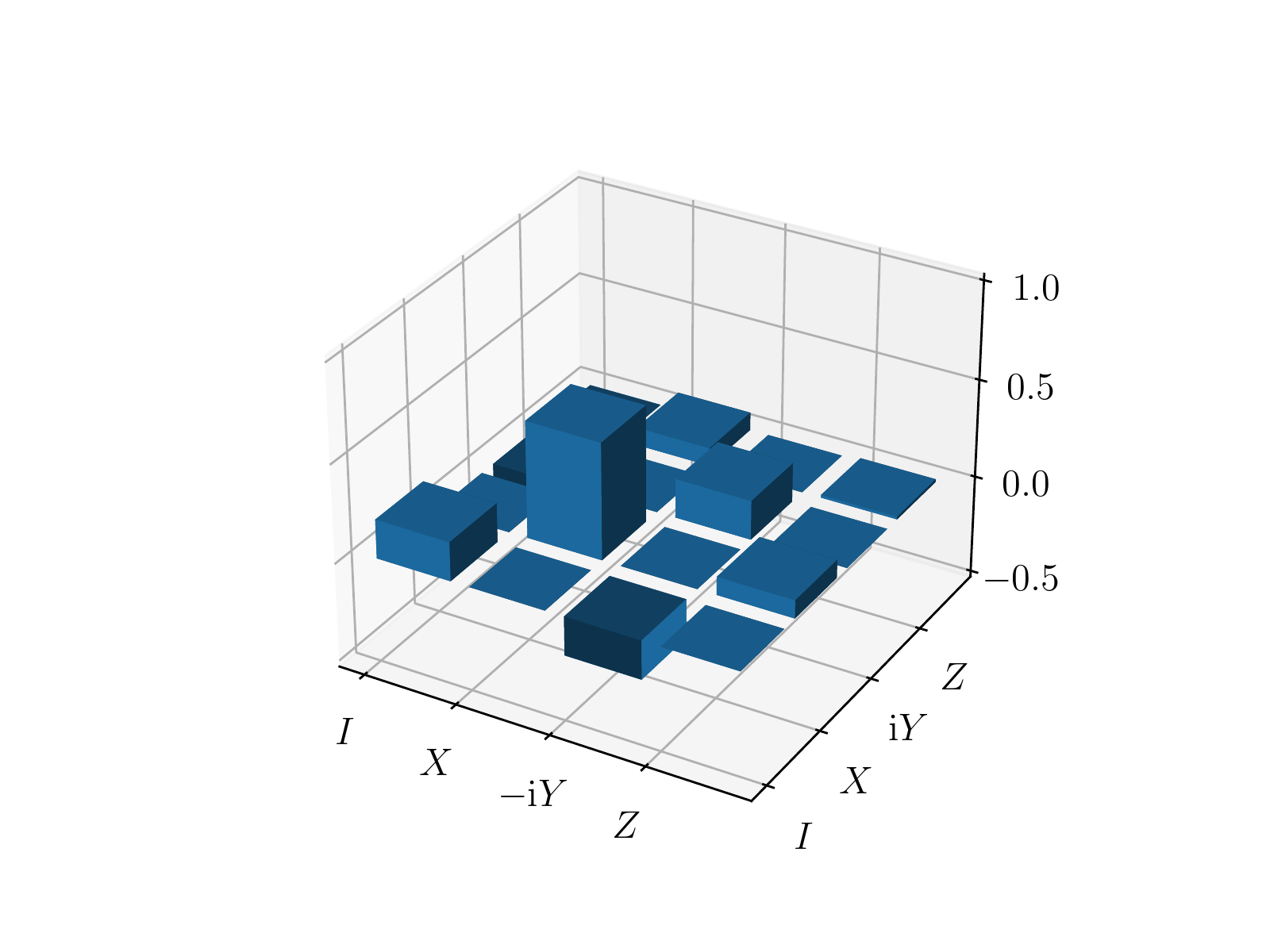}
    \includegraphics[trim=3.5cm 1cm 2.5cm 1.4cm, clip, scale=0.5]{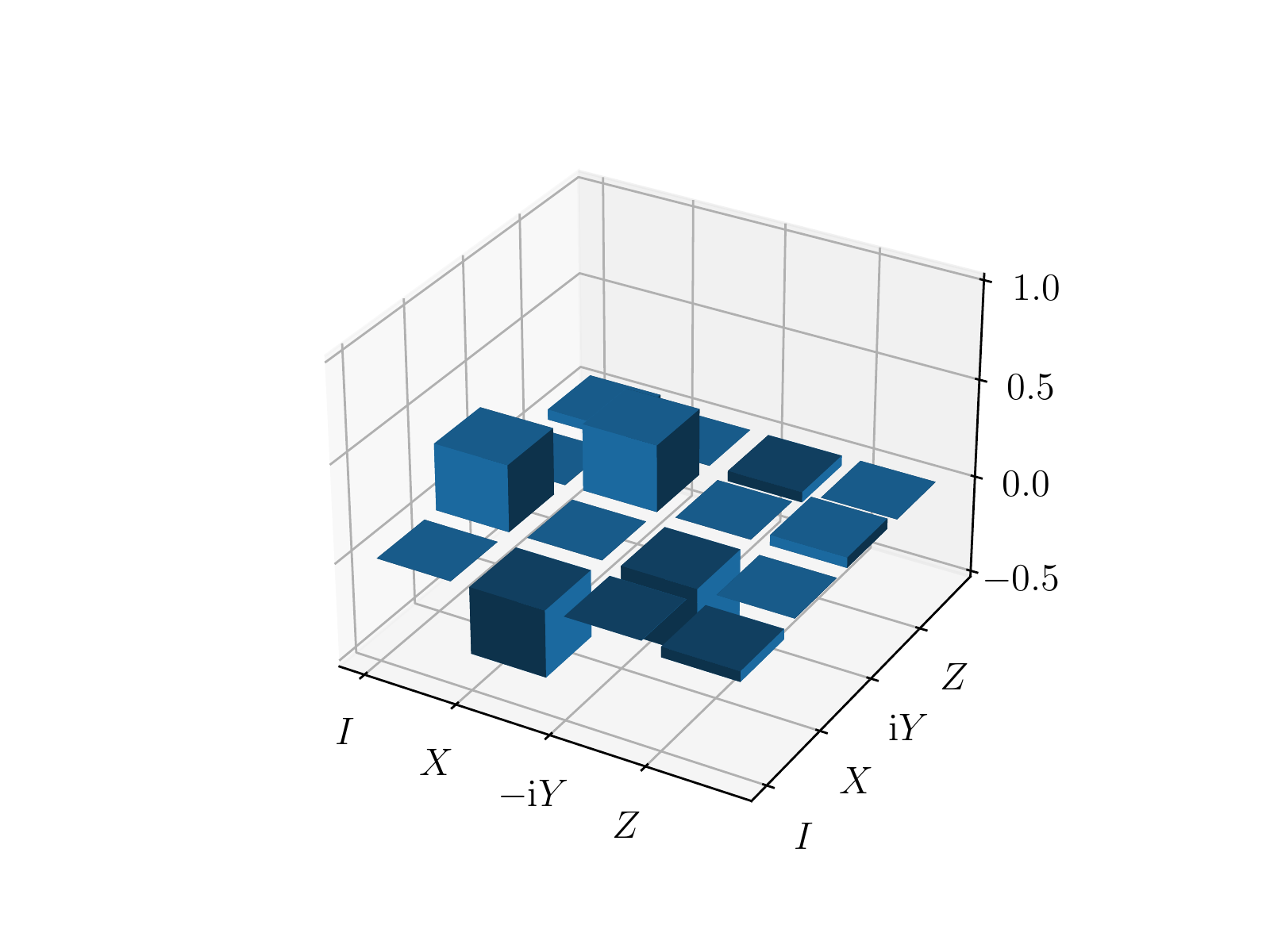}
    \end{minipage}
    \caption{Example of channel tomography results for the random unitary that agreed most with the ideal case generated for measurement outcome $\boldsymbol{m}=11001$ with the exact 3-design on the \textit{ibmq\_toronto} quantum processor.  The diagram at the top shows the entangled 6-qubit cluster state with the measurements performed on each qubit.  The $\chi$ matrix obtained without quantum readout error mitigation is shown on the left, the $\chi$ matrix obtained with quantum readout error mitigation is shown in the middle and the ideal $\chi$ matrix is shown on the right.  The real part of each matrix is shown above and the imaginary part of each matrix is shown below.}
    \label{fig:best}
\end{figure*}

%%%%%%%%%%%%%%%%%%%%%%%%%%%%
%%%%%%%%%%%%%%%%%%%%%%%%%%%%
\subsection{Channel tomography results}\label{sec:ctresults}

\begin{figure}
    \centering
    \begin{subfigure}[b]{.48\textwidth}
        \centering
        \includegraphics[scale=0.5]{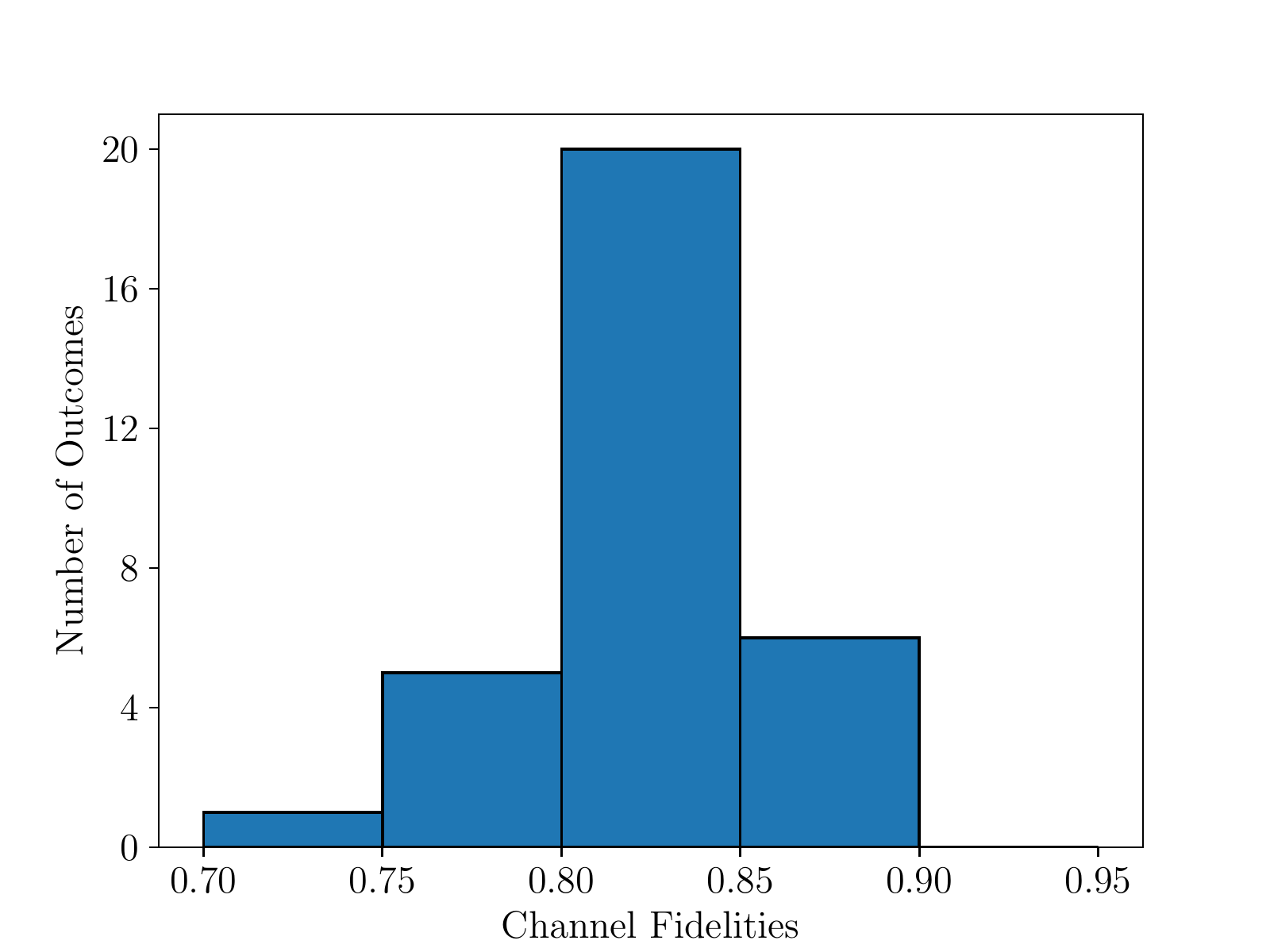}
        \caption{Raw}
    \end{subfigure}
    \begin{subfigure}[b]{.48\textwidth}
        \centering
        \includegraphics[scale=0.5]{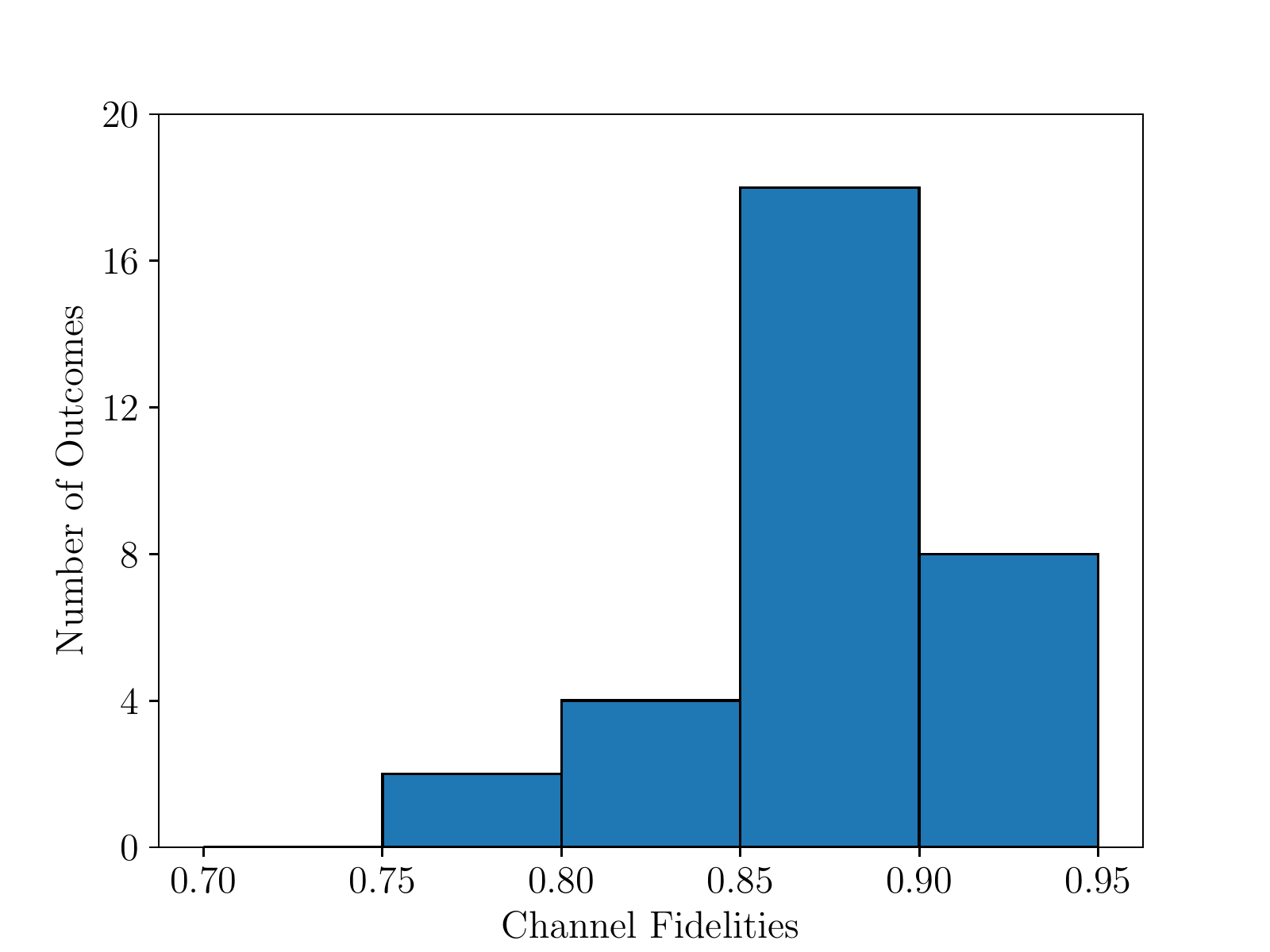}
        \caption{Processed}
    \end{subfigure}
    \caption{Distribution of channel fidelities for the 32 random unitaries generated with the exact 3-design on the \textit{ibmq\_toronto} quantum processor.  (a) Raw shows the distribution without quantum readout error mitigation.  (b) Processed shows the distribution with quantum readout error mitigation.}
    \label{fig:fidelities}
\end{figure}

Channel tomography results obtained for the two random unitaries, generated with the exact 3-design on the \textit{ibmq\_toronto} quantum processor, which showed the least and most agreement with theoretical predictions, are shown in Figs.~\ref{fig:worst} and \ref{fig:best} respectively.  Channel fidelities for each of the 32 different random unitaries generated with the exact 3-design implementation are given in \tabref{tab:fidelities} and the distribution of these channel fidelities is displayed in \figref{fig:fidelities}.  The average channel fidelity is (0.8220$\pm$0.0325) without quantum readout error mitigation and (0.8754$\pm$0.0361) with quantum readout error mitigation.  Quantum readout error mitigation improved all the channel fidelities, which confirms that classical measurement errors were responsible for a significant amount of noise in the exact measurement-based 3-design implementation, and would have resulted in channel fidelities which greatly underestimate the reliability with which the expected unitary operations are realised in the implementation, if left uncorrected.

\begin{table}[]
\begin{tabular}{|l|l|l|}
\hline
\textbf{Outcome} & \textbf{Fidelity (Raw)} & \textbf{Fidelity (Processed)} \\ \hline
00000            & 0.7246$\pm$0.0064      & 0.7670$\pm$0.0073            \\ \hline
00001            & 0.8090$\pm$0.0098      & 0.8688$\pm$0.0106            \\ \hline
00010            & 0.7820$\pm$0.0051      & 0.8397$\pm$0.0057            \\ \hline
00011            & 0.8013$\pm$0.0106      & 0.8596$\pm$0.0119            \\ \hline
00100            & 0.8061$\pm$0.0077      & 0.8657$\pm$0.0084            \\ \hline
00101            & 0.8506$\pm$0.0084      & 0.9192$\pm$0.0096            \\ \hline
00110            & 0.8498$\pm$0.0060      & 0.9157$\pm$0.0073            \\ \hline
00111            & 0.8358$\pm$0.0086      & 0.8949$\pm$0.0095            \\ \hline
01000            & 0.7819$\pm$0.0106      & 0.8295$\pm$0.0115            \\ \hline
01001            & 0.8356$\pm$0.0082      & 0.8919$\pm$0.0090            \\ \hline
01010            & 0.7865$\pm$0.0035      & 0.8295$\pm$0.0041            \\ \hline
01011            & 0.8194$\pm$0.0066      & 0.8715$\pm$0.0074            \\ \hline
01100            & 0.8215$\pm$0.0061      & 0.8755$\pm$0.0071            \\ \hline
01101            & 0.8426$\pm$0.0068      & 0.8920$\pm$0.0075            \\ \hline
01110            & 0.8272$\pm$0.0056      & 0.8708$\pm$0.0066            \\ \hline
01111            & 0.8538$\pm$0.0059      & 0.9042$\pm$0.0065            \\ \hline
10000            & 0.7503$\pm$0.0088      & 0.7953$\pm$0.0101            \\ \hline
10001            & 0.8423$\pm$0.0084      & 0.9056$\pm$0.0096            \\ \hline
10010            & 0.8135$\pm$0.0072      & 0.8656$\pm$0.0080            \\ \hline
10011            & 0.8203$\pm$0.0069      & 0.8830$\pm$0.0081            \\ \hline
10100            & 0.7992$\pm$0.0096      & 0.8569$\pm$0.0103            \\ \hline
10101            & 0.8597$\pm$0.0066      & 0.9202$\pm$0.0069            \\ \hline
10110            & 0.8645$\pm$0.0054      & 0.9240$\pm$0.0060            \\ \hline
10111            & 0.8460$\pm$0.0067      & 0.9035$\pm$0.0079            \\ \hline
11000            & 0.8104$\pm$0.0078      & 0.8593$\pm$0.0088            \\ \hline
11001            & 0.8774$\pm$0.0053      & 0.9353$\pm$0.0060            \\ \hline
11010            & 0.8238$\pm$0.0084      & 0.8689$\pm$0.0088            \\ \hline
11011            & 0.8419$\pm$0.0066      & 0.8876$\pm$0.0076            \\ \hline
11100            & 0.8013$\pm$0.0070      & 0.8447$\pm$0.0077            \\ \hline
11101            & 0.8287$\pm$0.0097      & 0.8765$\pm$0.0102            \\ \hline
11110            & 0.8513$\pm$0.0047      & 0.8936$\pm$0.0052            \\ \hline
11111            & 0.8467$\pm$0.0071      & 0.8959$\pm$0.0081            \\ \hline
\end{tabular}
\caption{Channel fidelities for the 32 random unitaries, corresponding to the 32 different measurement outcomes, generated with the exact 3-design on the \textit{ibmq\_toronto} quantum processor.  `Raw' shows the channel fidelities without quantum readout error mitigation.  `Processed' shows the channel fidelities with quantum readout error mitigation.}
\label{tab:fidelities}
\end{table}

%%%%%%%%%%%%%%%%%%%%%%%%%%%%
%%%%%%%%%%%%%%%%%%%%%%%%%%%%
\subsection{Relative frequencies}\label{sec:relfreqs}

\begin{table}[]
\begin{tabular}{|l|l|l|}
\hline
\textbf{Outcome} & \textbf{Frequency (Raw)} & \textbf{Frequency (Processed)} \\ \hline
00000            & 0.03836$\pm$0.00095     & 0.03664$\pm$0.00105           \\ \hline
00001            & 0.03130$\pm$0.00057     & 0.02987$\pm$0.00059           \\ \hline
00010            & 0.03372$\pm$0.00075     & 0.03196$\pm$0.00083           \\ \hline
00011            & 0.02737$\pm$0.00064     & 0.02632$\pm$0.00074           \\ \hline
00100            & 0.02861$\pm$0.00097     & 0.02702$\pm$0.00108           \\ \hline
00101            & 0.03335$\pm$0.00095     & 0.03183$\pm$0.00108           \\ \hline
00110            & 0.03447$\pm$0.00091     & 0.03367$\pm$0.00104           \\ \hline
00111            & 0.03305$\pm$0.00082     & 0.03253$\pm$0.00091           \\ \hline
01000            & 0.03535$\pm$0.00094     & 0.03567$\pm$0.00107           \\ \hline
01001            & 0.02721$\pm$0.00076     & 0.02699$\pm$0.00086           \\ \hline
01010            & 0.03530$\pm$0.00085     & 0.03636$\pm$0.00094           \\ \hline
01011            & 0.02716$\pm$0.00040     & 0.02750$\pm$0.00048           \\ \hline
01100            & 0.03079$\pm$0.00081     & 0.03089$\pm$0.00098           \\ \hline
01101            & 0.03349$\pm$0.00069     & 0.03428$\pm$0.00080           \\ \hline
01110            & 0.03477$\pm$0.00111     & 0.03628$\pm$0.00131           \\ \hline
01111            & 0.02986$\pm$0.00087     & 0.03124$\pm$0.00096           \\ \hline
10000            & 0.03788$\pm$0.00110     & 0.03733$\pm$0.00123           \\ \hline
10001            & 0.02789$\pm$0.00068     & 0.02666$\pm$0.00074           \\ \hline
10010            & 0.03399$\pm$0.00069     & 0.03379$\pm$0.00078           \\ \hline
10011            & 0.02408$\pm$0.00051     & 0.02289$\pm$0.00057           \\ \hline
10100            & 0.02548$\pm$0.00065     & 0.02432$\pm$0.00075           \\ \hline
10101            & 0.03348$\pm$0.00089     & 0.03325$\pm$0.00103           \\ \hline
10110            & 0.03032$\pm$0.00066     & 0.03009$\pm$0.00075           \\ \hline
10111            & 0.03238$\pm$0.00073     & 0.03287$\pm$0.00081           \\ \hline
11000            & 0.03188$\pm$0.00075     & 0.03221$\pm$0.00086           \\ \hline
11001            & 0.02757$\pm$0.00050     & 0.02828$\pm$0.00057           \\ \hline
11010            & 0.03254$\pm$0.00131     & 0.03376$\pm$0.00148           \\ \hline
11011            & 0.02753$\pm$0.00081     & 0.02886$\pm$0.00095           \\ \hline
11100            & 0.03008$\pm$0.00094     & 0.03121$\pm$0.00112           \\ \hline
11101            & 0.03017$\pm$0.00096     & 0.03116$\pm$0.00110           \\ \hline
11110            & 0.03430$\pm$0.00104     & 0.03673$\pm$0.00125           \\ \hline
11111            & 0.02631$\pm$0.00111     & 0.02759$\pm$0.00127           \\ \hline
\end{tabular}
\caption{Relative frequencies with which the 32 random unitaries, corresponding to the 32 different measurement outcomes, are generated with the exact 3-design on the \textit{ibmq\_toronto} quantum processor.  `Raw' shows the relative frequencies without quantum readout error mitigation.  `Processed' shows the relative frequencies with quantum readout error mitigation.}
\label{tab:frequencies}
\end{table}

\begin{figure}
    \centering
    \begin{subfigure}[b]{.48\textwidth}
        \centering
        \includegraphics[scale=0.5]{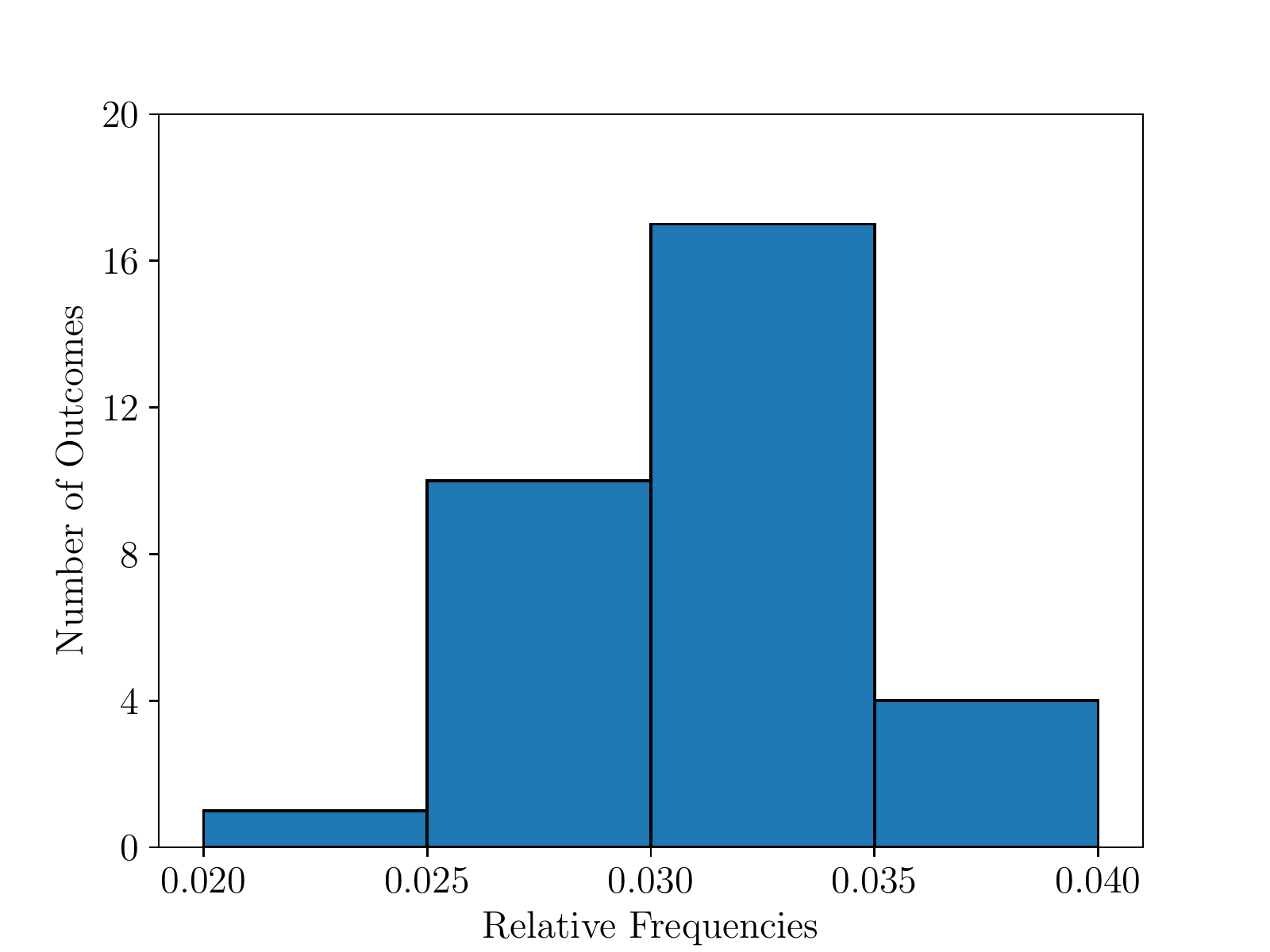}
        \caption{Raw}
    \end{subfigure}
    \begin{subfigure}[b]{.48\textwidth}
        \centering
        \includegraphics[scale=0.5]{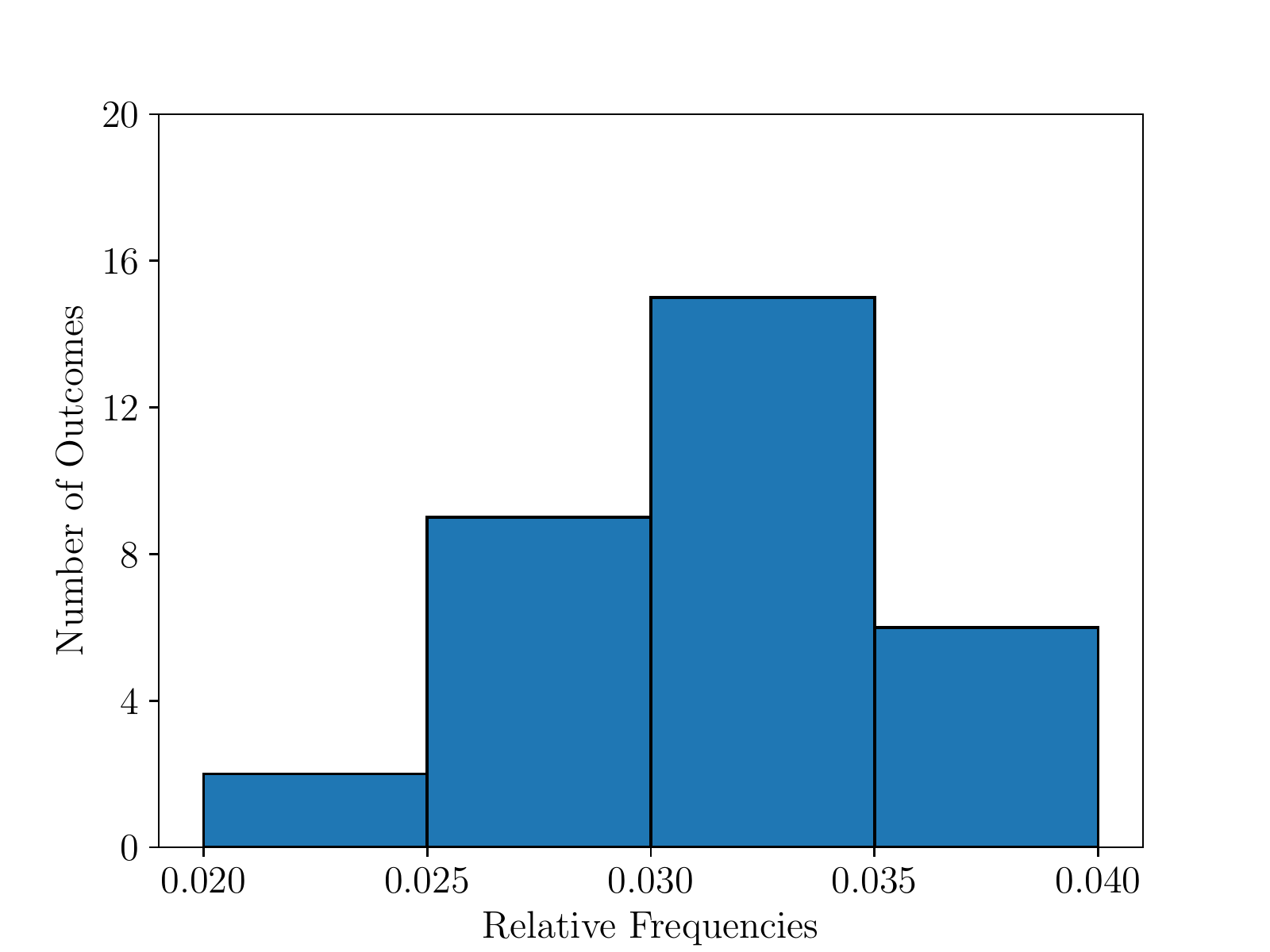}
        \caption{Processed}
    \end{subfigure}
    \caption{Distribution of relative frequencies with which the 32 random unitaries are generated with the exact 3-design on the \textit{ibmq\_toronto} quantum processor.  (a) Raw shows the distribution without quantum readout error mitigation.  (b) Processed shows the distribution with quantum readout error mitigation.}
    \label{fig:frequencies}
\end{figure}

Due to errors that occur when gates are applied and measurements are made, the relative frequencies with which the 32 random unitaries are generated, with the exact 3-design on the \textit{ibmq\_toronto} quantum processor, do not exactly match the expected uniform probabilities of $\frac{1}{32}=0.03125$.  To determine the relative frequency with which each unitary is generated, a separate investigation was conducted in which the 6-qubit linear cluster state was prepared on the same 6 physical qubits of the \textit{ibmq\_toronto} quantum processor as was used for the 3-design implementation (see \appendref{append:qubits3}) and the first 5 qubits were measured in the same way as in the 3-design implementation.  The relative frequency of each set of measurement outcomes is the relative frequency with which the random unitary corresponding to that set of measurement outcomes is generated.  The required circuit was run 5 times with 8000 shots each and counts were once again combined to obtain an effective run with 40000 shots.  This was repeated 10 times, so that relative frequencies quoted are an average of 10 repetitions and the errors quoted are the standard deviations.  The relative frequencies with which each of the 32 different random unitaries are generated with the exact 3-design on the \textit{ibmq\_toronto} quantum processor are presented in \tabref{tab:frequencies} and the distribution of these relative frequencies is displayed in \figref{fig:frequencies}.  The average relative frequency is (0.03125$\pm$0.00355) without quantum readout error mitigation and (0.03125$\pm$0.00375) with quantum readout error mitigation.  We note that even though the relative frequencies with which the different unitaries are generated deviate from the expected uniform probabilities, the average relative frequency is equal to the expected probability.

%%%%%%%%%%%%%%%%%%%%%%%%%%%%
%%%%%%%%%%%%%%%%%%%%%%%%%%%%
\subsection{Testing for a t-design}\label{sec:test}

The definition of an approximate $t$-design as given by inequality (\ref{eq:def}) in \secref{sec:mbbackground} leads naturally to a simple method for testing whether the ensemble of 32 unitaries generated with the 6-qubit cluster state on the \textit{ibmq\_toronto} quantum processor is at least an approximate $t$-design.  Although this definition applies to any density matrix acting on the tensor product space $(\mathbb{C}^2)^{\otimes t}$, we restrict ourselves to density matrices which are $t$ copies of an arbitrary single-qubit density matrix for the purposes of testing.  This is sufficient for quantifying the extent to which the unitaries are able to randomise single-qubit states, which is our primary interest here, and can at least provide a lower bound on the $\epsilon$ for which the ensemble of unitaries is an $\epsilon$-approximate $t$-design with more general states included.  The restriction to copies of single-qubit density matrices has two major advantages, namely that the test is computationally feasible for all $t$, since the number of parameters that need to be varied when creating samples of density matrices would otherwise grow exponentially with $t$, and that the results can be interpreted geometrically, since single-qubit states can be represented by points in the Bloch sphere.  For the purposes of testing, we therefore consider the adapted inequality
\begin{equation}
(1-\epsilon)\mathbb{E}^t_H(\rho^{\otimes t})\leq\sum_{i}p_i{\rho_i'}^{\otimes t}\leq(1+\epsilon)\mathbb{E}^t_H(\rho^{\otimes t}),
\label{eq:master}
\end{equation}
where $p_i$ are the experimentally determined relative frequencies and
\begin{equation}
\rho_i'=\sum_{mn}E_m\rho E_n^{\dagger}\chi^{(i)}_{mn},
\end{equation}
where $\chi^{(i)}$ are the $\chi$ matrices determined by doing channel tomography for the different unitaries.  The test amounts to generating a sample of single-qubit density matrices and finding, for each density matrix in the sample, the smallest possible $\epsilon$ such that inequality (\ref{eq:master}) is satisfied.  The largest $\epsilon$ found is the one which ensures that inequality (\ref{eq:master}) is satisfied for all density matrices in the sample and is therefore the test result.

We set the passing criterion for the test to $\epsilon \leq 0.5$, by which we simply mean that, for the purposes of this paper, we consider the quality of an approximate $t$-design acceptable if $\epsilon \leq 0.5$.  In practice, some applications of approximate $t$-designs may require a smaller value of $\epsilon$.  We also note that, given an ensemble of unitaries which is an exact $t$-design or an approximate $t$-design with $\epsilon_t \leq 0.5$, it is generally not possible to say whether this ensemble of unitaries is also an approximate $(t+1)$-design with $\epsilon_{t+1} \leq 0.5$, as this depends on the ensemble.  Therefore an ensemble of unitaries which passes our test for an approximate $t$-design, for a given $t$, may or may not pass our test for an approximate $(t+1)$-design.

To generate a sample of single-qubit density matrices, we first generate a representative sample of points in the Bloch sphere using spherical coordinates.  We generate 10 evenly spaced values of $r$ in the range $[0, 1]$, 10 evenly spaced values of $\phi$ in the range $[0, 2\pi)$ and 10 evenly spaced values of $\theta$ in the range $[0, \pi]$.  Using the standard conversions from spherical to cartesian coordinates, we compute $r_x$, $r_y$ and $r_z$ for all combinations of the sampled values of $r$, $\phi$ and $\theta$, thereby obtaining 1000 points in the Bloch sphere.  Using
\begin{equation}
\rho=\frac{1}{2}\begin{pmatrix}1+r_z&r_x-\text{i}r_y\\r_x+\text{i}r_y&1-r_z\\\end{pmatrix}
\end{equation}
we obtain a sample of 1000 density matrices.

\begin{table*}[]
\begin{tabular}{|l|l|l|l|l|l|l|}
\hline
\textbf{}     & \multicolumn{3}{l|}{\textbf{Raw}}                                   & \multicolumn{3}{l|}{\textbf{Processed}}                             \\ \cline{2-7} 
\textbf{Test} & \textbf{Radius} & $\boldsymbol{\epsilon}$ & $\boldsymbol{\epsilon}$ \textbf{(uniform)} & \textbf{Radius} & $\boldsymbol{\epsilon}$ & $\boldsymbol{\epsilon}$ \textbf{(uniform)} \\ \hline
1-design      & 1.00           & 0.0777$\pm$0.0066   & 0.0760$\pm$0.0072            & 1.00           & 0.0683$\pm$0.0054   & 0.0677$\pm$0.0072            \\ \hline
2-design      & 0.68           & 0.4543$\pm$0.0074   & 0.4559$\pm$0.0063            & 0.75           & 0.4538$\pm$0.0188   & 0.4464$\pm$0.0179            \\ \hline
3-design      & 0.66           & 0.4590$\pm$0.0061   & 0.4592$\pm$0.0070            & 0.69           & 0.4814$\pm$0.0062   & 0.4696$\pm$0.0058            \\ \hline
\end{tabular}
\caption{Summary of test results for the ensemble of unitaries generated using the \textit{ibmq\_toronto} quantum processor.  `Raw' shows the results without quantum readout error mitigation.  `Processed' shows the results with quantum readout error mitigation.  `Radius' is the truncation radius considered for a test.  The column with `uniform' shows the values of $\epsilon$ obtained when replacing the experimentally determined relative frequencies with uniform probabilities.}
\label{tab:results}
\end{table*}

\begin{table*}[]
\begin{tabular}{|l|l|l|l|l|}
\hline
\textbf{}     & \multicolumn{2}{l|}{\textbf{Raw}}          & \multicolumn{2}{l|}{\textbf{Processed}}    \\ \cline{2-5} 
\textbf{Test} & \textbf{Frac}     & \textbf{Frac (uniform)} & \textbf{Frac}     & \textbf{Frac (uniform)} \\ \hline
1-design      & 1.0000            & 1.0000                 & 1.0000            & 1.0000                 \\ \hline
2-design      & 0.3834$\pm$0.0027 & 0.3858$\pm$0.0034      & 0.5648$\pm$0.0079 & 0.5768$\pm$0.0081      \\ \hline
3-design      & 0.3473$\pm$0.0048 & 0.3534$\pm$0.0054      & 0.5315$\pm$0.0061 & 0.5454$\pm$0.0073      \\ \hline
\end{tabular}
\caption{Fraction of states for which the ensemble of unitaries generated using the \textit{ibmq\_toronto} quantum processor passed the different tests.  `Raw' shows the fractions without quantum readout error mitigation.  `Processed' shows the fractions with quantum readout error mitigation.  The column with `uniform' shows the fractions obtained when replacing the experimentally determined relative frequencies with uniform probabilities.}
\label{tab:volume}
\end{table*}

Using our sample of 1000 density matrices, we applied the test for the 1-design, the 2-design and the 3-design to the ensemble of unitaries generated using the \textit{ibmq\_toronto} quantum processor.  The expected unitary operations of the exact 3-design described in \secref{sec:mbbackground} were used to compute $\mathbb{E}^t_H(\rho^{\otimes t})$ for $t=1,2,3$.  The test for the 1-design passed.  The tests for the 2-design and the 3-design did not pass, as $\epsilon$ diverged for states close to the surface of the Bloch sphere.  Nevertheless, inequality (\ref{eq:master}) could be satisfied for states close to the centre of the Bloch sphere.  This was investigated further by re-applying the tests for the 2-design and the 3-design, this time truncating the values of $r$ considered when generating density matrices so that $\epsilon$ did not exceed 0.5.  The test results are summarised in \tabref{tab:results}.  Applying quantum readout error mitigation improved the test results.  The changes in the values of $\epsilon$ resulting from replacing the experimentally determined relative frequencies with uniform probabilities, are mostly within the error margins.  This suggests that non-uniformity did not significantly impair the quality of the ensemble.

The divergence in $\epsilon$ observed for states close to the surface of the Bloch sphere is likely a result of pure states becoming inaccessible due to depolarising noise and was investigated further by applying the test for the 1-design, the 2-design and the 3-design to an exact 3-design combined with a depolarising channel.  The full study is presented in \appendref{append:depolarising} and shows that depolarising noise is a very good noise model for the data.  Our measurement-based implementation of the identity operation (see \appendref{append:identity}) shows that depolarising noise is the predominant type of noise in measurement-based processes on IBM processors, providing further confirmation that depolarising noise is indeed what prevented the tests for the 2-design and the 3-design from passing.  Urbanek \textit{et al.}~recently proposed a method for mitigating depolarising noise in quantum computations where the final outcome is an expectation value~\cite{depnoisemit}.  Since the final outcome of our implementation is a quantum process or channel, and not an expectation value, this method is unfortunately not applicable here.  However, their method may have potential for improving results in applications of measurement-based $t$-designs where the final outcome is an expectation value.

To determine the fraction of the Bloch sphere for which a test passes, we consider 8000 evenly spaced points in a cube which encloses the Bloch sphere.  Points in the Bloch sphere then correspond to valid states.  For each valid state, we determine whether inequality (\ref{eq:master}) can be satisfied with $\epsilon \leq 0.5$.  The fraction of the Bloch sphere for which a test passes is given by the number of states for which the inequality can be satisfied divided by the number of states considered.  The fraction of states for which the ensemble of unitaries generated using the \textit{ibmq\_toronto} quantum processor passed the test for the 1-design, the 2-design and the 3-design are given in \tabref{tab:volume}.  The fraction of states for which the test for the 2-design and the 3-design pass is substantially improved by quantum readout error mitigation.  This confirms that classical measurement noise is responsible for many states failing to satisfy inequality~(\ref{eq:master}) and, if left uncorrected, would result in test results which greatly underestimate the extent to which the various designs are realised by our implementation on the \textit{ibmq\_toronto} quantum processor.

%%%%%%%%%%%%%%%%%%%%%%%%%%%%
%%%%%%%%%%%%%%%%%%%%%%%%%%%%
\subsection{Approximate 2-design}\label{sec:2-design}

Turner and Markham~\cite{MBT1} show that there is no set of measurement angles such that the ensemble of unitaries generated by performing single-qubit measurements on the 5-qubit linear cluster state is an exact 2-design.  However, applying our test for an approximate 2-design to the expected ensemble of unitaries generated for the 5-qubit cluster state and the measurement angles $\phi_1=0$, $\phi_2=\frac{\pi}{4}$, $\phi_3=\frac{\pi}{4}$ and $\phi_4=0$ yields $\epsilon=0.5$.  Hence this ensemble is an approximate 2-design, with our passing criterion, although it must be noted that the ensemble does not resemble an exact 2-design closely and that the passing criterion is satisfied only for the subset of density matrices in $\mathcal{B}(\mathbb{C}^2\otimes\mathbb{C}^2)$ which are tensor products of pairs of single-qubit states.  We implemented this approximate measurement-based 2-design on 5 physical qubits of the \textit{ibmq\_sydney} quantum processor.  \appendref{append:qubits2} provides more detail on the \textit{ibmq\_sydney} quantum processor, the qubits that were used and why the \textit{ibmq\_sydney} quantum processor was used for this experiment instead of the \textit{ibmq\_toronto} quantum processor.  Generation of channel tomography results for the 16 different unitary operations corresponding to the 16 different measurement outcomes, determining of relative frequencies, combining of counts to reduce statistical noise and construction of calibration matrices for quantum readout error mitigation were all done in the same way as for the exact 3-design implementation on the \textit{ibmq\_toronto} quantum processor.

\begin{table}[]
\begin{tabular}{|l|l|l|}
\hline
\textbf{Outcome} & \textbf{Fidelity (Raw)} & \textbf{Fidelity (Processed)} \\ \hline
0000            & 0.7931$\pm$0.0033      & 0.8947$\pm$0.0038            \\ \hline
0001            & 0.8871$\pm$0.0042      & 0.9851$\pm$0.0047            \\ \hline
0010            & 0.8382$\pm$0.0051      & 0.9220$\pm$0.0061            \\ \hline
0011            & 0.8506$\pm$0.0040      & 0.9242$\pm$0.0047            \\ \hline
0100            & 0.8231$\pm$0.0062      & 0.8900$\pm$0.0069            \\ \hline
0101            & 0.8912$\pm$0.0053      & 0.9649$\pm$0.0053            \\ \hline
0110            & 0.8399$\pm$0.0051      & 0.8978$\pm$0.0058            \\ \hline
0111            & 0.8885$\pm$0.0041      & 0.9378$\pm$0.0047            \\ \hline
1000            & 0.7991$\pm$0.0039      & 0.8978$\pm$0.0044            \\ \hline
1001            & 0.9039$\pm$0.0053      & 0.9944$\pm$0.0063            \\ \hline
1010            & 0.8365$\pm$0.0058      & 0.9061$\pm$0.0062            \\ \hline
1011            & 0.8885$\pm$0.0059      & 0.9527$\pm$0.0061            \\ \hline
1100            & 0.8487$\pm$0.0052      & 0.9183$\pm$0.0054            \\ \hline
1101            & 0.9052$\pm$0.0055      & 0.9639$\pm$0.0055            \\ \hline
1110            & 0.8571$\pm$0.0030      & 0.9053$\pm$0.0031            \\ \hline
1111            & 0.9027$\pm$0.0060      & 0.9448$\pm$0.0066            \\ \hline
\end{tabular}
\caption{Channel fidelities for the 16 random unitaries, corresponding to the 16 different measurement outcomes, generated with the approximate 2-design on the \textit{ibmq\_sydney} quantum processor.  `Raw' shows the channel fidelities without quantum readout error mitigation.  `Processed' shows the channel fidelities with quantum readout error mitigation.}
\label{tab:fidelities2}
\end{table}

\begin{table}[]
\begin{tabular}{|l|l|l|}
\hline
\textbf{Outcome} & \textbf{Frequency (Raw)} & \textbf{Frequency (Processed)} \\ \hline
0000            & 0.07835$\pm$0.00104     & 0.06486$\pm$0.00113           \\ \hline
0001            & 0.06705$\pm$0.00151     & 0.06268$\pm$0.00169           \\ \hline
0010            & 0.07319$\pm$0.00100     & 0.06585$\pm$0.00122           \\ \hline
0011            & 0.05817$\pm$0.00074     & 0.05930$\pm$0.00091           \\ \hline
0100            & 0.07856$\pm$0.00119     & 0.07513$\pm$0.00138           \\ \hline
0101            & 0.05442$\pm$0.00073     & 0.05646$\pm$0.00096           \\ \hline
0110            & 0.07869$\pm$0.00115     & 0.08351$\pm$0.00144           \\ \hline
0111            & 0.04837$\pm$0.00095     & 0.05567$\pm$0.00121           \\ \hline
1000            & 0.06820$\pm$0.00120     & 0.05981$\pm$0.00137           \\ \hline
1001            & 0.05429$\pm$0.00100     & 0.05296$\pm$0.00120           \\ \hline
1010            & 0.06837$\pm$0.00120     & 0.06680$\pm$0.00135           \\ \hline
1011            & 0.04801$\pm$0.00088     & 0.05098$\pm$0.00107           \\ \hline
1100            & 0.06347$\pm$0.00120     & 0.06286$\pm$0.00150           \\ \hline
1101            & 0.04988$\pm$0.00149     & 0.05518$\pm$0.00191           \\ \hline
1110            & 0.06642$\pm$0.00137     & 0.07384$\pm$0.00168           \\ \hline
1111            & 0.04458$\pm$0.00125     & 0.05411$\pm$0.00164           \\ \hline
\end{tabular}
\caption{Relative frequencies with which the 16 random unitaries, corresponding to the 16 different measurement outcomes, are generated with the approximate 2-design on the \textit{ibmq\_sydney} quantum processor.  `Raw' shows the relative frequencies without quantum readout error mitigation.  `Processed' shows the relative frequencies with quantum readout error mitigation.}
\label{tab:frequencies2}
\end{table}

Channel fidelities for each of the 16 different random unitaries are given in \tabref{tab:fidelities2}.  The average channel fidelity is (0.8596$\pm$0.0356) without quantum readout error mitigation and (0.9312$\pm$0.0323) with quantum readout error mitigation.  The average channel fidelity is larger than the average channel fidelity for the exact 3-design implementation, reflecting reduced noise in the implementation with the smaller cluster state with fewer qubits.  The relative frequencies with which each of the 16 different random unitaries are generated are given in \tabref{tab:frequencies2}.  The average relative frequency is (0.06250$\pm$0.01127) without quantum readout error mitigation and (0.06250$\pm$0.00871) with quantum readout error mitigation.  The average relative frequency is once again equal to the expected uniform probability of $\frac{1}{16}=0.0625$.

\begin{table*}[]
\begin{tabular}{|l|l|l|l|l|l|l|l|l|}
\hline
\textbf{}     & \multicolumn{4}{l|}{\textbf{Raw}}                                                                 & \multicolumn{4}{l|}{\textbf{Processed}}                                                           \\ \cline{2-9} 
\textbf{Test} & \textbf{Radius} & $\boldsymbol{\epsilon}$ & $\boldsymbol{\epsilon}$ \textbf{(uniform)} & $\boldsymbol{\epsilon}$ \textbf{(ideal)} & \textbf{Radius} & $\boldsymbol{\epsilon}$ & $\boldsymbol{\epsilon}$ \textbf{(uniform)} & $\boldsymbol{\epsilon}$ \textbf{(ideal)} \\ \hline
1-design      & 1.00            & 0.1505$\pm$0.0056   & 0.1468$\pm$0.0073            & 0.0000                     & 1.00            & 0.1505$\pm$0.0056   & 0.1397$\pm$0.0065            & 0.0000                     \\ \hline
2-design      & 0.69            & 0.4488$\pm$0.0035   & 0.4690$\pm$0.0052            & 0.2739                     & 0.81            & 0.4623$\pm$0.0103   & 0.4865$\pm$0.0073            & 0.3589                     \\ \hline
\end{tabular}
\caption{Summary of test results for the ensemble of unitaries generated using the \textit{ibmq\_sydney} quantum processor.  `Raw' shows the results without quantum readout error mitigation.  `Processed' shows the results with quantum readout error mitigation.  `Radius' is the truncation radius considered for a test.  The column with `uniform' shows the values of $\epsilon$ obtained when replacing the experimentally determined relative frequencies with uniform probabilities.  The column with `ideal' shows the expected values of $\epsilon$ for the approximate 2-design for the truncation radii considered.}
\label{tab:results2}
\end{table*}

\begin{table*}[]
\begin{tabular}{|l|l|l|l|l|}
\hline
\textbf{}     & \multicolumn{2}{l|}{\textbf{Raw}}          & \multicolumn{2}{l|}{\textbf{Processed}}    \\ \cline{2-5} 
\textbf{Test} & \textbf{Frac}     & \textbf{Frac (uniform)} & \textbf{Frac}     & \textbf{Frac (uniform)} \\ \hline
1-design      & 1.0000            & 1.0000                 & 1.0000            & 1.0000                 \\ \hline
2-design      & 0.3984$\pm$0.0017 & 0.4104$\pm$0.0020      & 0.6773$\pm$0.0095 & 0.6925$\pm$0.0050      \\ \hline
\end{tabular}
\caption{Fraction of states for which the ensemble of unitaries generated using the \textit{ibmq\_sydney} quantum processor passed the different tests.  `Raw' shows the fractions without quantum readout error mitigation.  `Processed' shows the fractions with quantum readout error mitigation.  The column with `uniform' shows the fractions obtained when replacing the experimentally determined relative frequencies with uniform probabilities.}
\label{tab:volume2}
\end{table*}

We applied our test for the 1-design and the 2-design to the ensemble of unitaries generated using the \textit{ibmq\_sydney} quantum processor.  The test for the 1-design passed, but the test for the 2-design did not.  Test results are summarised in \tabref{tab:results2} and the fraction of states for which each test passed is conveyed in \tabref{tab:volume2}.  The values of $\epsilon$ obtained for the ensemble of unitaries generated using the \textit{ibmq\_sydney} quantum processor, for a given truncation radius, are not much larger than the expected values for the approximate 2-design.  This suggests that inherent deviations from an exact 2-design, present in the approximate 2-design considered, had a more significant effect on the quality of the ensemble of unitaries than noise on the \textit{ibmq\_sydney} quantum processor.

The fraction of states for which the test for the 2-design passed with quantum readout error mitigation is almost double that without quantum readout error mitigation.  This confirms that the noise in this implementation was also predominantly classical measurement errors.  The effect of readout errors was more pronounced in this implementation, likely because gate errors of the qubits used are much smaller (see \appendref{append:qubits2}).  We note that the fraction of states for which the ensemble of unitaries generated with the 5-qubit cluster state on the \textit{ibmq\_sydney} quantum processor passed the test for the 2-design, especially with quantum readout error mitigation, is larger than that of the ensemble of unitaries generated with the 6-qubit cluster state on the \textit{ibmq\_toronto} quantum processor.  Hence, even though the approximate 2-design considered does not closely resemble an exact 2-design, the ensemble of unitaries generated with this approximate 2-design implementation more closely resembles a 2-design than the ensemble of unitaries generated with the exact 3-design implementation.  This is as a result of significantly reduced noise in the implementation with the smaller 5-qubit cluster state, compared to the implementation with the larger 6-qubit cluster state.

%%%%%%%%%%%%%%%%%%%%%%%%%%%%
%%%%%%%%%%%%%%%%%%%%%%%%%%%%
%%%%%%%%%%%%%%%%%%%%%%%%%%%%
%%%%%%%%%%%%%%%%%%%%%%%%%%%%
\section{Conclusion}\label{sec:conclusion} 

The exact measurement-based 3-design of Ref.~\cite{MBT1} was implemented by performing single-qubit measurements on a 6-qubit linear cluster state, prepared on the \textit{ibmq\_toronto} quantum processor.  To infer the ensemble of unitaries realised in the implementation, we performed channel tomography for all possible measurement outcomes.  This ensemble of unitaries passed our test for a 1-design, but not for a 2-design or a 3-design.  Further studies, presented in Appendices~\ref{append:depolarising} and \ref{append:identity}, strongly suggest that depolarising noise prevented the tests for the 2-design and the 3-design from passing.  Therefore, for measurement-based $t$-designs to be effectively realised for $t>1$ on superconducting systems, such as IBM quantum processors, a significant amount of work will need to be done to reduce or mitigate depolarising noise in these devices.

The noteworthy improvement in results obtained by applying quantum readout error mitigation confirms that classical measurement errors are indeed responsible for a substantial amount of noise on IBM quantum processors in this instance.  It also shows the importance of mitigating these errors, as not doing so would lead to results that give an inaccurate account of the actual implementations realised on these processors.  The ensemble of unitaries realised by our approximate measurement-based 2-design implementation, in which single-qubit measurements were performed on a 5-qubit linear cluster state prepared on the \textit{ibmq\_sydney} quantum processor, showed improved results for the 2-design test as a result of reduced noise for the smaller 5-qubit cluster state.  This clearly demonstrates the advantage of keeping entangled resource states used in measurement-based processes small.  It also shows that in experimental realisations (where noise is present), the quality of a noisy approximate $t$-design may be better than the quality of a noisy exact $t$-design, if the implementation of the approximate $t$-design is significantly less sensitive to noise than the implementation of the exact $t$-design.

%%%%%%%%%%%%%%%%%%%%%%%%%%%%
%%%%%%%%%%%%%%%%%%%%%%%%%%%%
%%%%%%%%%%%%%%%%%%%%%%%%%%%%
%%%%%%%%%%%%%%%%%%%%%%%%%%%%
\section*{Data availability} 

The datasets generated during and/or analysed during the study are available from the corresponding author
on reasonable request.

\begin{acknowledgements}
We acknowledge the use of IBM Quantum services for this work. The views expressed are those of the authors, and do not reflect the official policy or position of IBM or the IBM Quantum team. We thank Taariq Surtee and Barry Dwolatzky at the University of Witwatersrand and Ismail Akhalwaya at IBM Research Africa for access to the IBM processors through the Q Network and Africa Research Universities Alliance.  We also thank Damian Markham for valuable comments and insights.  This research was supported by the South African National Research Foundation, the University of Stellenbosch, and the South African Research Chair Initiative of the Department of Science and Technology and National Research Foundation.
\end{acknowledgements}

\section*{Author contributions}

CS and MT conceived the idea and designed the experiments. CS performed the experiments. Both authors analysed the results, contributed to the discussions and interpretations. CS and MT wrote the manuscript.

\section*{Competing interests} 

The authors declare no competing interests.

%%%%%%%%%%%%%%%%%%%%%%%%%%%%
%%%%%%%%%%%%%%%%%%%%%%%%%%%%
%%%%%%%%%%%%%%%%%%%%%%%%%%%%
%%%%%%%%%%%%%%%%%%%%%%%%%%%%

\appendix

\onecolumngrid

\newpage

%%%%%%%%%%%%%%%%%%%%%%%%%%%%
%%%%%%%%%%%%%%%%%%%%%%%%%%%%
%%%%%%%%%%%%%%%%%%%%%%%%%%%%
%%%%%%%%%%%%%%%%%%%%%%%%%%%%
\section{Qubits used for the 3-design}\label{append:qubits3} 

\begin{figure*}[h]
    \centering
    \begin{tikzpicture}
    \draw (3, 1) circle[radius=0.25] node {6};
    \draw (7, 1) circle[radius=0.25] node {17};
    \draw (0, 0) circle[radius=0.25] node {0};
    \draw (1, 0) circle[radius=0.25] node {1};
    \draw (2, 0) circle[radius=0.25] node {4};
    \draw (3, 0) circle[radius=0.25] node {7};
    \draw (4, 0) circle[radius=0.25] node {10};
    \draw (5, 0) circle[radius=0.25] node {12};
    \draw (6, 0) circle[radius=0.25] node {15};
    \draw (7, 0) circle[radius=0.25] node {18};
    \draw (8, 0) circle[radius=0.25] node {21};
    \filldraw[fill=gray, draw=black] (9, 0) circle[radius=0.25] node {23};
    \draw (1, -1) circle[radius=0.25] node {2};
    \draw (5, -1) circle[radius=0.25] node {13};
    \filldraw[fill=gray, draw=black] (9, -1) circle[radius=0.25] node {24};
    \draw (1, -2) circle[radius=0.25] node {3};
    \draw (2, -2) circle[radius=0.25] node {5};
    \draw (3, -2) circle[radius=0.25] node {8};
    \draw (4, -2) circle[radius=0.25] node {11};
    \draw (5, -2) circle[radius=0.25] node {14};
    \filldraw[fill=gray, draw=black] (6, -2) circle[radius=0.25] node {16};
    \filldraw[fill=gray, draw=black] (7, -2) circle[radius=0.25] node {19};
    \filldraw[fill=gray, draw=black] (8, -2) circle[radius=0.25] node {22};
    \filldraw[fill=gray, draw=black] (9, -2) circle[radius=0.25] node {25};
    \draw (10, -2) circle[radius=0.25] node {26};
    \draw (3, -3) circle[radius=0.25] node {9};
    \draw (7, -3) circle[radius=0.25] node {20};
    \draw (3, 0.25) -- (3, 0.75);
    \draw (7, 0.25) -- (7, 0.75);
    \draw (0.25, 0) -- (0.75, 0);
    \draw (1.25, 0) -- (1.75, 0);
    \draw (2.25, 0) -- (2.75, 0);
    \draw (3.25, 0) -- (3.75, 0);
    \draw (4.25, 0) -- (4.75, 0);
    \draw (5.25, 0) -- (5.75, 0);
    \draw (6.25, 0) -- (6.75, 0);
    \draw (7.25, 0) -- (7.75, 0);
    \draw (8.25, 0) -- (8.75, 0);
    \draw (1, -0.25) -- (1, -0.75);
    \draw (5, -0.25) -- (5, -0.75);
    \draw (9, -0.25) -- (9, -0.75);
    \draw (1, -1.25) -- (1, -1.75);
    \draw (5, -1.25) -- (5, -1.75);
    \draw (9, -1.25) -- (9, -1.75);
    \draw (1.25, -2) -- (1.75, -2);
    \draw (2.25, -2) -- (2.75, -2);
    \draw (3.25, -2) -- (3.75, -2);
    \draw (4.25, -2) -- (4.75, -2);
    \draw (5.25, -2) -- (5.75, -2);
    \draw (6.25, -2) -- (6.75, -2);
    \draw (7.25, -2) -- (7.75, -2);
    \draw (8.25, -2) -- (8.75, -2);
    \draw (9.25, -2) -- (9.75, -2);
    \draw (3, -2.25) -- (3, -2.75);
    \draw (7, -2.25) -- (7, -2.75);
    \end{tikzpicture}
    \caption{Qubit topology of the \textit{ibmq\_toronto} quantum processor.  The connecting lines between qubits indicate the qubit pairs for which the $CX$ gate is supported at the hardware level.  The qubits used for the exact 3-design implementation are shaded gray.}
    \label{fig:topology}
\end{figure*}
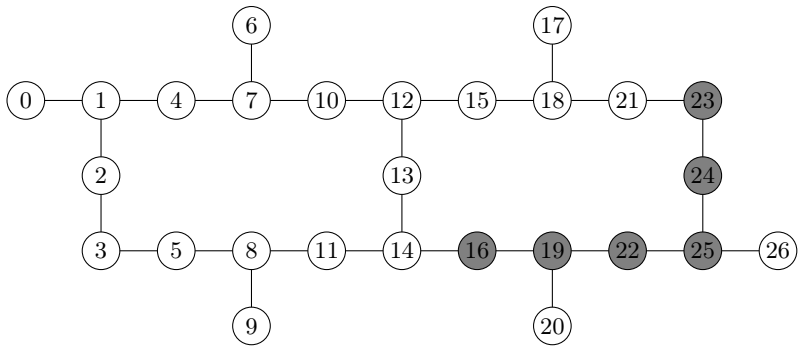

The exact measurement-based 3-design was implemented on 6 physical qubits of the \textit{ibmq\_toronto} quantum processor.  This processor was chosen for its low error rates compared to other processors.  The qubit topology of the \textit{ibmq\_toronto} quantum processor is shown in \figref{fig:topology}.  The qubits 1 to 6 of the 6-qubit linear cluster state in the 3-design implementation were mapped onto the physical qubits 16, 19, 22, 25, 24 and 23 of the \textit{ibmq\_toronto} quantum processor, in such a way that the input state was prepared on qubit 16 and the output state was retrieved from qubit 23.  These qubits were chosen as they form one of the few sets of 6 connected qubits (see \figref{fig:topology}) in which all the qubits generally have relatively low error rates.  The relevant calibration information as obtained at the time of running the circuits for the exact 3-design implementation, is shown in \tabref{tab:calibration3}.

\begin{table}[h]
\centering
\begin{minipage}{9.5cm}
\centering
\begin{tabular}{|l|l|l|l|l|}
\hline
\textbf{Qubit} & $\boldsymbol{T_1}\ \boldsymbol{(}\boldsymbol{\mu}\mathbf{s}\boldsymbol{)}$ & $\boldsymbol{T_2}\ \boldsymbol{(}\boldsymbol{\mu}\mathbf{s}\boldsymbol{)}$ & $\boldsymbol{\sqrt{X}}$ \textbf{Error} & \textbf{Readout Error} \\ \hline
16             & 123.42           & 135.48           & 0.000297         & 0.0154                 \\ \hline
19             & 114.98           & 123.10           & 0.000434         & 0.0116                 \\ \hline
22             & 110.50           & 148.90           & 0.000330         & 0.0191                 \\ \hline
25             & 125.41           & 114.64           & 0.000323         & 0.0108                 \\ \hline
24             & 121.49           & 155.26           & 0.000180         & 0.0093                 \\ \hline
23             & $\,\,\,$98.92            & $\,\,\,$40.59            & 0.000441         & 0.0553                 \\ \hline
\end{tabular}
\end{minipage}%
\begin{minipage}{6.5cm}
\centering
\begin{tabular}{|l|l|}
\hline
\textbf{Qubit Pair} & $\boldsymbol{C}\boldsymbol{X}$ \textbf{Error} \\ \hline
16--19              & 0.00758           \\ \hline
19--22              & 0.01024           \\ \hline
22--25              & 0.01053           \\ \hline
25--24              & 0.01099           \\ \hline
24--23              & 0.00892           \\ \hline
\end{tabular}
\end{minipage}
\caption{Calibration information for the \textit{ibmq\_toronto} quantum processor as obtained at the time of running the circuits for the exact 3-design.  The single-qubit calibration information for the relevant qubits is shown on the left.  $T_1$ and $T_2$ are the relaxation and dephasing times respectively of the qubits.  The $CX$ error rates for relevant qubit pairs are shown on the right.}
\label{tab:calibration3}
\end{table}

%%%%%%%%%%%%%%%%%%%%%%%%%%%%
%%%%%%%%%%%%%%%%%%%%%%%%%%%%
%%%%%%%%%%%%%%%%%%%%%%%%%%%%
%%%%%%%%%%%%%%%%%%%%%%%%%%%%
\section{Depolarising noise}\label{append:depolarising}

The action of the depolarising channel on a single-qubit state $\rho$ is described by
\begin{equation}
\rho' = \frac{p}{2}I + (1-p)\rho,
\end{equation}
that is, the state is replaced by the maximally mixed state with probability $p$.  We investigated the effect of depolarising noise on an exact $t$-design's ability to accurately reproduce the moments of the uniform Haar ensemble.  In particular, we carried out the test for an approximate $t$-design with the middle term in inequality (\ref{eq:master}) replaced by $\mathbb{E}^t_H({\rho'}^{\otimes t})$, the expectation of the uniform Haar ensemble computed from a state to which depolarising noise has been applied.

\begin{figure*}
    \centering
    \includegraphics[trim=1cm 0.3cm 0cm 1.5cm, clip, scale=0.5]{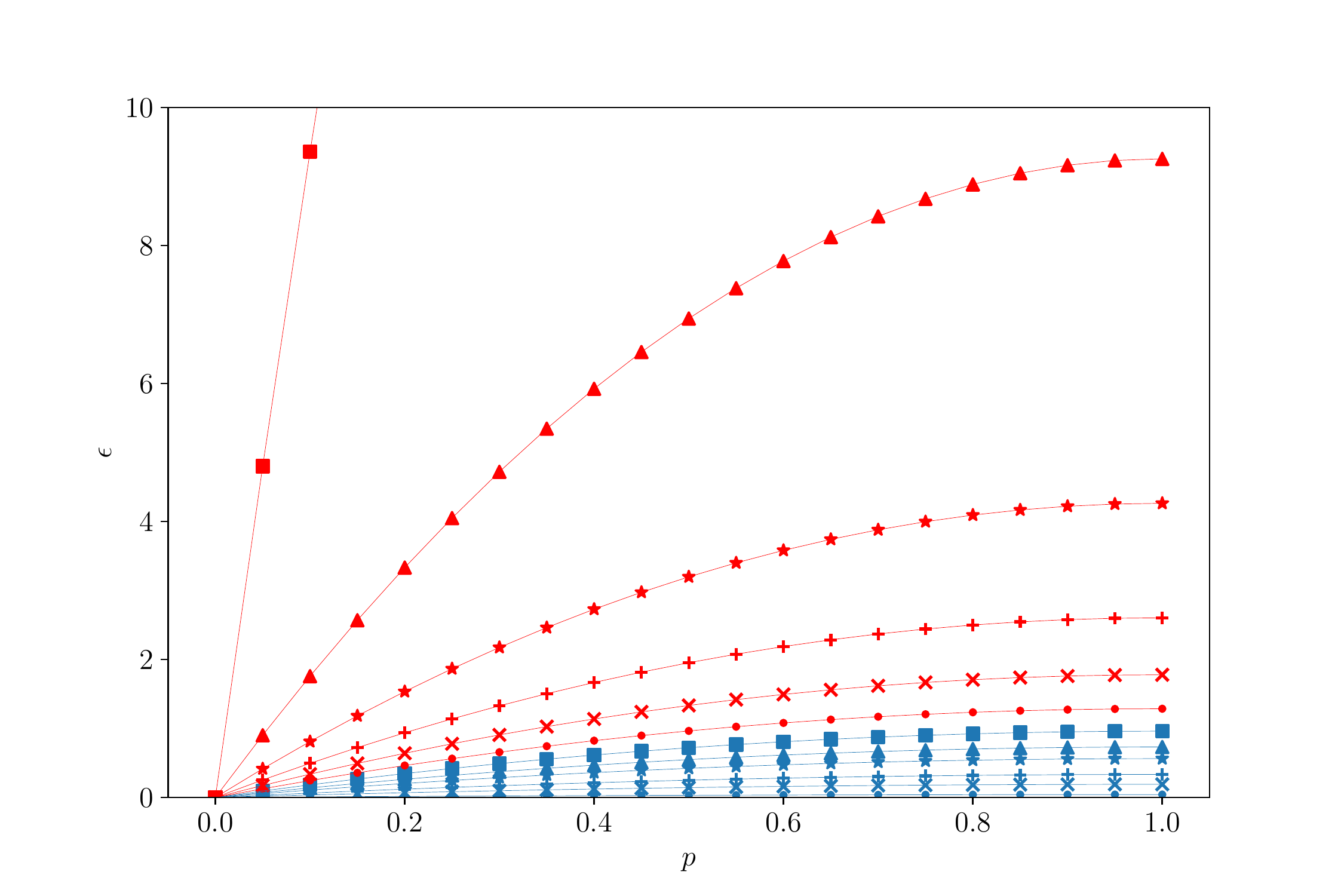}
    \vspace{0.5cm}
    \includegraphics[trim=0cm 0.3cm 0cm 1.5cm, clip, scale=0.5]{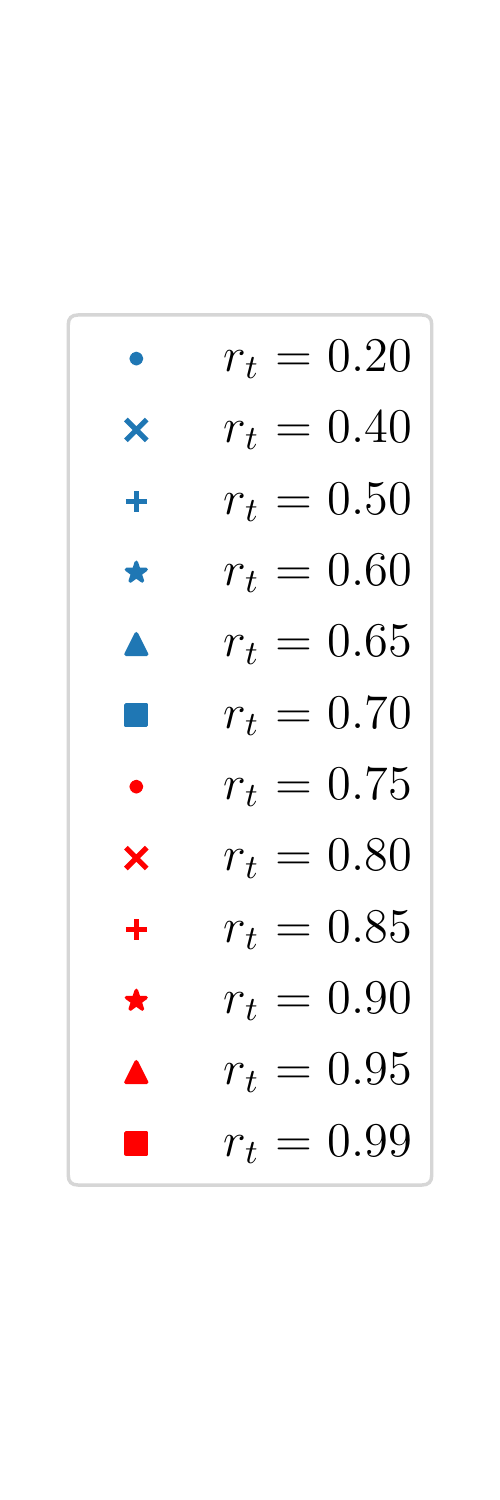}
    \includegraphics[trim=1cm 0.3cm 0cm 1.5cm, clip, scale=0.5]{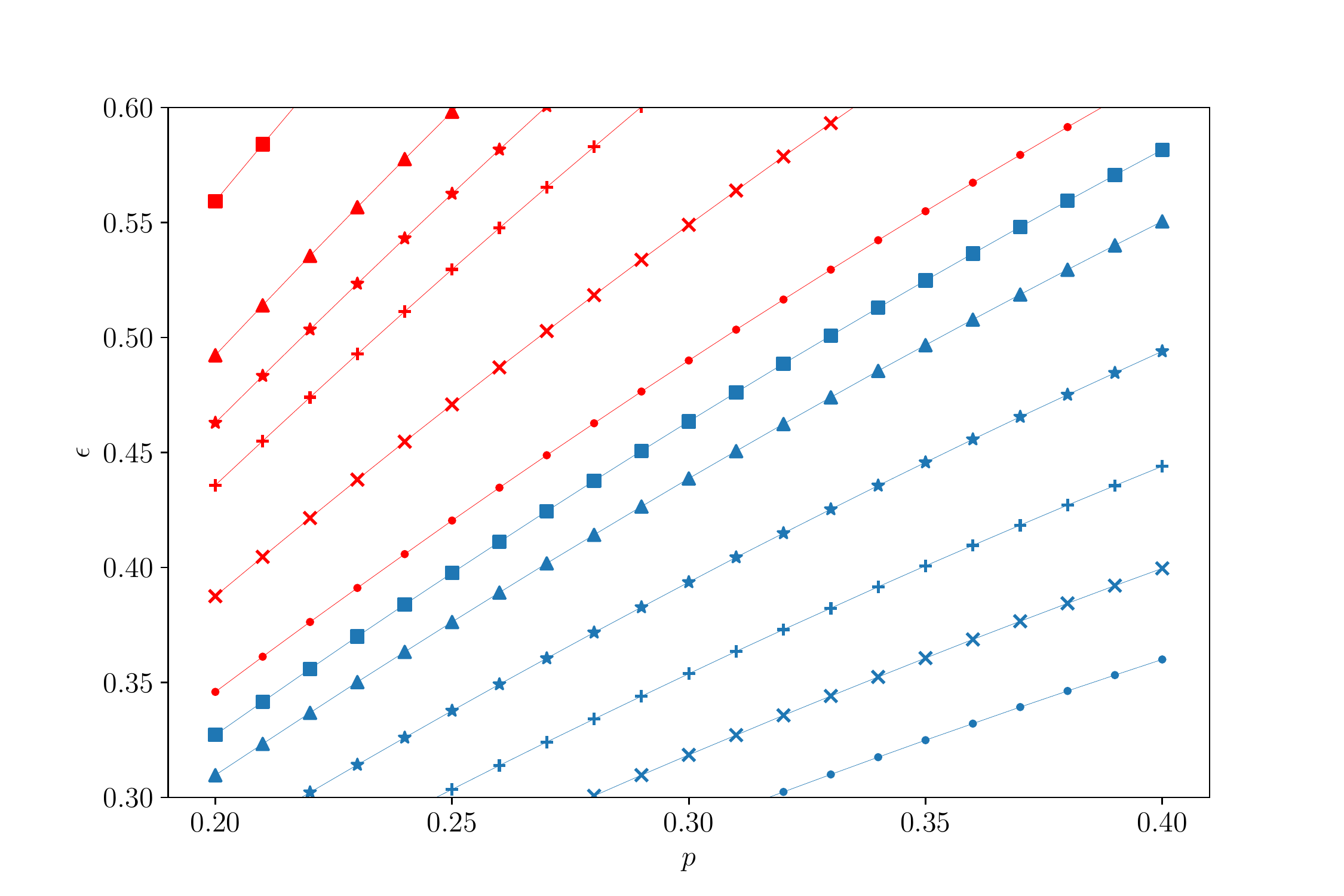}
    \includegraphics[trim=0cm 0.3cm 0cm 1.5cm, clip, scale=0.5]{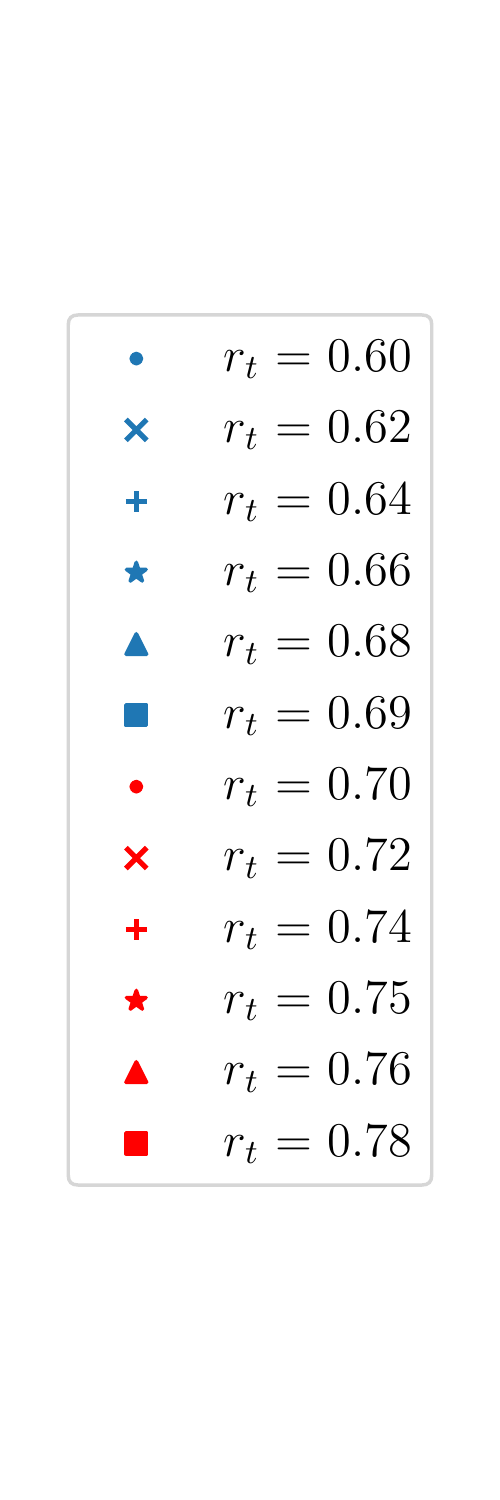}
    \caption{$\epsilon$ versus $p$ for the 2-design test with the middle term in inequality (\ref{eq:master}) replaced by $\mathbb{E}^2_H({\rho'}^{\otimes 2})$, for different truncation radii $r_t$.  A plot of the full set of results is shown above and a plot focused on the region of interest is shown below.}
    \label{fig:depolarisingsingle}
\end{figure*}

We managed to obtain test results analytically for the 1-design.  Using the Pauli 1-design, $\{I, X, Y, Z\}$, we showed that $\mathbb{E}^1_H(\rho)=\frac{1}{2}I$ for all states $\rho$, which we used to show that $\epsilon=0$ for all $p \in [0,1]$.  Hence depolarising noise has no effect on a 1-design's ability to reproduce the first moment of the uniform Haar ensemble.  Since $\mathbb{E}^2_H(\rho^{\otimes 2})$ and $\mathbb{E}^3_H(\rho^{\otimes 3})$ depend on the state $\rho$, analytic test results for the 2-design and 3-design are much harder to find.  Using a sample of 1000 density matrices, obtained as described in \secref{sec:test}, and computing $\mathbb{E}^2_H(\rho^{\otimes 2})$ and $\mathbb{E}^3_H(\rho^{\otimes 3})$ for each density matrix $\rho$ using the exact 3-design described in \secref{sec:mbbackground}, we obtained results numerically for the 2-design and the 3-design.  Test results obtained for the 2-design for different values of $p$ and different truncation radii $r_t$ are plotted in \figref{fig:depolarisingsingle}.  For all truncation radii, $\epsilon$ increases linearly with $p$, up to about $p=0.4$, after which the increase becomes more gradual.  The values of $\epsilon$ obtained for states close to the surface of the Bloch sphere are very large, even for small $p$.  This shows that the second moment of the uniform Haar ensemble is very sensitive to depolarising noise.  Test results for the 3-design are identical to that of the 2-design, shown in \figref{fig:depolarisingsingle}.  This suggests that the third moment of the uniform Haar ensemble is unaffected by depolarising noise.

To determine whether this depolarising noise model is a good noise model for the exact 3-design implementation on the \textit{ibmq\_toronto} quantum processor, we attempt to infer a consistent value for the parameter $p$ from the test results in \tabref{tab:results}.  Given $r_t$ and $\epsilon$ obtained in a test for the 2-design or the 3-design, we simply read off the corresponding value of $p$ from the plot in \figref{fig:depolarisingsingle}.  Using the results for the 2-design test (without quantum readout error mitigation) we infer $p=0.31$ and using the results for the 3-design test we infer $p=0.36$.  For the results with quantum readout error mitigation, we infer $p=0.20$ using the 2-design test results and $p=0.32$ using the 3-design test results.  Since the values inferred from the 2-design test and the 3-design test are different in both cases, we conclude that this depolarising noise model is not the correct noise model for the 3-design implementation.

\begin{figure*}
    \centering
    \includegraphics[trim=1cm 0.3cm 0cm 1.5cm, clip, scale=0.5]{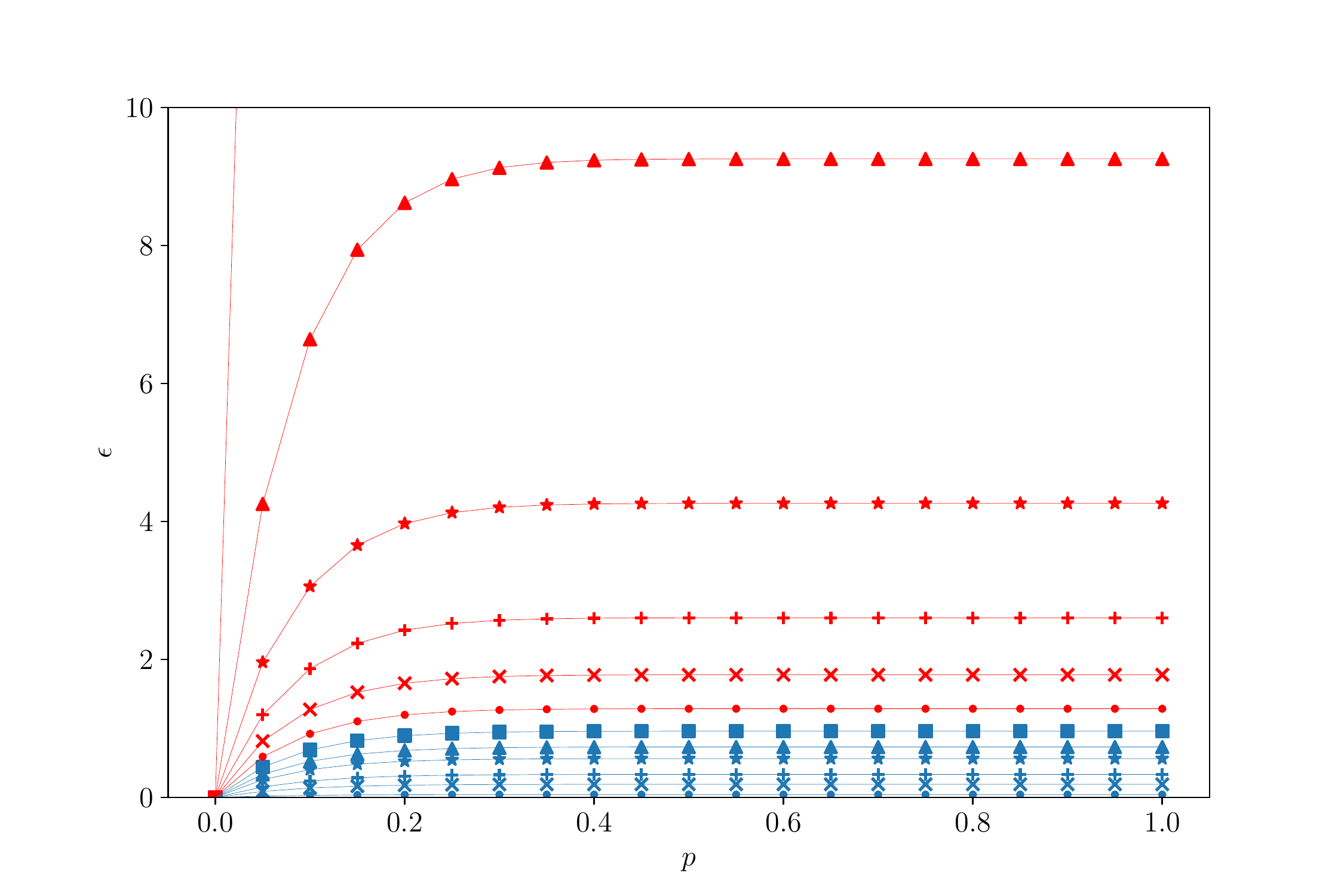}
    \vspace{0.5cm}
    \includegraphics[trim=0cm 0.3cm 0cm 1.5cm, clip, scale=0.5]{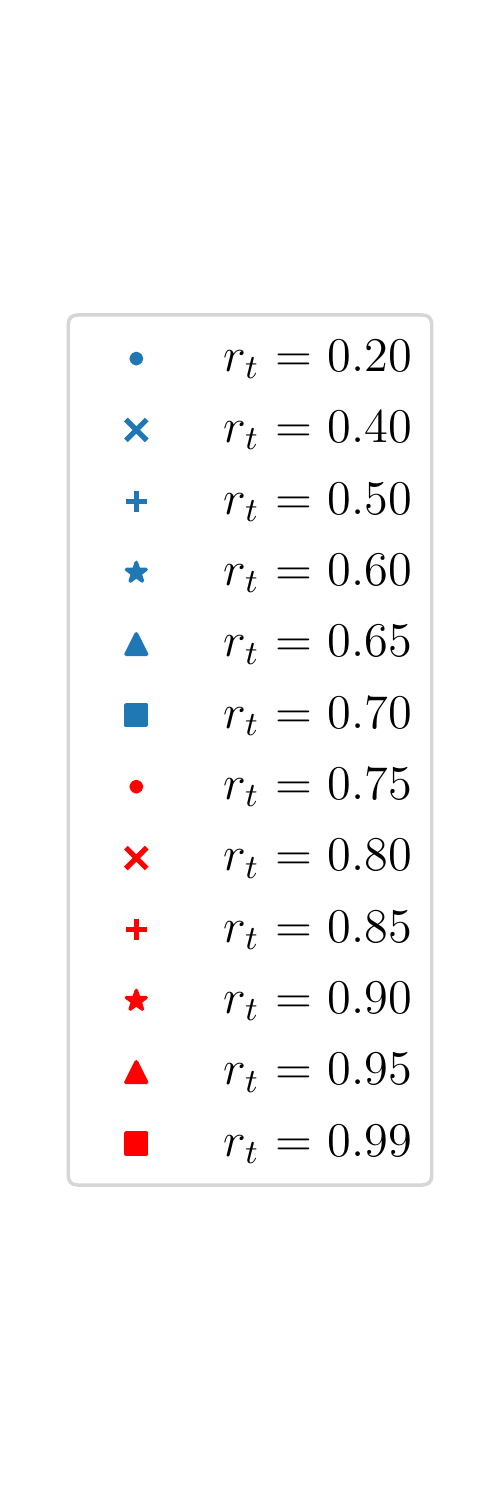}
    \includegraphics[trim=1cm 0.3cm 0cm 1.5cm, clip, scale=0.5]{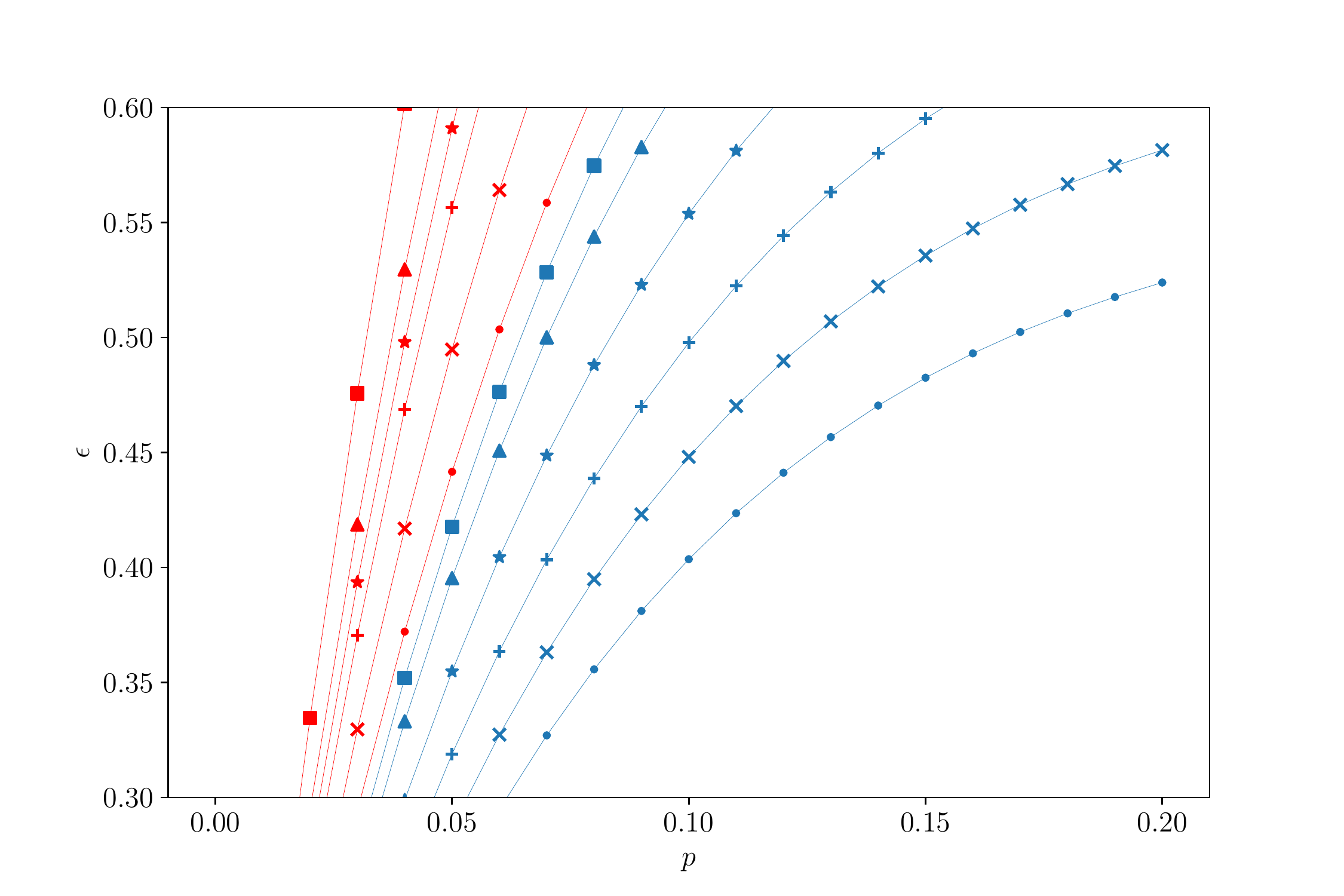}
    \includegraphics[trim=0cm 0.3cm 0cm 1.5cm, clip, scale=0.5]{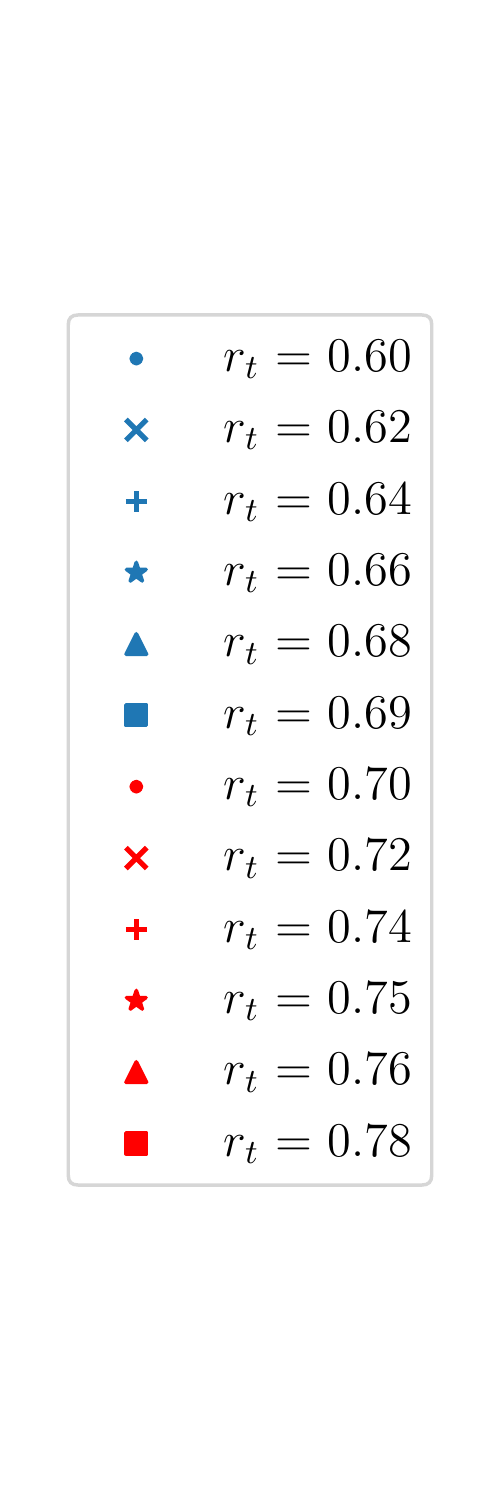}
    \caption{$\epsilon$ versus $p$ for the 2-design test with the middle term in inequality (\ref{eq:master}) calculated by repeatedly applying depolarising noise to a state, in between the 5 individual unitary operations that are applied to the input state in the exact 3-design described in \secref{sec:mbbackground}, for different truncation radii $r_t$.  A plot of the full set of results is shown above and a plot focused on the region of interest is shown below.}
    \label{fig:depolarisingcontinuous}
\end{figure*}

We now consider a depolarising noise model which more closely resembles the way in which depolarising noise occurs in $t$-designs generated using a measurement-based approach.  In this model, the test for an approximate $t$-design is performed with the middle term in inequality (\ref{eq:master}) calculated by repeatedly applying depolarising noise to a state, in between the 5 individual unitary operations that are applied to the input state in the exact 3-design described in \secref{sec:mbbackground}.  Our analytic results for the 1-design carry over to this model.  Test results obtained numerically for the 2-design are shown in \figref{fig:depolarisingcontinuous}.  For this model, $\epsilon$ versus $p$ starts to plateau at much smaller values of $p$.  As a result of repeated applications of depolarising noise, the largest possible $\epsilon$, for a given truncation radius, is reached for much smaller $p$.  Results for the 3-design test are once again identical to that of the 2-design.  Considering this model and using \figref{fig:depolarisingcontinuous}, we infer $p=0.06$ using the results for the 2-design test (without quantum readout error mitigation) and $p=0.07$ using the results for the 3-design test.  For the results with quantum readout error mitigation, we infer $p=0.04$ using the 2-design test results and $p=0.06$ using the 3-design test results.  The values of $p$ inferred here from the two test results are close enough to be considered consistent in both cases.  We therefore conclude that this is a very good noise model for the exact 3-design implementation on the \textit{ibmq\_toronto} quantum processor.  From the values of $p$ inferred, it is also clear that quantum readout error mitigation has reduced depolarising noise in the implementation.

%%%%%%%%%%%%%%%%%%%%%%%%%%%%
%%%%%%%%%%%%%%%%%%%%%%%%%%%%
%%%%%%%%%%%%%%%%%%%%%%%%%%%%
%%%%%%%%%%%%%%%%%%%%%%%%%%%%
\section{Identity implementation}\label{append:identity} 

Measurement-based processing using $n$-qubit linear cluster states, as summarised in \secref{sec:mbbackground}, can be used to implement the identity operation for odd $n$.  For measurements in the Pauli $X$-basis, that is in the direction $\phi=0$, \equatref{eq:unitary} reduces to $U_m(0)=HZ^m$.  Hence when all measurements on a $n$-qubit cluster state are performed in the Pauli $X$-basis, \equatref{eq:unitarycluster} reduces to $U_{\boldsymbol{m}}(\boldsymbol{0})=I$ for odd $n$ when all measurement outcomes are 0.  When some of the measurement outcomes are non-zero, the identity can still be implemented by performing the appropriate corrective (Pauli) operations on the output state~\cite{Nielsen}.  For example, for the 3-qubit cluster state and the measurement outcome $\boldsymbol{m}=10$, the operation $HZH=X$ will be implemented and the identity can be implemented by applying the Pauli $X$ operation to the output state.  Measurement-based implementations of the identity operation can be used to determine the type of noise on a set of qubits.  For depolarising noise, we expect non-zero real entries along the diagonals of $\chi$ matrices obtained by channel tomography.

We implemented the identity operation by performing single-qubit measurements on 3-qubit, 5-qubit and 7-qubit linear cluster states prepared on the same qubits of the \textit{ibmq\_toronto} quantum processor as was used for the exact 3-design implementation.  \appendref{append:qubitsidentity} provides more detail on the qubits used.  Generation of channel tomography results, combining of counts to reduce statistical noise and construction of calibration matrices for quantum readout error mitigation were done in much the same way as for the exact 3-design implementation on the \textit{ibmq\_toronto} quantum processor.  The only significant difference is that we applied the appropriate corrective operations to the density matrices of the output states constructed by state tomography before using them to do channel tomography.  The different $\chi$ matrices obtained by doing channel tomography for the different measurement outcomes were used to calculate an average $\chi$ matrix for each cluster state.

\begin{figure*}
    \centering
    \vspace{-0.5cm}
    \begin{minipage}{17cm}
    \begin{tikzpicture}
    \draw (0, 0) circle[radius=0.25] node {1};
    \draw (1, 0) circle[radius=0.25] node {2};
    \draw (2, 0) circle[radius=0.25] node {3};
    \node[] at (0, -0.6) {$X$};
    \node[] at (1, -0.6) {$X$};
    \draw (0.25, 0) -- (0.75, 0);
    \draw (1.25, 0) -- (1.75, 0);
    \end{tikzpicture}
    \end{minipage}
            
    \vspace{0.5cm}
    
    \begin{minipage}{1cm}
    Re($\chi$)
    
    \vspace{4.5cm}
    
    Im($\chi$)
    \end{minipage}
    \begin{minipage}{5.2cm}
    Raw
    \includegraphics[trim=3.5cm 1cm 2.5cm 1.4cm, clip, scale=0.5]{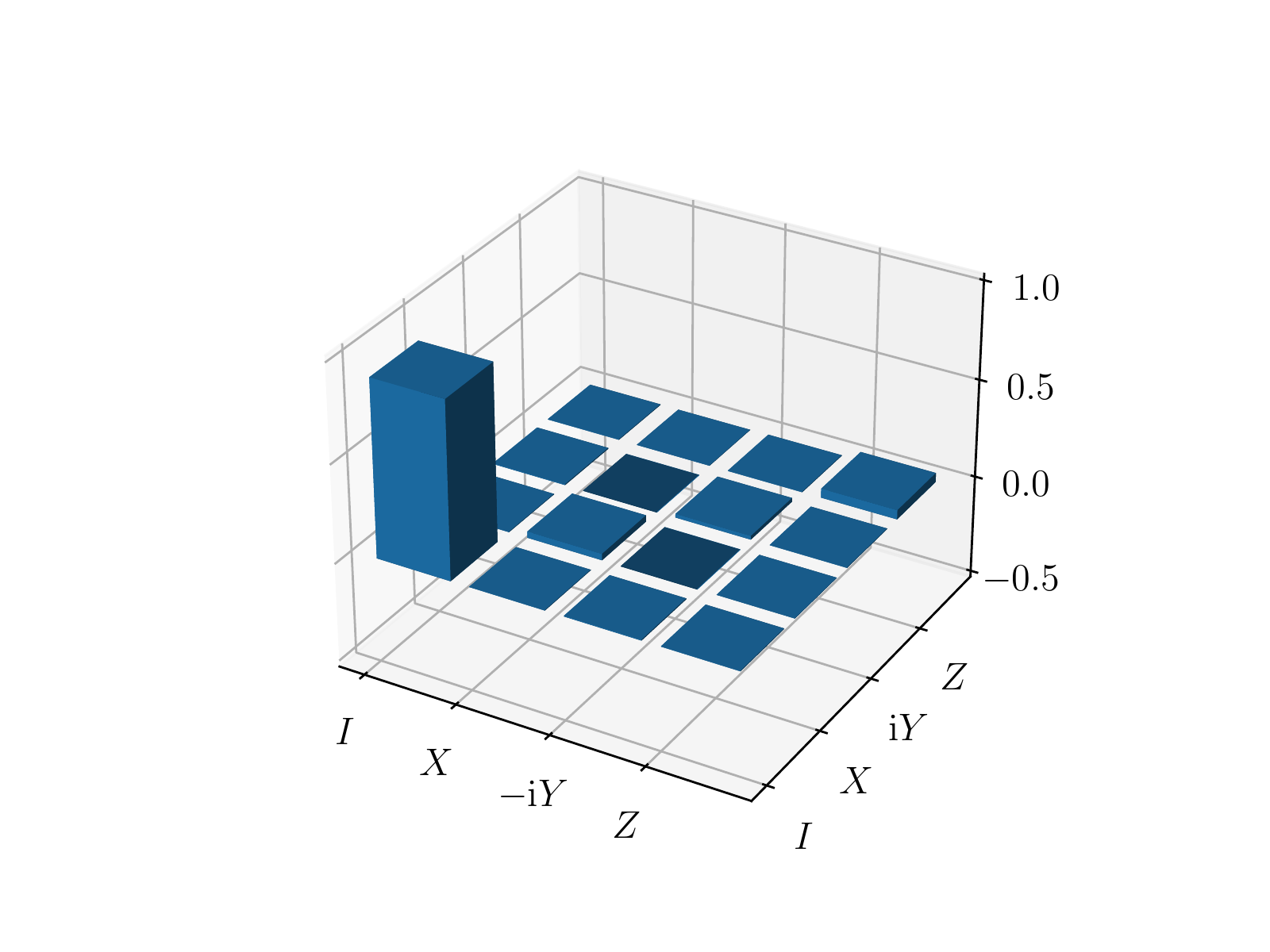}
    \includegraphics[trim=3.5cm 1cm 2.5cm 1.4cm, clip, scale=0.5]{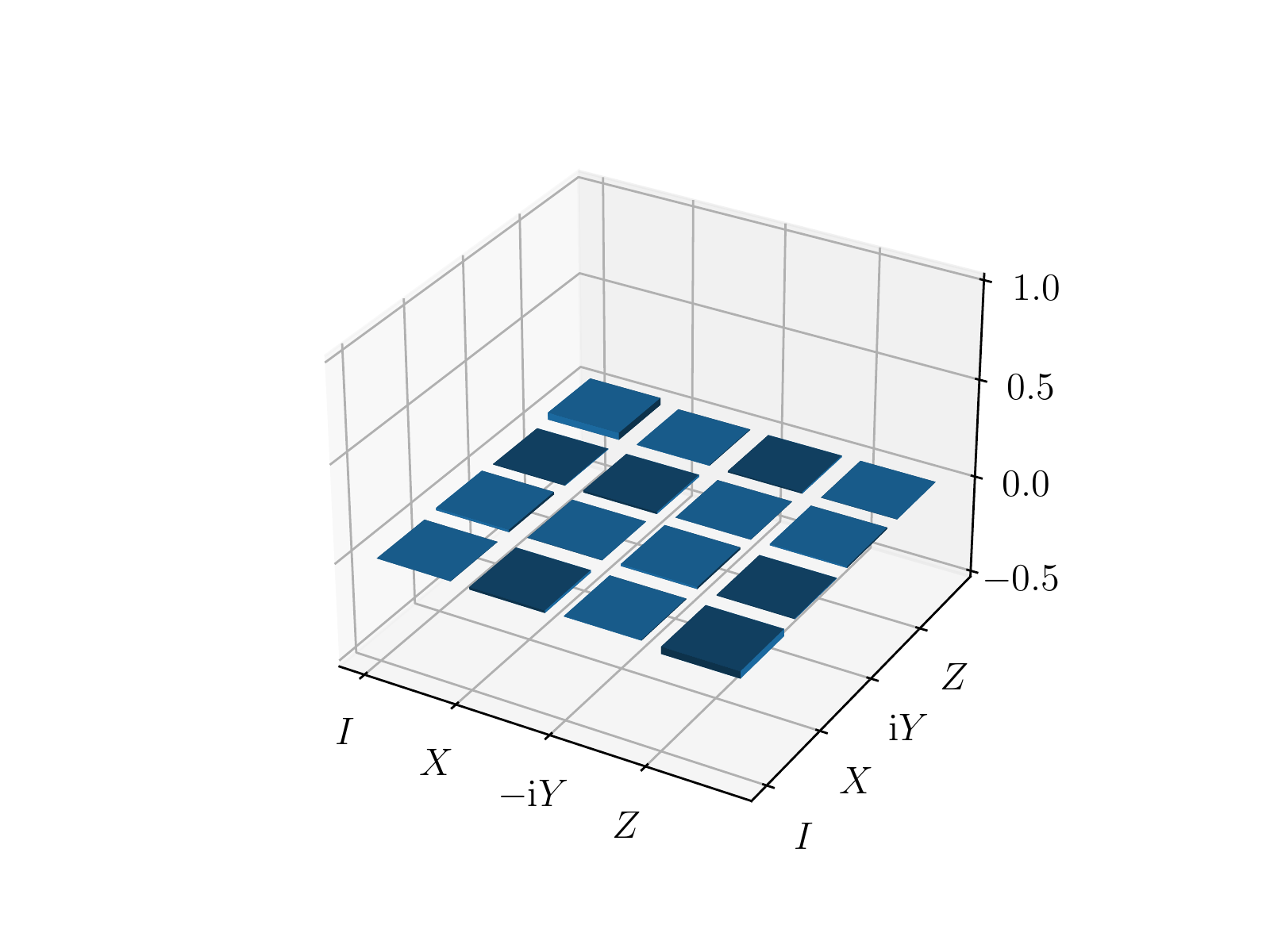}
    \end{minipage}
    \begin{minipage}{0.1cm}
    \end{minipage}
    \begin{minipage}{0.1cm}
    \begin{tikzpicture}
    \draw[gray, dashed] (0, 0) -- (0, -10);
    \end{tikzpicture}
    \end{minipage}
    \begin{minipage}{5.2cm}
    Processed
    \includegraphics[trim=3.5cm 1cm 2.5cm 1.4cm, clip, scale=0.5]{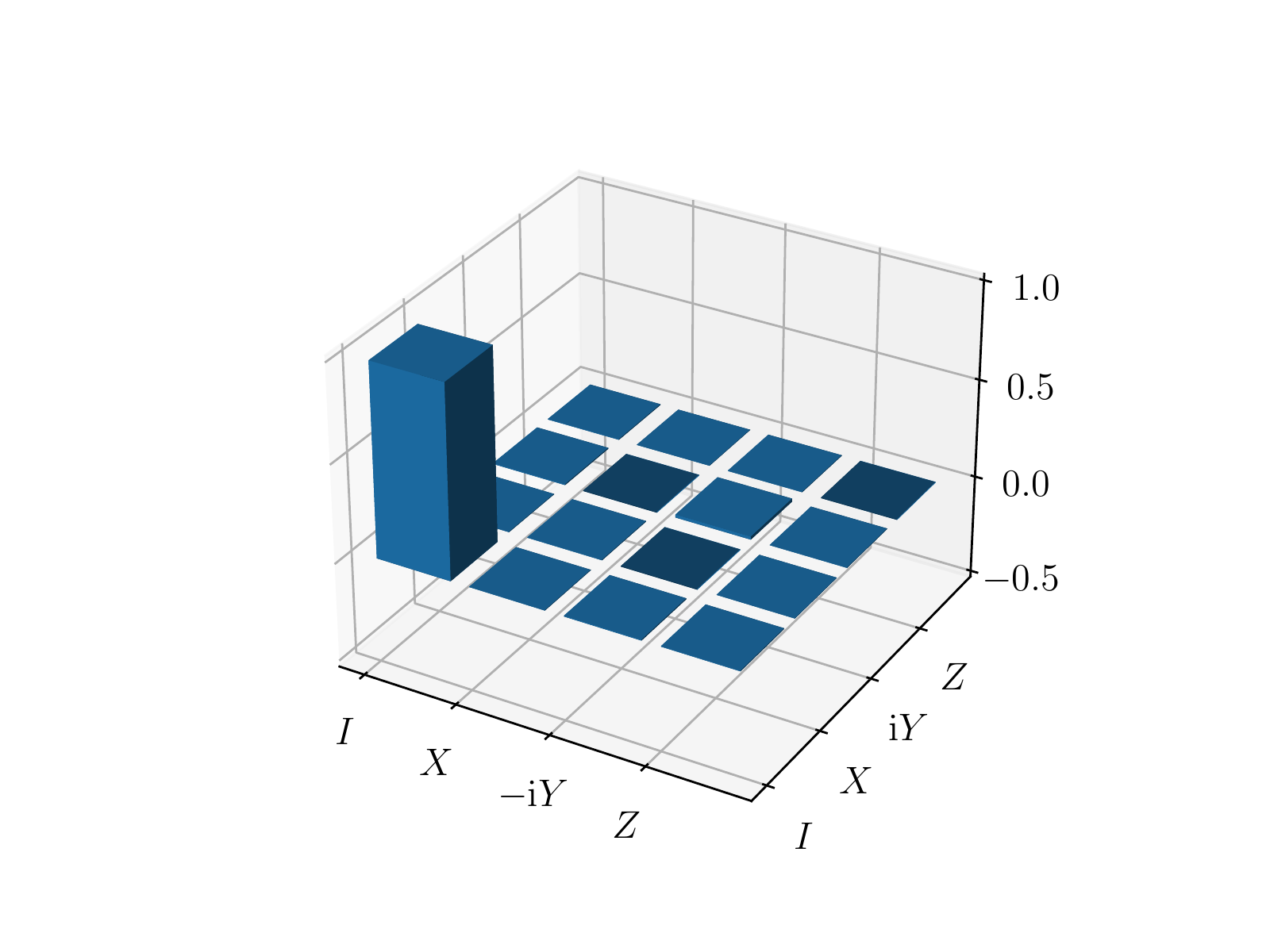}
    \includegraphics[trim=3.5cm 1cm 2.5cm 1.4cm, clip, scale=0.5]{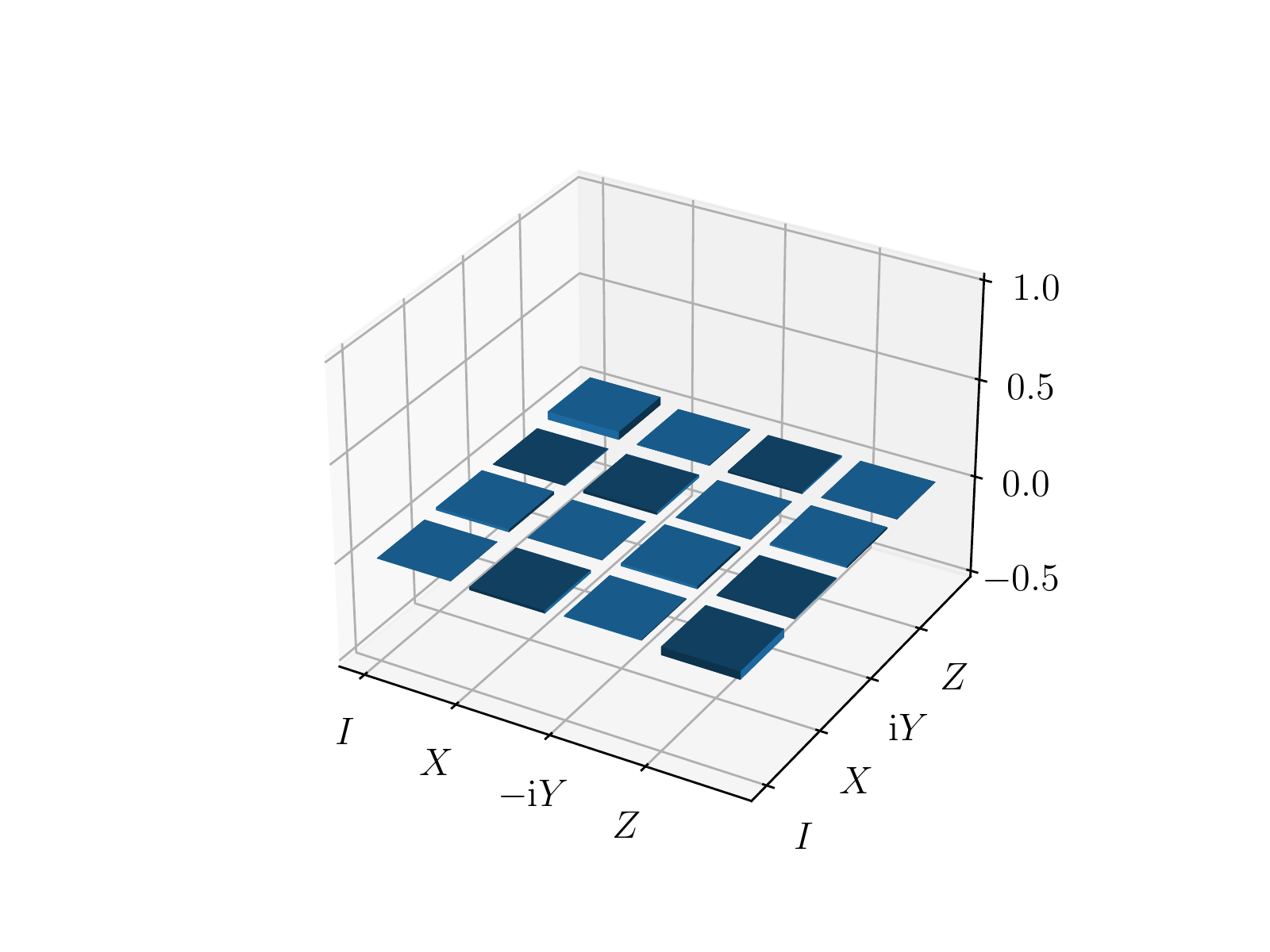}
    \end{minipage}
    \begin{minipage}{0.1cm}
    \end{minipage}
    \begin{minipage}{0.1cm}
    \begin{tikzpicture}
    \draw[gray, dashed] (0, 0) -- (0, -10);
    \end{tikzpicture}
    \end{minipage}
    \begin{minipage}{5.2cm}
    Ideal
    \includegraphics[trim=3.5cm 1cm 2.5cm 1.4cm, clip, scale=0.5]{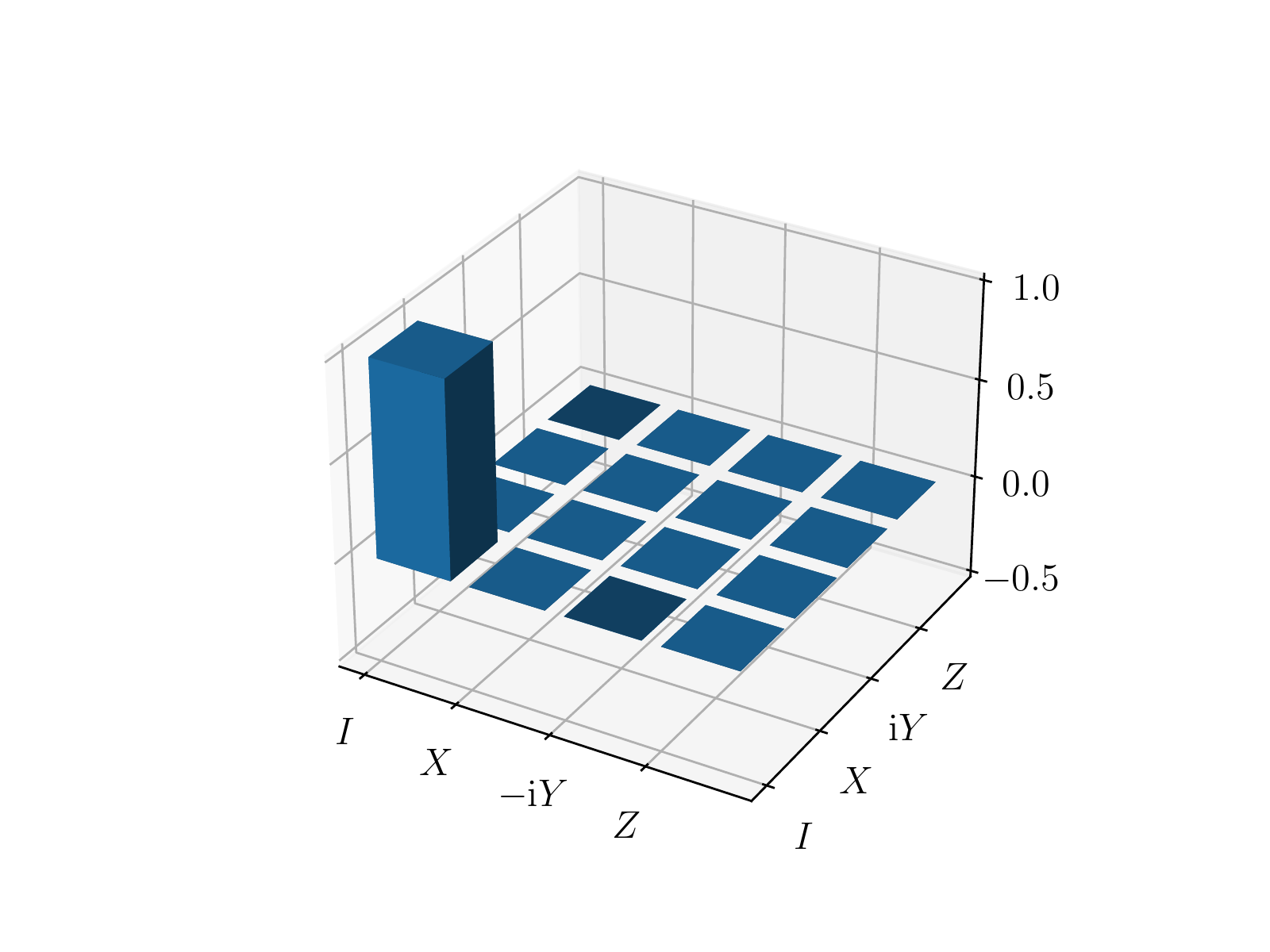}
    \includegraphics[trim=3.5cm 1cm 2.5cm 1.4cm, clip, scale=0.5]{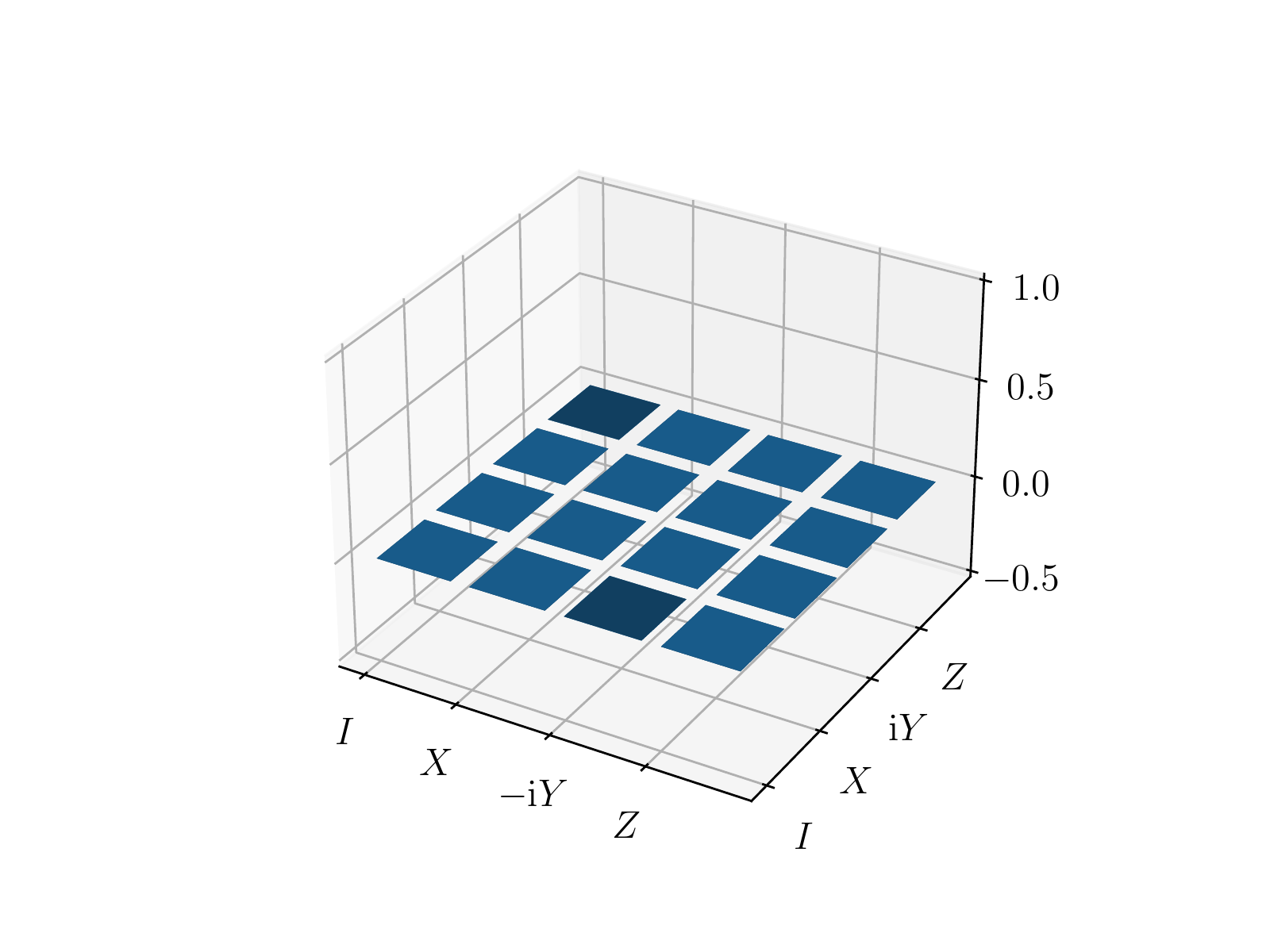}
    \end{minipage}
    \caption{Channel tomography results (average $\chi$ matrix) for the implementation of the identity operation with a 3-qubit cluster state on the \textit{ibmq\_toronto} quantum processor.  The diagram at the top shows the entangled 3-qubit cluster state with the measurements performed on each qubit.  The $\chi$ matrix obtained without quantum readout error mitigation is shown on the left, the $\chi$ matrix obtained with quantum readout error mitigation is shown in the middle and the ideal $\chi$ matrix is shown on the right.  The real part of each matrix is shown above and the imaginary part of each matrix is shown below.}
    \label{fig:identity3}
\end{figure*}
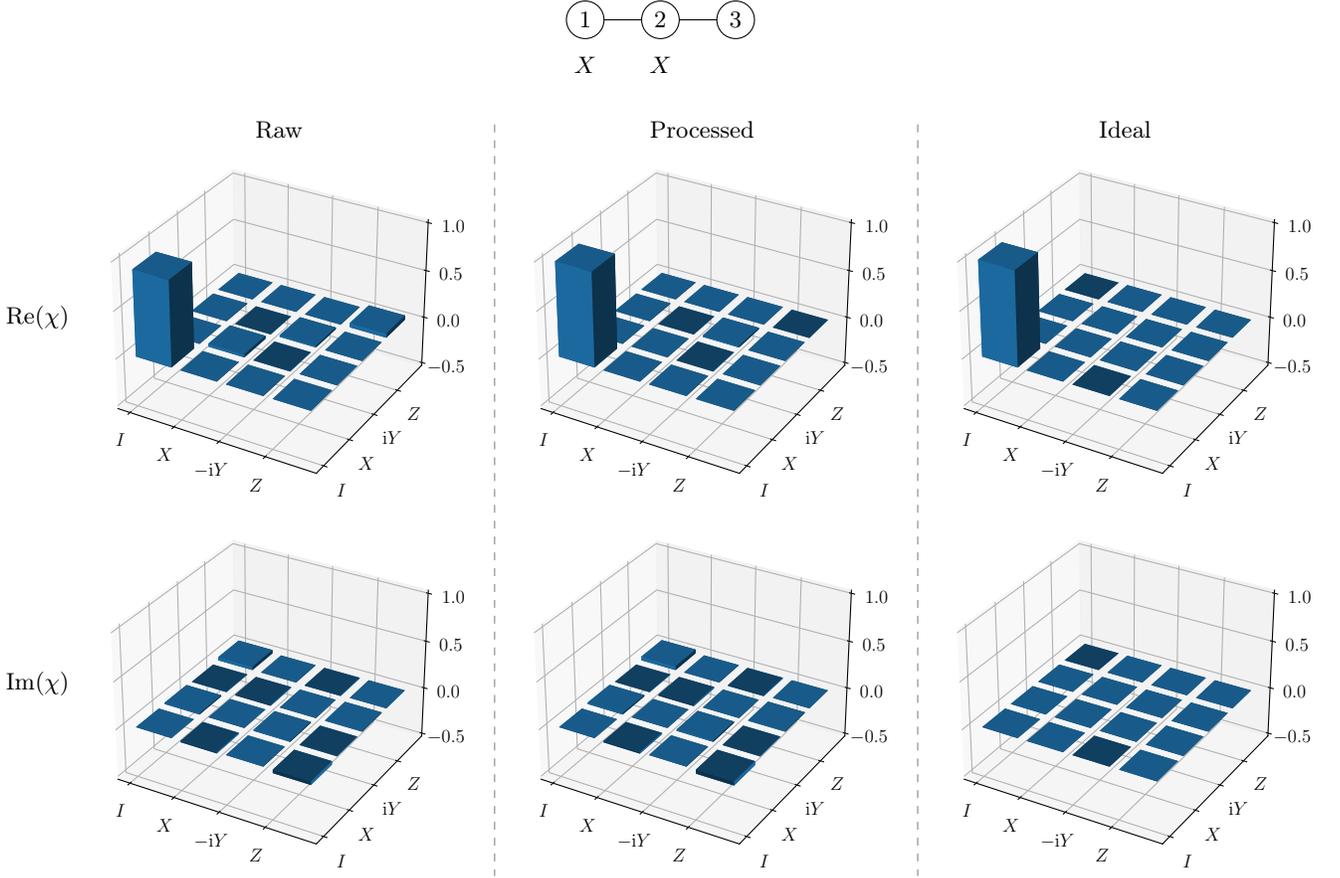

\begin{figure*}
    \centering
    \vspace{-0.5cm}
    \begin{minipage}{17cm}
    \begin{tikzpicture}
    \draw (0, 0) circle[radius=0.25] node {1};
    \draw (1, 0) circle[radius=0.25] node {2};
    \draw (2, 0) circle[radius=0.25] node {3};
    \draw (3, 0) circle[radius=0.25] node {4};
    \draw (4, 0) circle[radius=0.25] node {5};
    \node[] at (0, -0.6) {$X$};
    \node[] at (1, -0.6) {$X$};
    \node[] at (2, -0.6) {$X$};
    \node[] at (3, -0.6) {$X$};
    \draw (0.25, 0) -- (0.75, 0);
    \draw (1.25, 0) -- (1.75, 0);
    \draw (2.25, 0) -- (2.75, 0);
    \draw (3.25, 0) -- (3.75, 0);
    \end{tikzpicture}
    \end{minipage}
                
    \vspace{0.5cm}
    
    \begin{minipage}{1cm}
    Re($\chi$)
    
    \vspace{4.5cm}
    
    Im($\chi$)
    \end{minipage}
    \begin{minipage}{5.2cm}
    Raw
    \includegraphics[trim=3.5cm 1cm 2.5cm 1.4cm, clip, scale=0.5]{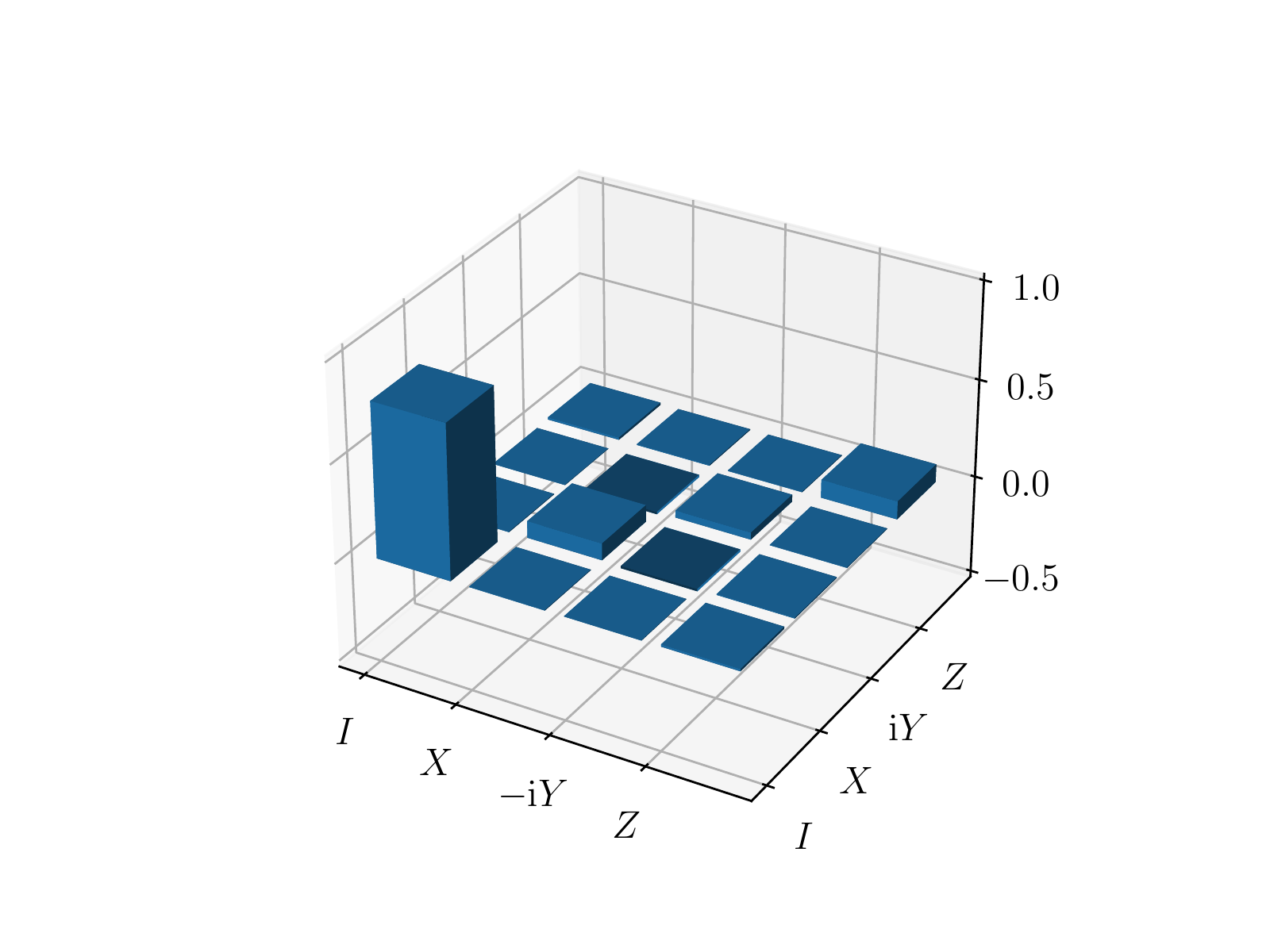}
    \includegraphics[trim=3.5cm 1cm 2.5cm 1.4cm, clip, scale=0.5]{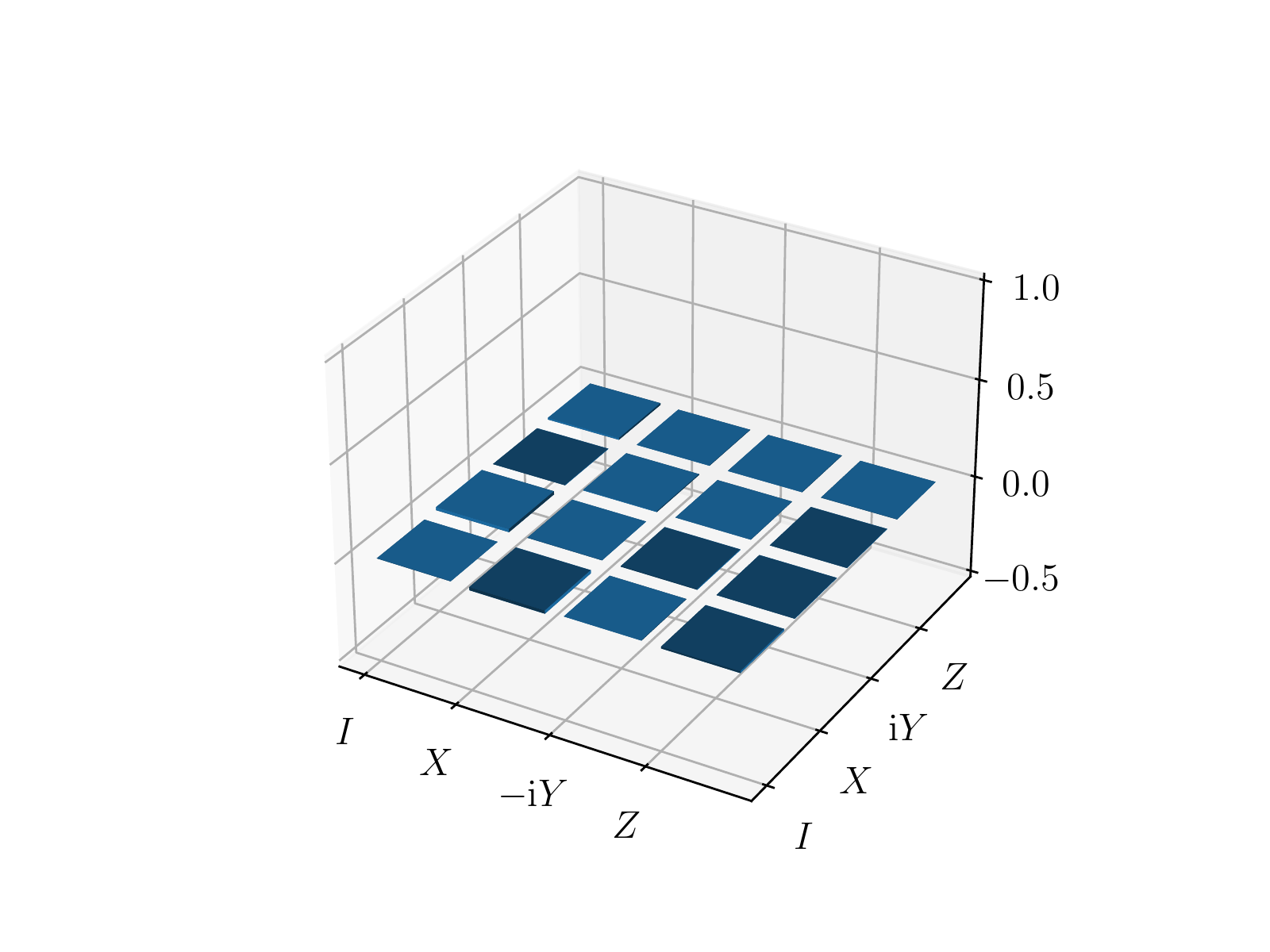}
    \end{minipage}
    \begin{minipage}{0.1cm}
    \end{minipage}
    \begin{minipage}{0.1cm}
    \begin{tikzpicture}
    \draw[gray, dashed] (0, 0) -- (0, -10);
    \end{tikzpicture}
    \end{minipage}
    \begin{minipage}{5.2cm}
    Processed
    \includegraphics[trim=3.5cm 1cm 2.5cm 1.4cm, clip, scale=0.5]{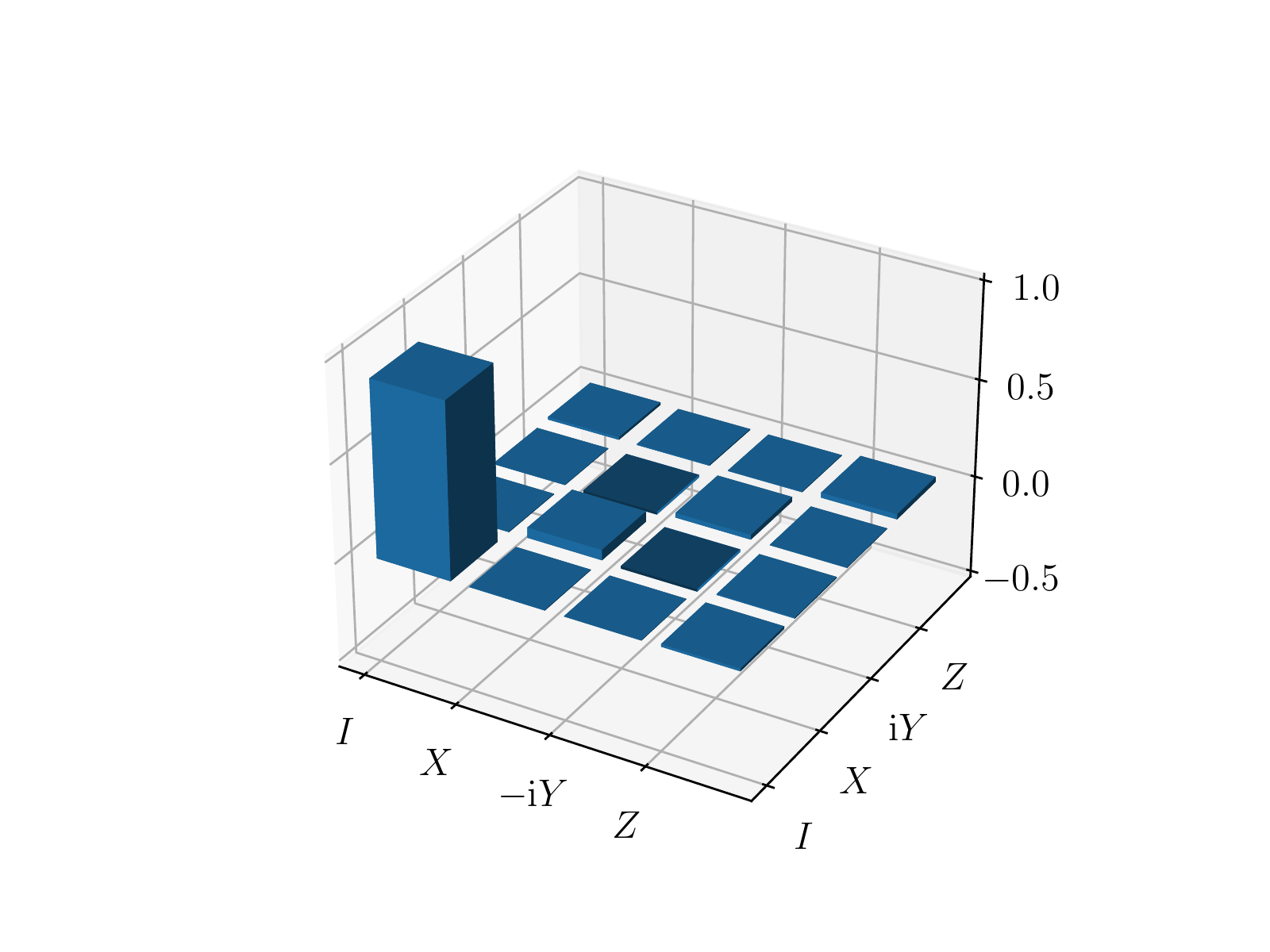}
    \includegraphics[trim=3.5cm 1cm 2.5cm 1.4cm, clip, scale=0.5]{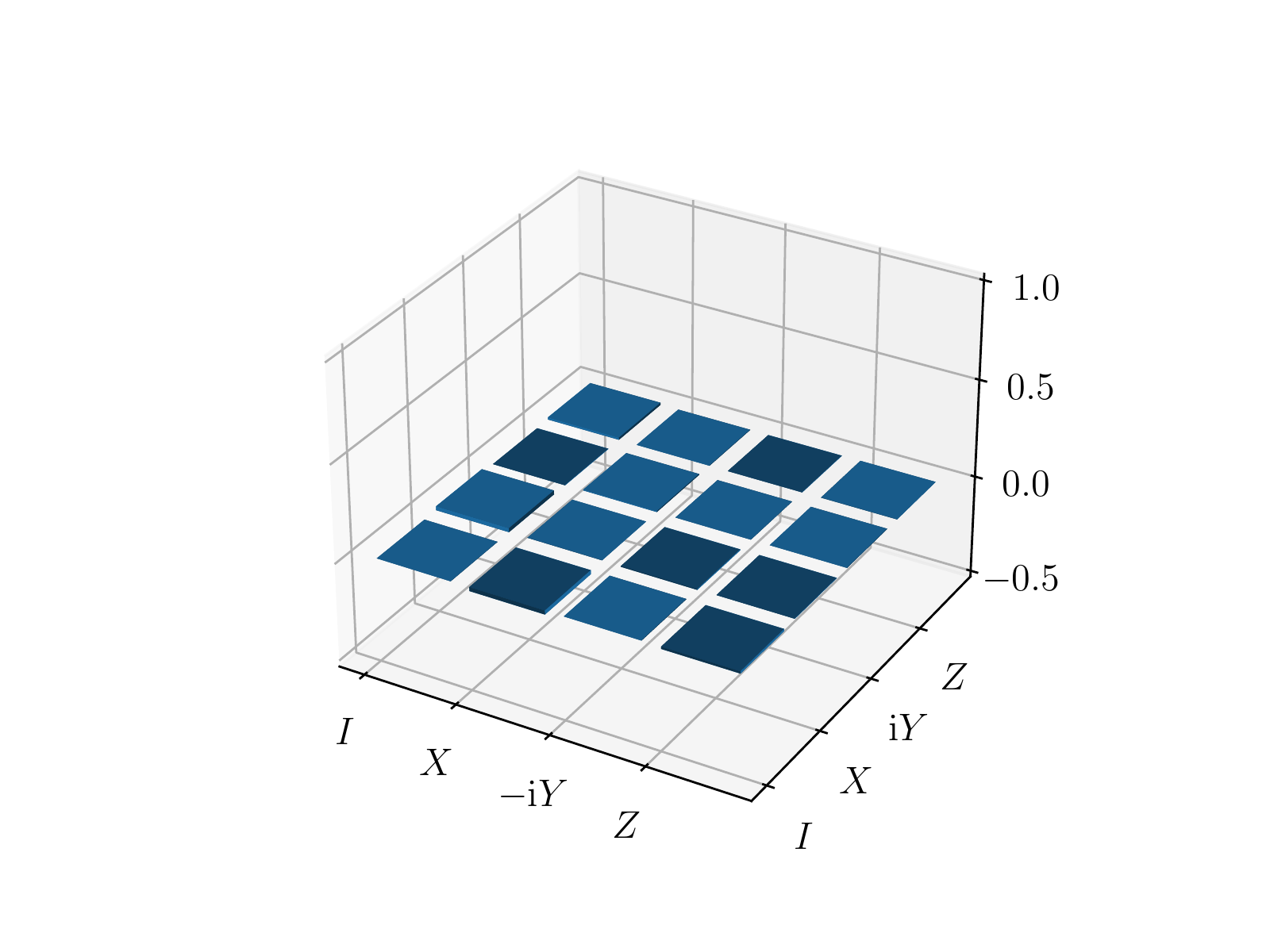}
    \end{minipage}
    \begin{minipage}{0.1cm}
    \end{minipage}
    \begin{minipage}{0.1cm}
    \begin{tikzpicture}
    \draw[gray, dashed] (0, 0) -- (0, -10);
    \end{tikzpicture}
    \end{minipage}
    \begin{minipage}{5.2cm}
    Ideal
    \includegraphics[trim=3.5cm 1cm 2.5cm 1.4cm, clip, scale=0.5]{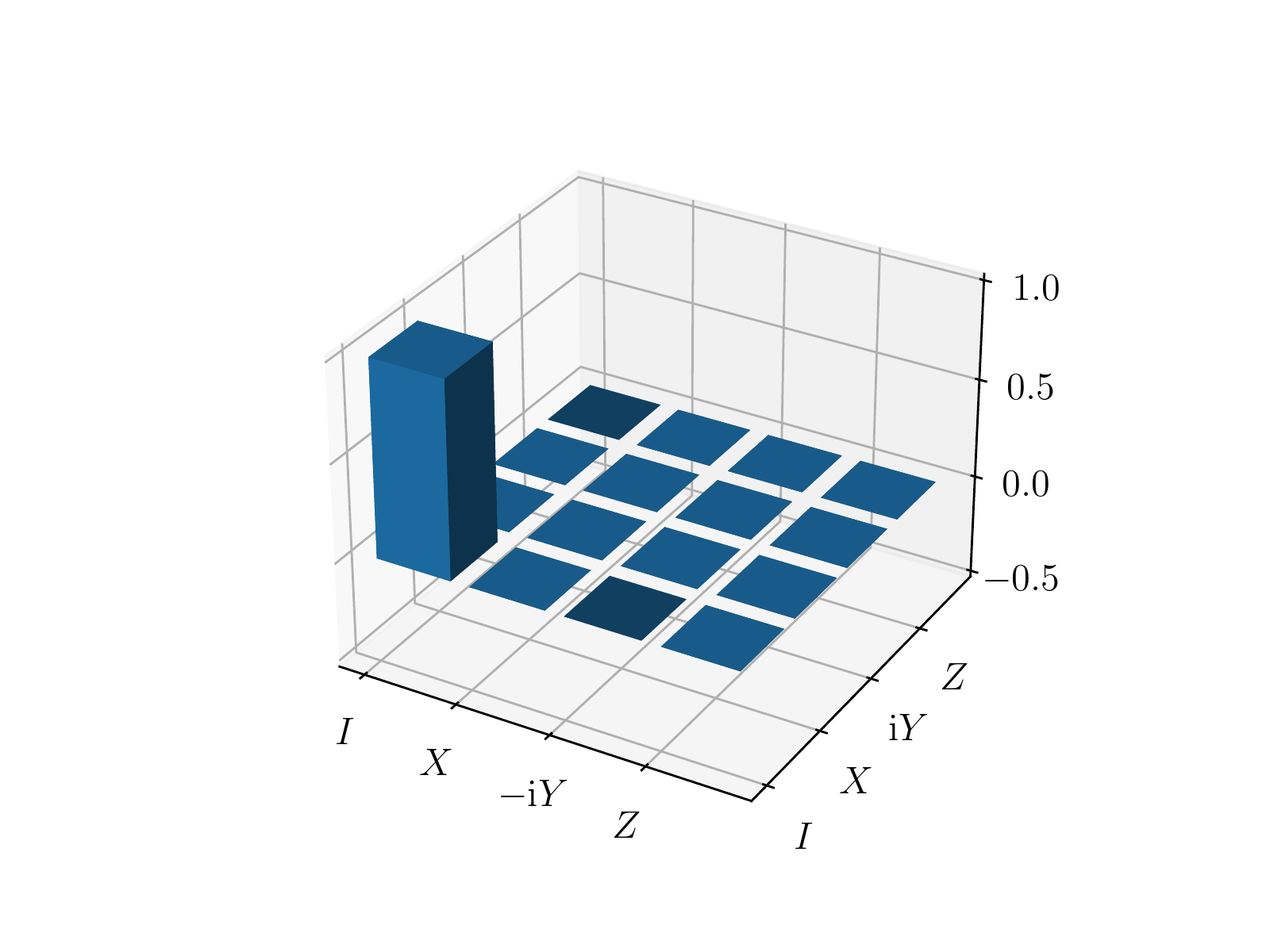}
    \includegraphics[trim=3.5cm 1cm 2.5cm 1.4cm, clip, scale=0.5]{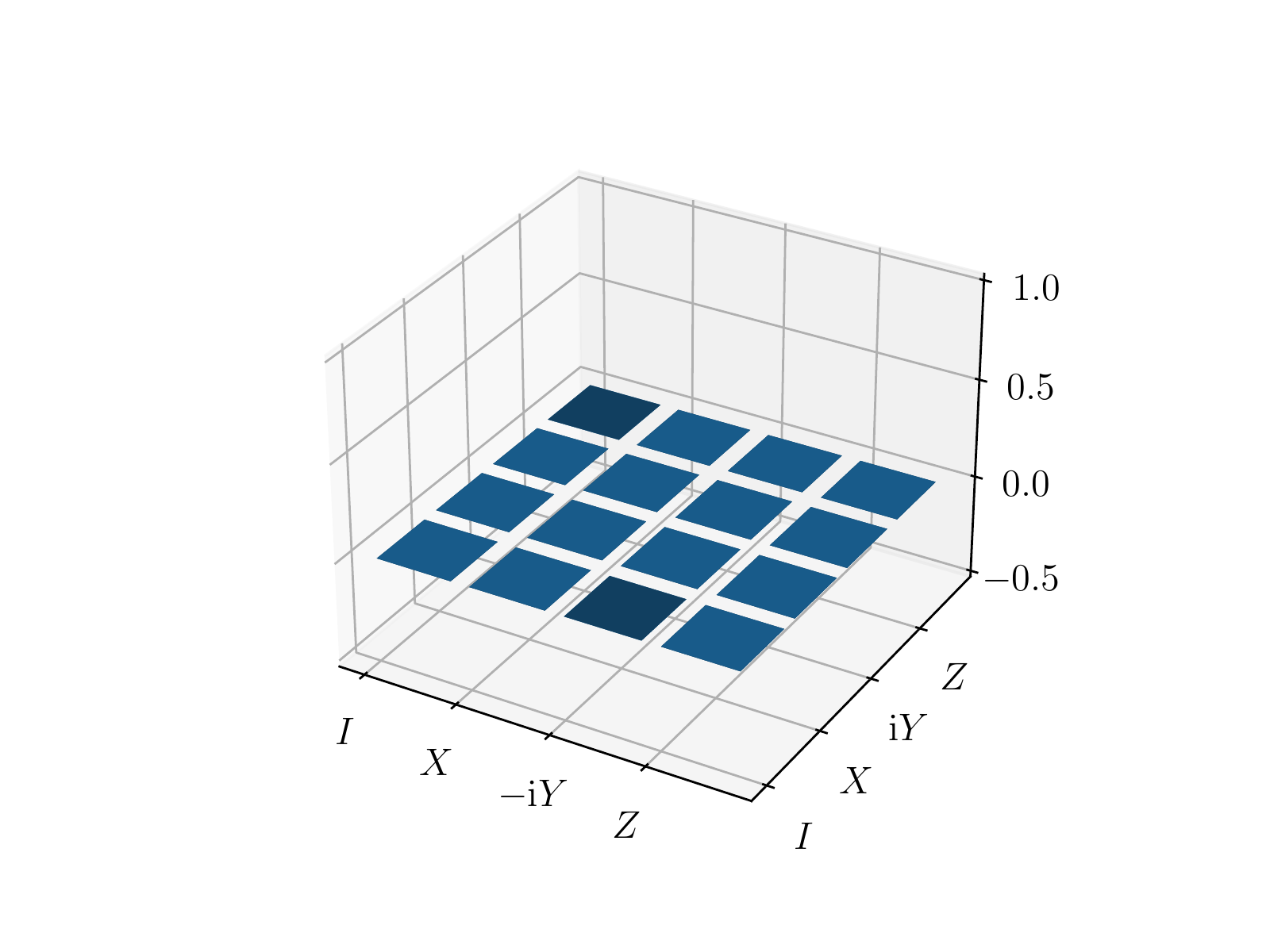}
    \end{minipage}
    \caption{Channel tomography results (average $\chi$ matrix) for the implementation of the identity operation with a 5-qubit cluster state on the \textit{ibmq\_toronto} quantum processor.  The diagram at the top shows the entangled 5-qubit cluster state with the measurements performed on each qubit.  The $\chi$ matrix obtained without quantum readout error mitigation is shown on the left, the $\chi$ matrix obtained with quantum readout error mitigation is shown in the middle and the ideal $\chi$ matrix is shown on the right.  The real part of each matrix is shown above and the imaginary part of each matrix is shown below.}
    \label{fig:identity5}
\end{figure*}

\begin{figure*}
    \centering
    \vspace{-0.5cm}
    \begin{minipage}{17cm}
    \begin{tikzpicture}
    \draw (0, 0) circle[radius=0.25] node {1};
    \draw (1, 0) circle[radius=0.25] node {2};
    \draw (2, 0) circle[radius=0.25] node {3};
    \draw (3, 0) circle[radius=0.25] node {4};
    \draw (4, 0) circle[radius=0.25] node {5};
    \draw (5, 0) circle[radius=0.25] node {6};
    \draw (6, 0) circle[radius=0.25] node {7};
    \node[] at (0, -0.6) {$X$};
    \node[] at (1, -0.6) {$X$};
    \node[] at (2, -0.6) {$X$};
    \node[] at (3, -0.6) {$X$};
    \node[] at (4, -0.6) {$X$};
    \node[] at (5, -0.6) {$X$};
    \draw (0.25, 0) -- (0.75, 0);
    \draw (1.25, 0) -- (1.75, 0);
    \draw (2.25, 0) -- (2.75, 0);
    \draw (3.25, 0) -- (3.75, 0);
    \draw (4.25, 0) -- (4.75, 0);
    \draw (5.25, 0) -- (5.75, 0);
    \end{tikzpicture}
    \end{minipage}
                
    \vspace{0.5cm}
    
    \begin{minipage}{1cm}
    Re($\chi$)
    
    \vspace{4.5cm}
    
    Im($\chi$)
    \end{minipage}
    \begin{minipage}{5.2cm}
    Raw
    \includegraphics[trim=3.5cm 1cm 2.5cm 1.4cm, clip, scale=0.5]{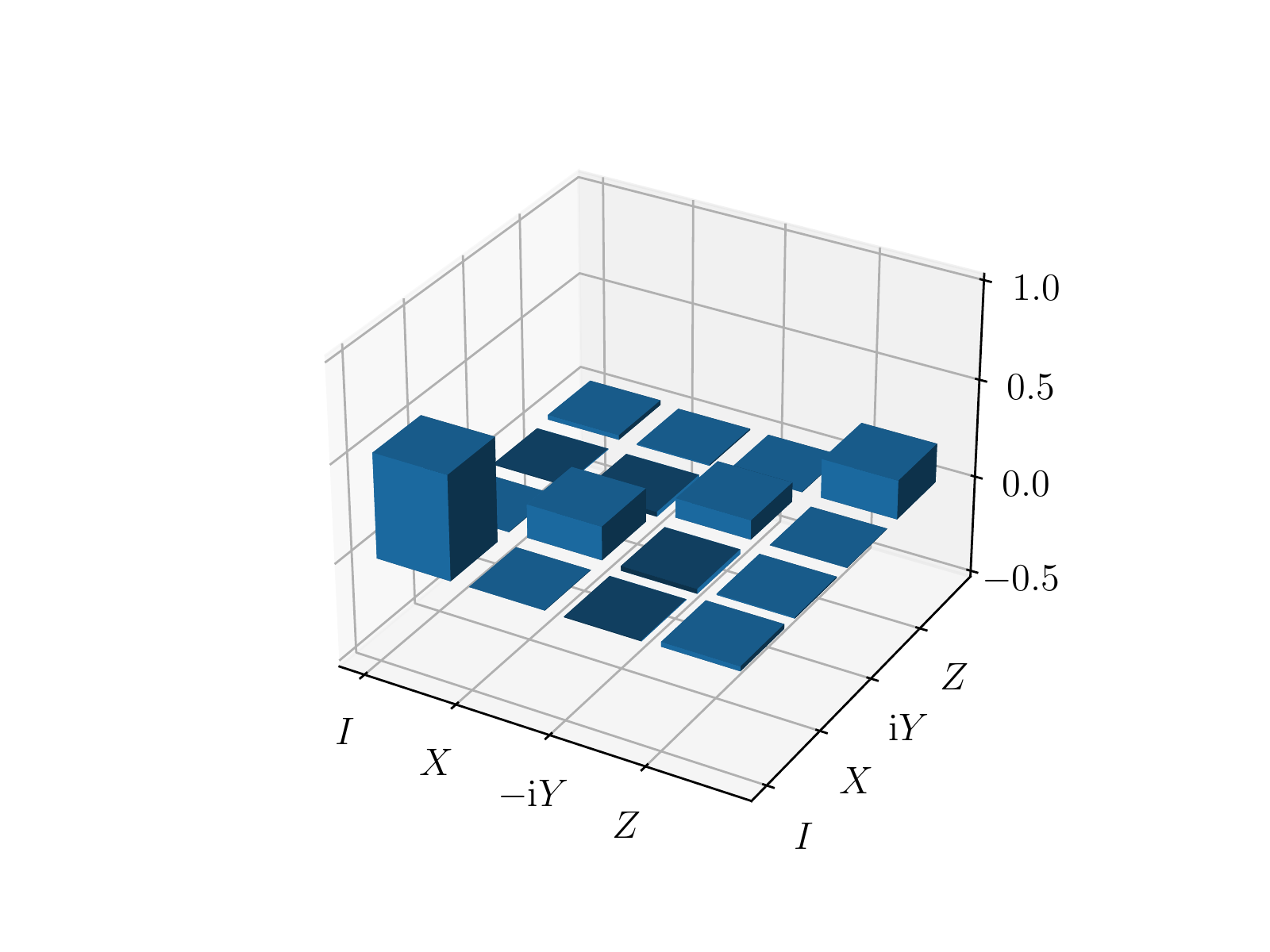}
    \includegraphics[trim=3.5cm 1cm 2.5cm 1.4cm, clip, scale=0.5]{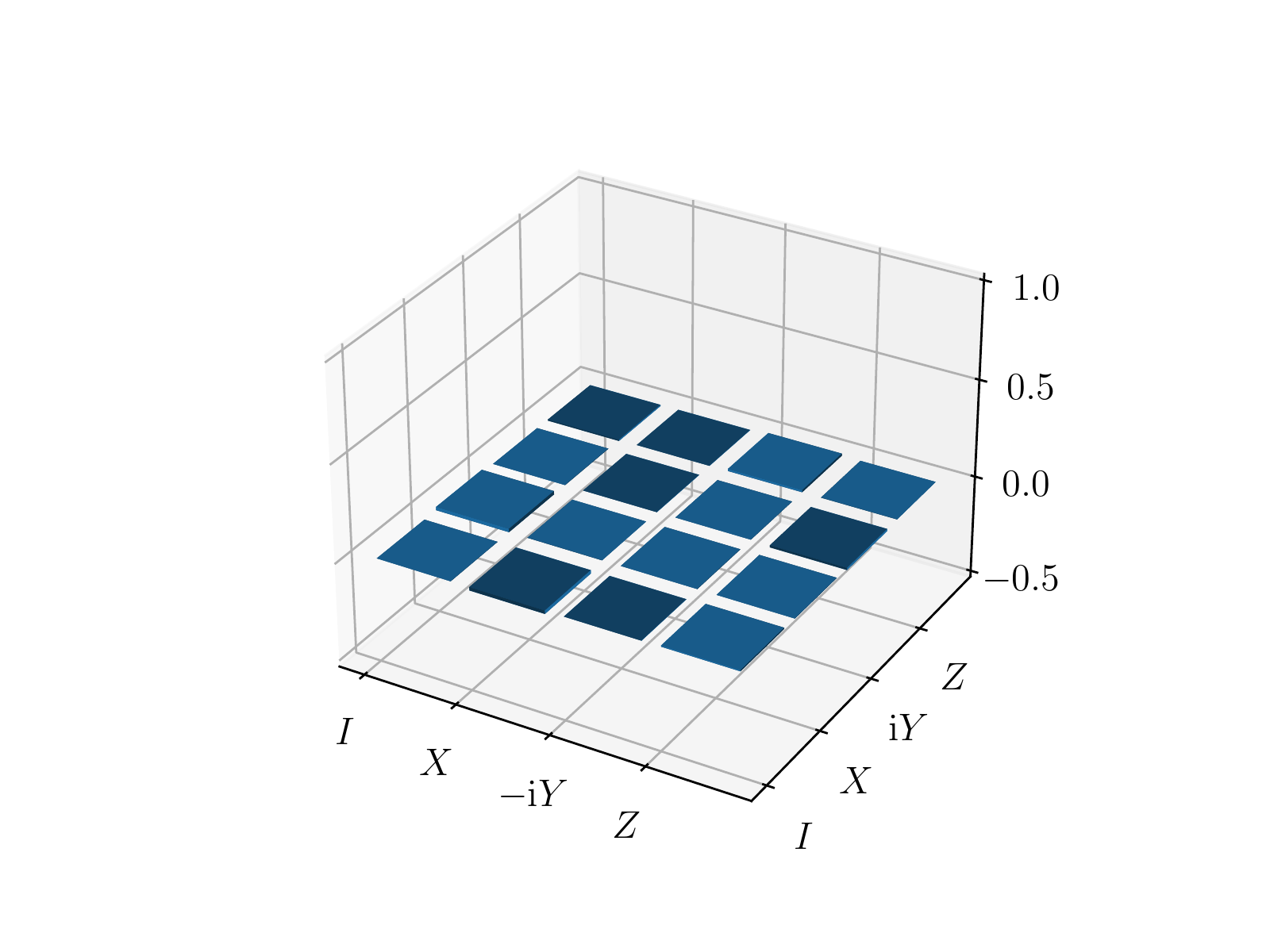}
    \end{minipage}
    \begin{minipage}{0.1cm}
    \end{minipage}
    \begin{minipage}{0.1cm}
    \begin{tikzpicture}
    \draw[gray, dashed] (0, 0) -- (0, -10);
    \end{tikzpicture}
    \end{minipage}
    \begin{minipage}{5.2cm}
    Processed
    \includegraphics[trim=3.5cm 1cm 2.5cm 1.4cm, clip, scale=0.5]{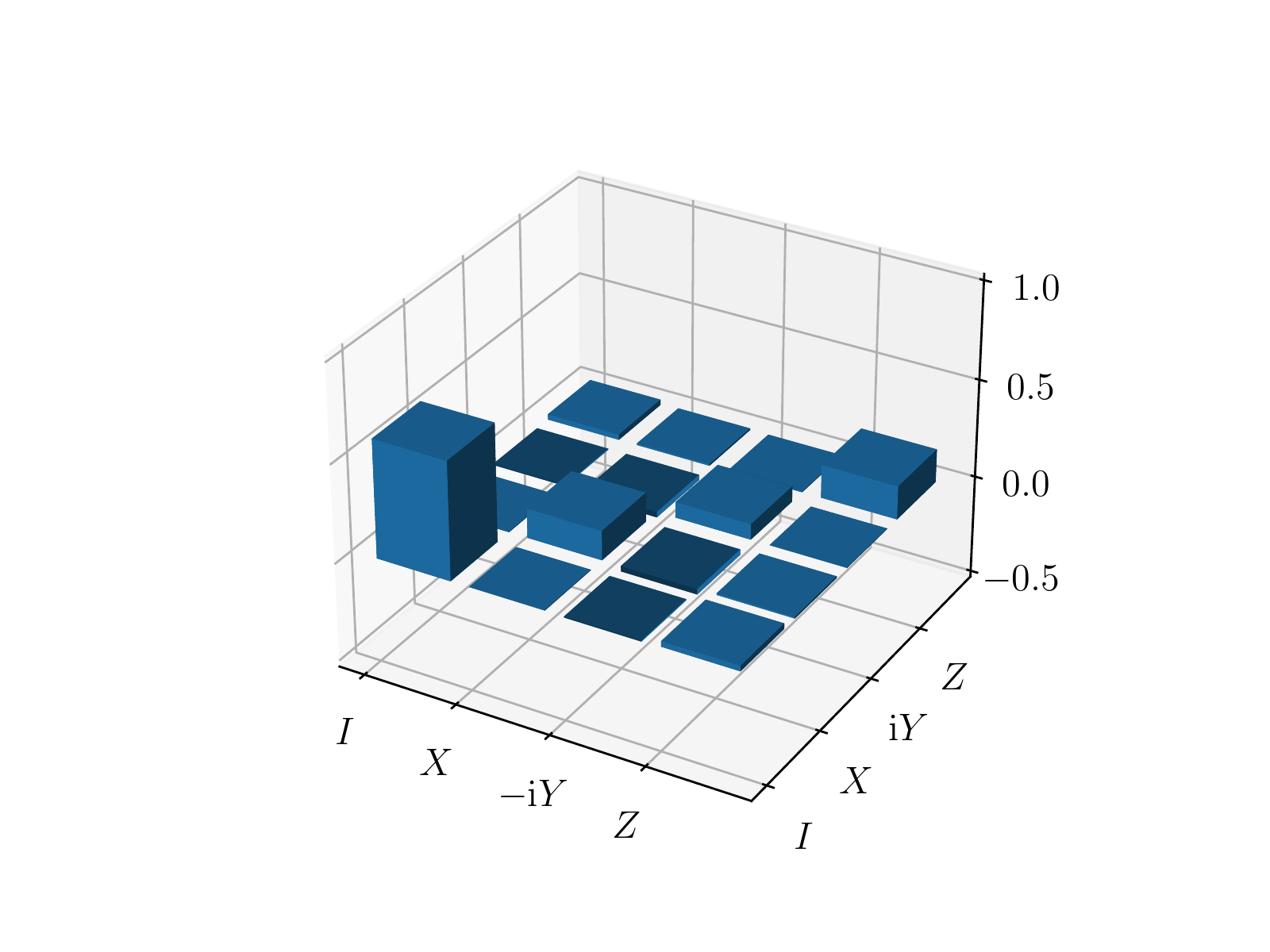}
    \includegraphics[trim=3.5cm 1cm 2.5cm 1.4cm, clip, scale=0.5]{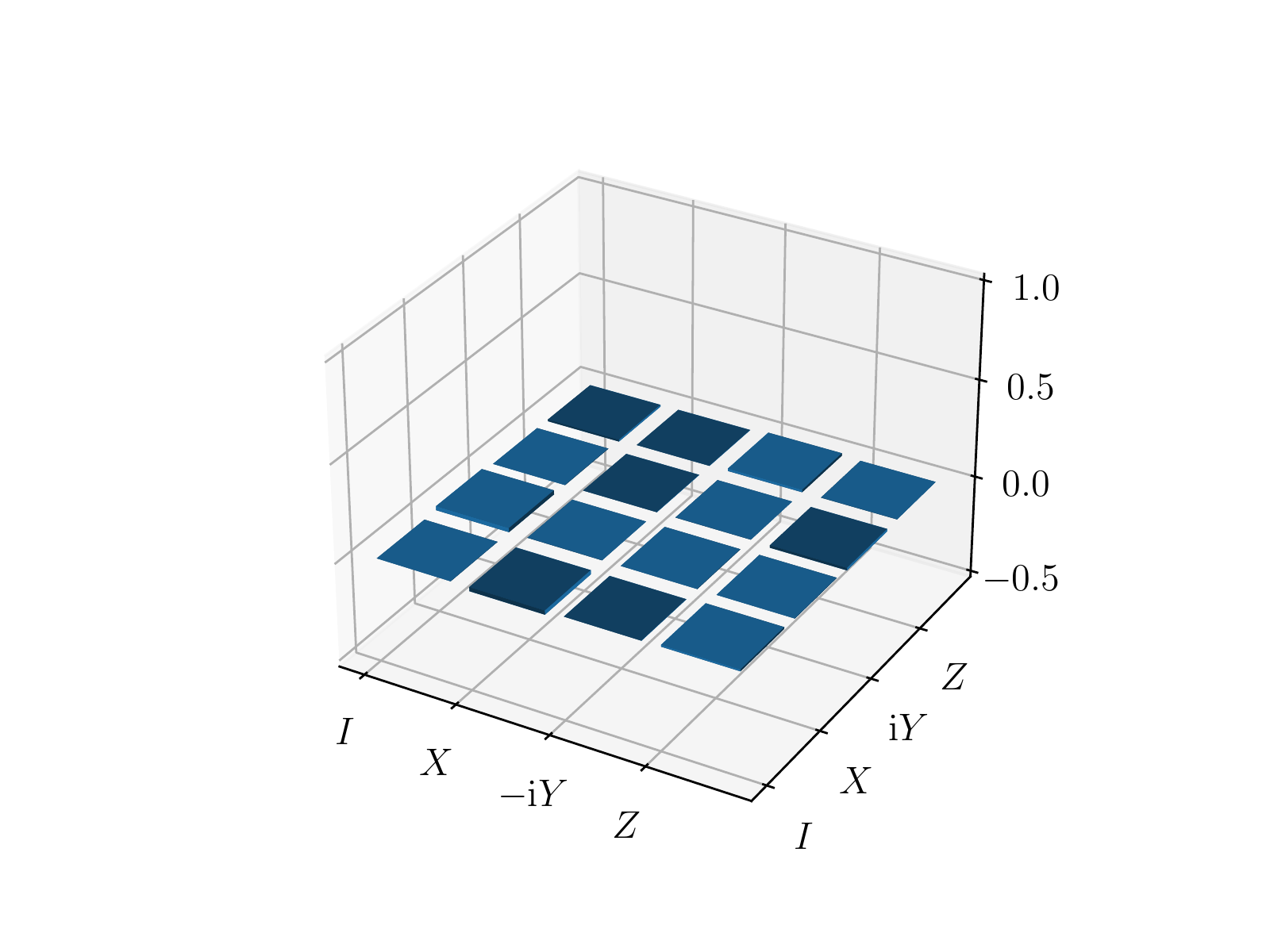}
    \end{minipage}
    \begin{minipage}{0.1cm}
    \end{minipage}
    \begin{minipage}{0.1cm}
    \begin{tikzpicture}
    \draw[gray, dashed] (0, 0) -- (0, -10);
    \end{tikzpicture}
    \end{minipage}
    \begin{minipage}{5.2cm}
    Ideal
    \includegraphics[trim=3.5cm 1cm 2.5cm 1.4cm, clip, scale=0.5]{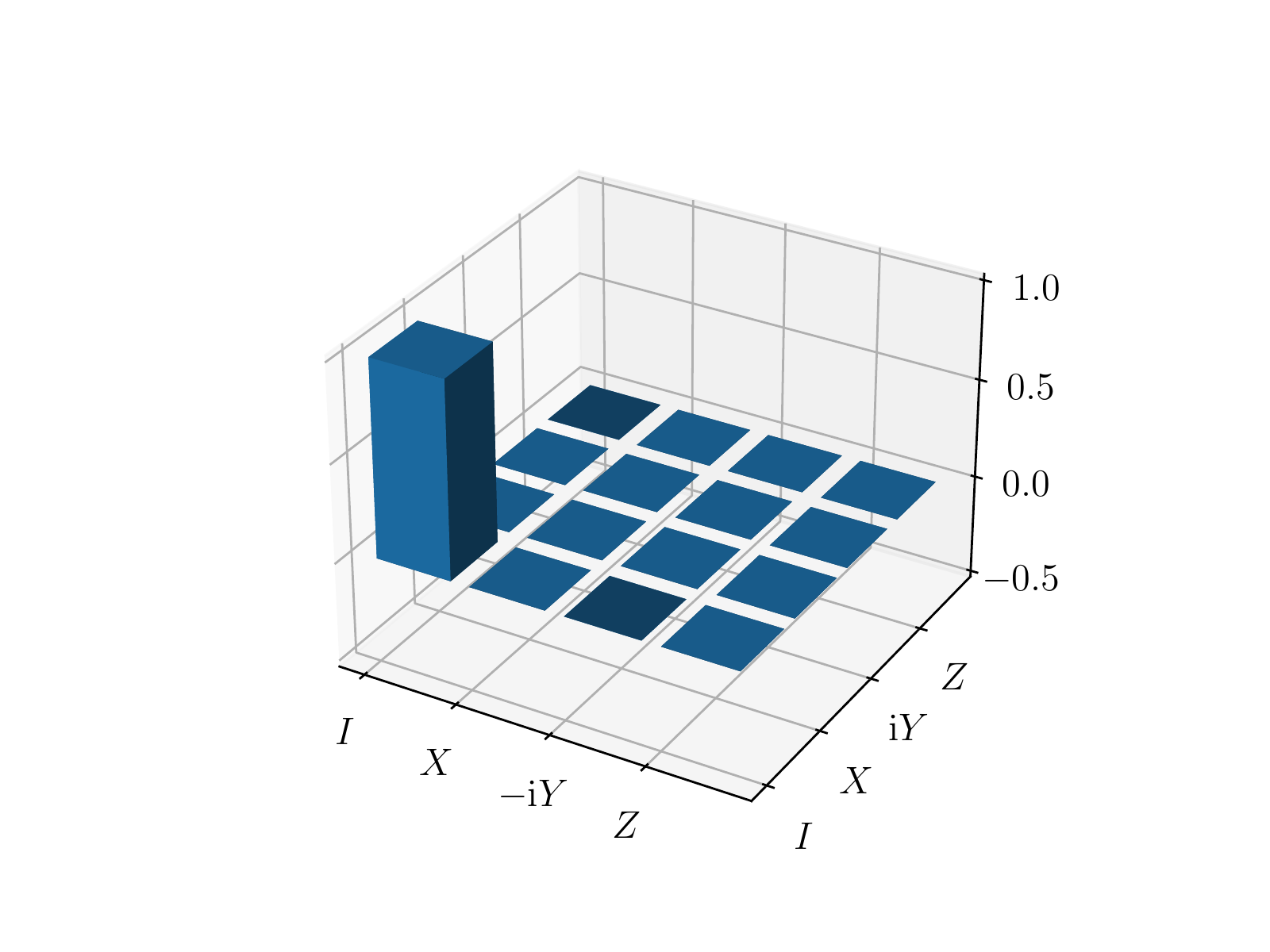}
    \includegraphics[trim=3.5cm 1cm 2.5cm 1.4cm, clip, scale=0.5]{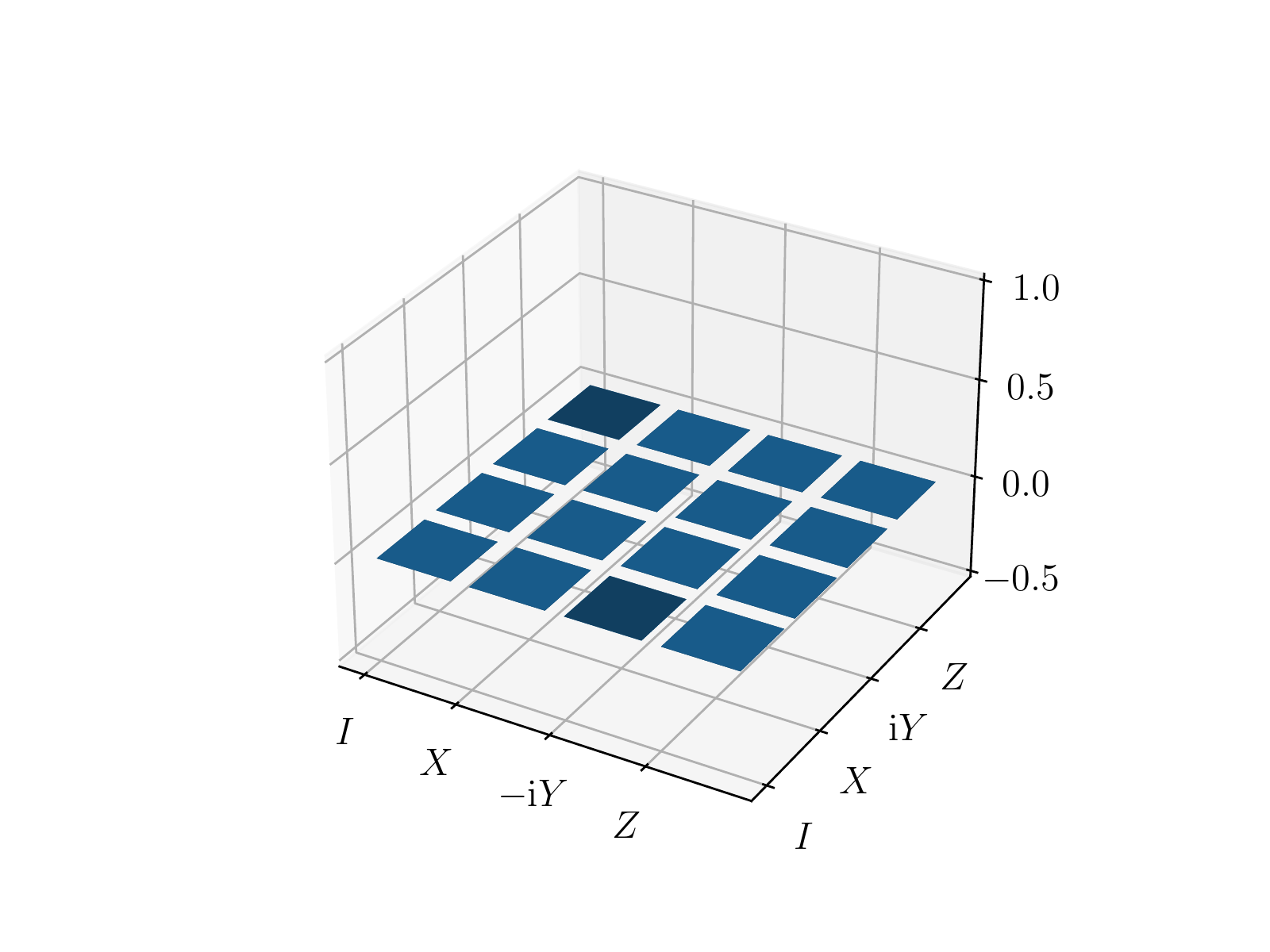}
    \end{minipage}
    \caption{Channel tomography results (average $\chi$ matrix) for the implementation of the identity operation with a 7-qubit cluster state on the \textit{ibmq\_toronto} quantum processor.  The diagram at the top shows the entangled 7-qubit cluster state with the measurements performed on each qubit.  The $\chi$ matrix obtained without quantum readout error mitigation is shown on the left, the $\chi$ matrix obtained with quantum readout error mitigation is shown in the middle and the ideal $\chi$ matrix is shown on the right.  The real part of each matrix is shown above and the imaginary part of each matrix is shown below.}
    \label{fig:identity7}
\end{figure*}

Channel tomography results (average $\chi$ matrices) obtained for the implementation of the identity operation with the 3-qubit, 5-qubit and 7-qubit linear cluster states are displayed in Figs.~\ref{fig:identity3}, \ref{fig:identity5} and \ref{fig:identity7} respectively.  Non-zero real entries are clearly visible along the diagonals of constructed $\chi$ matrices, which confirms that depolarising noise was indeed the predominant type of noise for these qubits.  Comparing the channel tomography results for the different cluster states, we see that as the length of the cluster used in the implementation increases, so does the depolarising noise in the implementation.

\begin{table}[]
\begin{tabular}{|l|l|l|}
\hline
\textbf{Cluster State} & $\boldsymbol{p}$ \textbf{(Raw)} & $\boldsymbol{p}$ \textbf{(Processed)} \\ \hline
3-qubit                & 0.042           & 0.004                 \\ \hline
5-qubit                & 0.062           & 0.027                 \\ \hline
7-qubit                & 0.127           & 0.099                 \\ \hline
\end{tabular}
\caption{Values of $p$ inferred for implementations of the identity with different cluster states on the \textit{ibmq\_toronto} quantum processor.  `Raw' shows the values of $p$ without quantum readout error mitigation.  `Processed' shows the values of $p$ with quantum readout error mitigation.}
\label{tab:p}
\end{table}

We now infer a value for the parameter $p$ for the depolarising noise present in each implementation of the identity.  To this end, we model depolarising noise in one of these measurement-based implementations as a channel in which depolarising noise is applied to the input state $n$ times for an implementation with a $n$-qubit cluster state.  For each implementation, we considered 10000 evenly spaced values of $p$ in the range 0 to 1.  For each $p$, we determined the $\chi$ matrix for the corresponding model channel and calculated the channel fidelity for the implemented channel using the average $\chi$ matrix.  Our inferred value of $p$ for a given implementation, is the one which yields the channel fidelity which is closest to 1.  The values of $p$ inferred for the different measurement-based implementations of the identity operation are given in \tabref{tab:p}.  These values quantify the increase in depolarising noise resulting from increasing the length of the cluster state.  The values of $p$ are greatly reduced by applying quantum readout error mitigation, which suggests that classical measurement errors are responsible for a substantial amount of depolarising noise in the implementations.  Finally, we note that the values of $p$ inferred for the identity implementation with the 5-qubit cluster state agree very well with the values of $p$ inferred for the exact 3-design implementation with a 6-qubit cluster state on the same set of qubits.  This shows that our methods used to infer the values of $p$ are consistent.

%%%%%%%%%%%%%%%%%%%%%%%%%%%%
%%%%%%%%%%%%%%%%%%%%%%%%%%%%
%%%%%%%%%%%%%%%%%%%%%%%%%%%%
%%%%%%%%%%%%%%%%%%%%%%%%%%%%
\section{Qubits used for the 2-design}\label{append:qubits2} 

\begin{table}[]
\centering
\begin{minipage}{9.5cm}
\centering
\begin{tabular}{|l|l|l|l|l|}
\hline
\textbf{Qubit} & $\boldsymbol{T_1}\ \boldsymbol{(}\boldsymbol{\mu}\mathbf{s}\boldsymbol{)}$ & $\boldsymbol{T_2}\ \boldsymbol{(}\boldsymbol{\mu}\mathbf{s}\boldsymbol{)}$ & $\boldsymbol{\sqrt{X}}$ \textbf{Error} & \textbf{Readout Error} \\ \hline
13             & 154.36           & 162.92           & 0.000172         & 0.0097                 \\ \hline
14             & $\,\,\,$94.78           & 208.81           & 0.000206         & 0.0295                 \\ \hline
16             & $\,\,\,$97.54           & 121.99           & 0.001255         & 0.0198                 \\ \hline
19             & 102.87           & $\,\,\,$89.78           & 0.000382         & 0.0271                 \\ \hline
22             & 114.71           & 170.64           & 0.000233         & 0.0441                 \\ \hline
\end{tabular}
\end{minipage}
\begin{minipage}{6.5cm}
\centering
\begin{tabular}{|l|l|}
\hline
\textbf{Qubit Pair} & $\boldsymbol{C}\boldsymbol{X}$ \textbf{Error} \\ \hline
13--14              & 0.00576           \\ \hline
14--16              & 0.00621           \\ \hline
16--19              & 0.01155           \\ \hline
19--22              & 0.01168           \\ \hline
\end{tabular}
\end{minipage}
\caption{Calibration information for the \textit{ibmq\_sydney} quantum processor as obtained at the time of running the circuits for the approximate 2-design.  The single-qubit calibration information for the relevant qubits is shown on the left.  $T_1$ and $T_2$ are the relaxation and dephasing times respectively of the qubits.  The $CX$ error rates for relevant qubit pairs are shown on the right.}
\label{tab:calibration2}
\end{table}

The approximate measurement-based 2-design was implemented on 5 physical qubits of the \textit{ibmq\_sydney} quantum processor.  Its qubit topology is identical to that of the \textit{ibmq\_toronto} quantum processor shown in \figref{fig:topology}.  The qubits 1 to 5 of the 5-qubit linear cluster state in the 2-design implementation were mapped onto the physical qubits 13, 14, 16, 19 and 22 of the \textit{ibmq\_sydney} quantum processor, in such a way that the input state was prepared on qubit 13 and the output state was retrieved from qubit 22.  The relevant calibration information as obtained at the time of running the circuits for the approximate 2-design implementation is shown in \tabref{tab:calibration2}.  These 5 connected qubits (see \figref{fig:topology}) were chosen as they have unusually low error rates (in particular $CX$ error rates) --- much lower than the error rates of any 5 connected qubits on the \textit{ibmq\_toronto} quantum processor.  This is why the \textit{ibmq\_sydney} quantum processor was used for this investigation instead of the \textit{ibmq\_toronto} quantum processor.  The \textit{ibmq\_sydney} quantum processor was not considered for the exact 3-design implementation, as it does not have any 6 connected qubits with lower error rates than the 6 connected qubits of \textit{ibmq\_toronto} quantum processor.  In particular, the error rates of qubit 25, which is connected to the qubits used for the approximate 2-design implementation, are typically large.

%%%%%%%%%%%%%%%%%%%%%%%%%%%%
%%%%%%%%%%%%%%%%%%%%%%%%%%%%
%%%%%%%%%%%%%%%%%%%%%%%%%%%%
%%%%%%%%%%%%%%%%%%%%%%%%%%%%
\section{Qubits used for the identity}\label{append:qubitsidentity} 

\begin{table}[]
\centering
\begin{minipage}{9.5cm}
\centering
\begin{tabular}{|l|l|l|l|l|}
\hline
\textbf{Qubit} & $\boldsymbol{T_1}\ \boldsymbol{(}\boldsymbol{\mu}\mathbf{s}\boldsymbol{)}$ & $\boldsymbol{T_2}\ \boldsymbol{(}\boldsymbol{\mu}\mathbf{s}\boldsymbol{)}$ & $\boldsymbol{\sqrt{X}}$ \textbf{Error} & \textbf{Readout Error} \\ \hline
16             & 123.42           & 135.48           & 0.000297         & 0.0154                 \\ \hline
19             & 114.98           & 123.10           & 0.000434         & 0.0116                 \\ \hline
22             & 110.50           & 148.90           & 0.000330         & 0.0191                 \\ \hline
25             & 125.41           & 114.64           & 0.000323         & 0.0108                 \\ \hline
24             & 121.49           & 155.26           & 0.000180         & 0.0093                 \\ \hline
23             & $\,\,\,$98.92            & $\,\,\,$40.59            & 0.000441         & 0.0553                 \\ \hline
21             & $\,\,\,$76.60            & $\,\,\,$56.22            & 0.000508         & 0.0177                 \\ \hline
\end{tabular}
\end{minipage}
\begin{minipage}{6.5cm}
\centering
\begin{tabular}{|l|l|}
\hline
\textbf{Qubit Pair} & $\boldsymbol{C}\boldsymbol{X}$ \textbf{Error} \\ \hline
16--19              & 0.00758           \\ \hline
19--22              & 0.01024           \\ \hline
22--25              & 0.01053           \\ \hline
25--24              & 0.01099           \\ \hline
24--23              & 0.00892           \\ \hline
23--21              & 0.01652           \\ \hline
\end{tabular}
\end{minipage}
\caption{Calibration information for the \textit{ibmq\_toronto} quantum processor as obtained at the time of running the circuits for the identity implementation.  The single-qubit calibration information for the relevant qubits is shown on the left.  $T_1$ and $T_2$ are the relaxation and dephasing times respectively of the qubits.  The $CX$ error rates for relevant qubit pairs are shown on the right.}
\label{tab:calibrationidentity}
\end{table}

The identity operation was implemented on the \textit{ibmq\_toronto} quantum processor (see \figref{fig:topology} for the qubit topology) by performing single-qubit measurements on cluster states of different lengths.  Qubits 1 to 3 of the 3-qubit linear cluster state were mapped onto the physical qubits 16, 19 and 22, qubits 1 to 5 of the 5-qubit linear cluster state were mapped onto the physical qubits 16, 19, 22, 25 and 24 and qubits 1 to 7 of the 7-qubit linear cluster state were mapped onto the physical qubits 16, 19, 22, 25, 24, 23 and 21.  Qubits were chosen in this way to ensure maximum possible overlap with the qubits used for the exact 3-design implementation.  This allowed us to compare depolarising noise parameters inferred for the two implementations.  The relevant calibration information as obtained at the time of running the circuits for the identity implementation is shown in \tabref{tab:calibrationidentity}.

\end{document}